\documentclass[11pt]{article}

\usepackage[T1]{fontenc}
\usepackage{amsfonts, amsmath}
\usepackage{mathtools}
\usepackage{mathrsfs}
\usepackage[dvipsnames]{xcolor}
\usepackage[a4paper, left=2.5cm, right=2cm, top=2.5cm, bottom=2.5cm]{geometry}
\usepackage{tikz}
	\usetikzlibrary{external, positioning, arrows.meta, calc, decorations.markings, shapes.geometric}
	\tikzexternaldisable{} 
\usepackage{pgfplots}
	\pgfplotsset{compat=newest}
	\usepgfplotslibrary{colorbrewer, statistics, groupplots, fillbetween}
	\newlength{\figurewidth}
	\newlength{\figureheight}
\usepackage{pgfplotstable}
\usepackage{colortbl} 
\usepackage{tabularx}
\usepackage{booktabs}
\usepackage[version=4]{mhchem} 
\usepackage{algorithm}
\usepackage{algorithmicx}
\usepackage{algpseudocode}
	\algnewcommand\algorithmicforeach{\textbf{for each}}
	\algdef{S}[FOR]{ForEach}[1]{\algorithmicforeach\ #1\ \algorithmicdo}
	\algnewcommand{\IfThen}[2]{\State \algorithmicif\ #1\ \algorithmicthen\ #2}
	\algnewcommand{\LineComment}[1]{\State \textcolor{Cyan}{\texttt{/*\;#1\;*/}}}
	\algrenewcommand\algorithmiccomment[1]{\hfill \textcolor{Cyan}{\texttt{//\;#1}}}
	\algrenewcommand{\textproc}{\texttt}
	\newlength{\WidthOfInput}
	\newlength{\WidthOfOutput}
\usepackage[colorlinks=true]{hyperref} 
\hypersetup{
    linkcolor=RedViolet,
    urlcolor=OliveGreen,
    citecolor=PineGreen
}
\usepackage[capitalise, noabbrev, nameinlink]{cleveref}
	\crefname{equation}{equation}{equations} 
\usepackage{autonum} 
\usepackage{tcolorbox} 
    \tcbuselibrary{most}
\usepackage{siunitx}
	\DeclareSIUnit\atm{atm} 
\usepackage{lscape} 
\usepackage{soul} 
\usepackage{caption} 
\usepackage{authblk}
\usepackage[title]{appendix}
\usepackage{cite}

\newcommand{\bss}{{\boldsymbol{s}}}

\newcommand{\bsu}{{\boldsymbol{u}}}

\newcommand{\bsx}{{\boldsymbol{x}}}

\newcommand{\bszero}{{\boldsymbol{0}}}

\newcommand{\bsalpha}{{\boldsymbol{\alpha}}}

\newcommand{\bsnu}{{\boldsymbol{\nu}}}
\newcommand{\bsxi}{{\boldsymbol{\xi}}}

\newcommand{\bbR}{{\mathbb{R}}}

\newcommand{\N}{{\mathbb{N}}} 

\DeclareSymbolFont{bbold}{U}{bbold}{m}{n}
\DeclareSymbolFontAlphabet{\mathbbold}{bbold}

\newcommand{\calD}{{\mathcal{D}}}

\newcommand{\calI}{{\mathcal{I}}}
\newcommand{\calJ}{{\mathcal{J}}}

\newcommand{\calL}{{\mathcal{L}}}

\newcommand{\calN}{{\mathcal{N}}}

\newcommand{\calQ}{{\mathcal{Q}}}

\newcommand{\calU}{{\mathcal{U}}}

\newcommand{\calZ}{{\mathcal{Z}}}

\DeclareMathOperator{\cov}{Cov}

\tikzset{%
	>={Stealth[length=2mm, width=1.75mm]}, 
	default line/.style={%
		thick,
		line cap=round,
	},
	default dashed line/.style={%
		default line,
		dashed,
	},
	default marker line/.style={%
		default line,
		mark=*,
		mark size=0.75pt,
	},
	default dotted line/.style={%
		default line,
		dashed,
		dash pattern=on 0pt off 2\pgflinewidth
	},
}

\pgfkeys{%
	/pgf/number format/.cd,1000 sep={}
}

\pgfplotsset{%
	default axis/.style={%
		width=\figurewidth,
		height=\figureheight,
		major tick length={2pt},
		minor tick length={2pt},
		every tick/.style={black, line cap=round},
		ticklabel style={font=\scriptsize},
		x label style={font=\small},
		y label style={font=\small},
		legend style={%
			draw=none,
			font=\scriptsize,
			at={(1.03, 1)},
			anchor=north west,
			fill=none,
			legend cell align=left
		},
		cycle list/Set1,
	},
	default error bar/.style={%
		error bars/.cd,
		y dir=both,
		y explicit,
	},
	default scatter/.style={%
		scatter,
		only marks,
		scatter src=explicit,
		mark size=1.75pt,
		mark options={%
			line width=0.1pt
		}
	},
	/pgfplots/numbered legend/.style 2 args={%
		legend image code/.code={%
			\draw[##1,no markers]
				plot coordinates {%
					(0cm, 0cm)
					(0.6cm, 0cm)
			};
			\node[anchor=center, #1] at (0.3cm, 0cm) {\tiny #2};
		}
  	},
	/pgfplots/shorter legend/.style={%
		legend image code/.code={%
			\draw[mark repeat=2, mark phase=2, ##1]
				plot coordinates {%
					(0cm,0cm)
					(0.2cm,0cm)
					(0.4cm,0cm)
			};
		}
	}
}

\tcbset{%
	outer arc=2pt,
	arc=1pt,
	attach boxed title to top left={yshift=-3mm, xshift=3mm},
	enhanced,
	boxed title style={%
		top=3pt,
		bottom=3pt
	},
	fonttitle=\sffamily,
	top=12pt
}

\newtcolorbox[auto counter, number within=section]{todobox}[1][]{%
	colframe=NavyBlue,
	colback=NavyBlue!20,
	boxed title style={%
		colback=NavyBlue
	},
  	title=TO DO,
}

\newtcolorbox[auto counter, number within=section]{infobox}[1][]{%
	colframe=OliveGreen,
	colback=OliveGreen!20,
	boxed title style={%
		colback=OliveGreen
	},
	title=INFO,
}

\newtcolorbox[auto counter, number within=section]{warningbox}[1][]{%
	colframe=Red,
	colback=Red!20,
	boxed title style={%
		colback=Red
	},
  	title=WARNING,
}

\newcommand{\Centipede}{\texttt{Centipede}}
\newcommand{\Xolotl}{\texttt{Xolotl}}
\newcommand{\Nyx}{\texttt{Nyx}}
\newcommand{\Bison}{\texttt{BISON}}
\newcommand{\Marmot}{\texttt{MARMOT}}
\newcommand{\Moose}{\texttt{MOOSE}}
\newcommand{\libMesh}{\texttt{libMesh}}
\newcommand{\PETSc}{\texttt{PETSc}}

\newcommand{\dutheq}{D^{\text{theq}}_{\ce{U}}}
\newcommand{\dxetheq}{D^{\text{theq}}_{\ce{Xe}}}
\newcommand{\duirr}{D^{\text{irr}}_{\ce{U}}}
\newcommand{\dxeirr}{D^{\text{irr}}_{\ce{Xe}}}
\newcommand{\nonstoich}{\ce{UO}_{2 \pm x}}
\newcommand{\pOtwo}{p_{\ce{O2}}}

\newcommand{\captionsquare}[1]{%
    \raisebox{-1pt}{\protect\tikz{\protect\fill[#1] (0, 0) rectangle (0.75em, 0.75em);}}%
}

\newcommand{\captionlegend}[1]{%
    \raisebox{1.75pt}{\protect\tikz{\protect\draw[#1] (0, 0) -- (0.55, 0);}}%
}

\title{Bayesian calibration with summary statistics for the prediction of xenon diffusion in \ce{UO2} nuclear fuel}

\author[1]{Pieterjan Robbe}
\author[2]{David Andersson}
\author[1]{Luc Bonnet}
\author[1]{Tiernan Casey}
\author[2]{Michael W.\ D.\ Cooper}
\author[2]{Christopher Matthews}
\author[1]{Khachik Sargsyan}
\author[1]{Habib N. Najm}

\affil[1]{Sandia National Laboratories, Livermore, CA 94551, USA}
\affil[2]{Los Alamos National Laboratory, Los Alamos, NM 37996, USA}

\begin{document}

\maketitle

\begin{abstract}
    The evolution and release of fission gas impacts the performance of \ce{UO2} nuclear fuel. We have created a Bayesian framework to calibrate a novel model for fission gas transport that predicts diffusion rates of uranium and xenon in \ce{UO2} under both thermal equilibrium and irradiation conditions. Data sets are taken from historical diffusion, gas release, and thermodynamic experiments. These data sets consist invariably of summary statistics, including a measurement value with an associated uncertainty. Our calibration strategy uses synthetic data sets in order to estimate the parameters in the model, such that the resulting model predictions agree with the reported summary statistics. In doing so, the reported uncertainties are effectively reflected in the inferred uncertain parameters. Furthermore, to keep our approach computationally tractable, we replace the fission gas evolution model by a polynomial surrogate model with a reduced number of parameters, which are identified using global sensitivity analysis. We discuss the efficacy of our calibration strategy, and investigate how the contribution of the different data sets, taken from multiple sources in the literature, can be weighted in the likelihood function constructed as part of our Bayesian calibration setup, in order to account for the different number of data points in each set of data summaries. Our results indicate a good match between the calibrated diffusivity and non-stoichiometry predictions and the given data summaries. We demonstrate a good agreement between the calibrated xenon diffusivity and the established fit from Turnbull et al.\ (1982), indicating that the dominant uranium vacancy diffusion mechanism in the model is able to capture the trends in the data.
\end{abstract}

\section{Introduction}\label{sec:introduction}

Uranium dioxide (\ce{UO2}) is the fuel of choice in \emph{light water reactors} (LWRs), the most common type of nuclear power plant in use~\cite{olander2009}. Inside the reactor, uranium atoms fission into lighter elements, including noble gases such as xenon and krypton. The diffusion of these fission gas atoms, of which xenon atoms constitute the highest concentration, leads to significant performance concerns, as they cause a reduction of the fuel thermal conductivity, provoke fuel swelling, and contribute to a pressure buildup in the plenum, see, e.g.,~\cite{andersson2014,andersson2015}. It is therefore critical to better understand the behavior of these fission gases through modeling and simulation, especially in light of the recently developed new fuel types, such as \ce{Cr}-doped \ce{UO2}, see, e.g.,~\cite{cardinaels2012}.

\begin{figure}
    \centering
	\tikzsetnextfilename{diffusivity_regions}
	\tikzexternalenable
	\newcommand{\plotfilled}[5]{%
    \addplot[filled area, #1, domain=#2, name path=lower, forget plot] {log10(#3)};
    \addplot[filled area, #1, domain=#4, name path=upper, forget plot] {log10(#5)};
    \addplot[%
        #1, opacity=0.6, fill opacity=0.2,
        draw=none,
        legend image code/.code={ \draw[#1] (0.2cm,-0.08cm) rectangle (0.4cm,0.12cm);},
    ] fill between[of=lower and upper];
}

\newcommand{\plotfilledfrompoints}[5]{%
    \addplot[filled area, #1, name path=lower, forget plot] coordinates {(#2, #4) (#3, #4)};
    \addplot[filled area, #1, name path=upper, forget plot] coordinates {(#2, #5) (#3, #5)};
    \addplot[%
        #1, opacity=0.6, fill opacity=0.2,
        draw=none,
        legend image code/.code={ \draw[#1] (0.2cm,-0.08cm) rectangle (0.4cm,0.12cm);},
    ] fill between[of=lower and upper];
}

\setlength{\figurewidth}{8.5cm}
\setlength{\figureheight}{6cm}
\pgfplotstableread[header=false, col sep=comma, skip first n=4]{dat/diffusivity_regions.csv}{\data}
\begin{tikzpicture}[%
        data/.style={black, default marker line, only marks, mark options={fill=black}},
        filled area/.style={opacity=0.4, thick, mark=none, samples=50},
        D1/.style={SpringGreen},
        D2/.style={MidnightBlue},
        D3/.style={Red}
    ]
    \begin{axis}[%
            default axis,
            clip mode=individual,
            axis on top,
            axis x line*=bottom,
            xlabel={$10^4/T$ [\si[per-mode=symbol]{\per\kelvin}]},
            ylabel={$\log_{10}$ xenon diffusivity [\si[per-mode=symbol]{\square\metre\per\second}]},
            xmin=5, xmax=13,
            ymin=-21, ymax=-17,
            axis x line*=bottom,
            max space between ticks=25,
            legend style={at={(0.98,0.97)}, anchor=north east},
            legend columns=1,
            mark options={fill=white},
        ]

        \addplot[default dashed line, samples=50, domain=4:8.5] {log10((6.48e-7*exp(-3.7/8.61733e-5*x/1e4))/(1+1.82e4*exp(-1.45/8.61733e-5*x/1e4)))};
        \addlegendentry{\emph{Perriot et al.\ '19}}

        \addplot[data] table [x index={46}, y expr=log10(\thisrowno{47})] {\data};
        \addlegendentry{\emph{Turnbull et al.\ '89}}
        \addplot[data, forget plot] table [x index={48}, y expr=log10(\thisrowno{49})] {\data};
        \addplot[data, forget plot] table [x index={50}, y expr=log10(\thisrowno{51})] {\data};

        \addplot[default line, smooth] table [x index={52}, y expr=log10(\thisrowno{53})] {\data};
        \addlegendentry{Turnbull fit}

        \plotfilled{D1}{4:8}{7.6e-10*exp(-3.04/8.61733e-5*x/1e4)}{4:8.5}{6e-7*exp(-3.522/8.61733e-5*x/1e4)}
        \addlegendentry{$D_1$ regime}

        \plotfilled{D2}{5.5:10.4}{1.784E-15*exp(-13800*x/1e4)*2.5}{5.5:9.1}{1.784E-15*exp(-13800*x/1e4)/2.5}
        \addlegendentry{$D_2$ regime}

        \plotfilledfrompoints{D3}{7.6}{20.5}{-19.6726410656}{-20.468521083}
        \addlegendentry{$D_3$ regime}

    \end{axis}
    \begin{axis}[%
            default axis,
            clip mode=individual,
            hide y axis,
            axis x line*=top,
            xlabel = {Temperature [K]},
            xmin=5, xmax=13,
            xticklabel pos=top,
            axis x line*=top,
            xtick = {5.555555556,6.25,7.142857143,8.333333333,10,12.5},
            domain=5:13,
            xticklabels = {1800,1600,1400,1200,1000,800},
        ]
        \addplot[draw=none] {x}; 
    \end{axis}
\end{tikzpicture}
	\tikzexternaldisable

    \caption{Experimental data and data fit for xenon diffusion in \ce{UO2}. The intrinsic diffusivity $D_1$ is bounded by the fits from Davies \& Long~\cite{davies1963} and Matzke~\cite{matzke1980}. The intermediate and athermal regimes $D_2$ and $D_3$ are bounded by the re-evaluated data fit from Turnbull et al.~\cite{turnbull1989}. Calculations from Perriot et al.~\cite{perriot2019} for the intrinsic diffusivity are also included.}
    \label{fig:diffusivity_regions}
\end{figure}
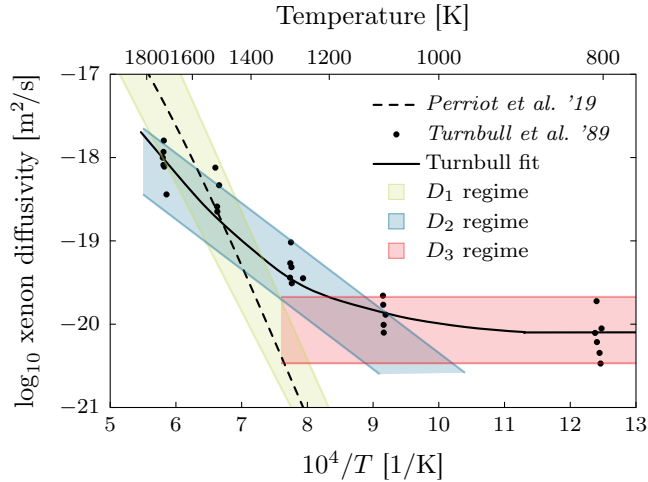

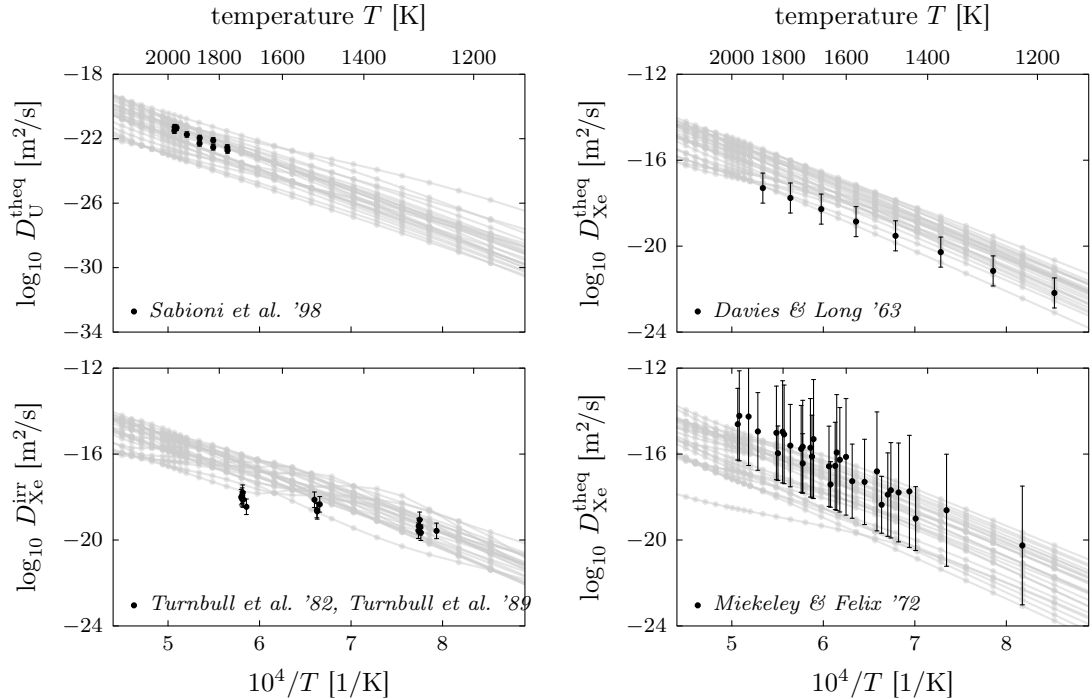
\begin{figure*}[t]
    \centering
	\tikzsetnextfilename{centipede_predictions_diffusivities}
	\tikzexternalenable
	\setlength{\figurewidth}{7cm}
\setlength{\figureheight}{5cm}
\begin{tikzpicture}[%
        sample/.style={default marker line, black!20, draw opacity=0.5},
        measurement/.style={black, default marker line, only marks},
    ]
    \begin{groupplot}[%
            group style={%
                group size=2 by 2,
                xticklabels at=edge bottom,
                xlabels at=edge bottom,
                vertical sep=\baselineskip,
                horizontal sep=2cm
            },
            default axis,
            axis on top,
            xtick={5, ..., 8},
            legend style={%
                at={(0.025, 0)},
                anchor=south west,
                /tikz/every odd column/.append style={column sep=1mm}
            },
            clip mode=individual,
            xlabel={$10^4/T$ [\si[per-mode=symbol]{\per\kelvin}]},
            axis x line*=bottom
        ]
        \nextgroupplot[%
            xmin=4.4, xmax=8.9, ymin=-34, ymax=-18,
            ytick={-34, -30, ..., -18},
            ylabel={$\log_{10}$ $\dutheq$ [\si[per-mode=symbol]{\square\metre\per\second}]},
	    ]
        \foreach \n in {1, ..., 25}{%
            \addplot[sample, forget plot] table[x index=0, y index=\n] {dat/Sabioni_predictions.dat};
        }
        \addplot[measurement, default error bar] table[x expr=1e4/\thisrowno{0}, y index=1, y error index=2] {dat/Sabioni_measurements.dat};
        \addlegendentry{\emph{Sabioni et al.\ '98}}
        \nextgroupplot[%
            xmin=4.4, xmax=8.9, ymin=-24, ymax=-12,
            ytick={-24, -20, ..., -12},
            ylabel={$\log_{10}$ $\dxetheq$ [\si[per-mode=symbol]{\square\metre\per\second}]}
        ]
        \foreach \n in {1, ..., 25}{%
            \addplot[sample, forget plot] table[x index=0, y index=\n] {dat/Davies_Long_predictions.dat};
        }
        \addplot[measurement, default error bar] table[x expr=1e4/\thisrowno{0}, y index=1, y error index=2] {dat/Davies_Long_measurements.dat};
        \addlegendentry{\emph{Davies \& Long '63}}
        \nextgroupplot[%
            xmin=4.4, xmax=8.9, ymin=-24, ymax=-12,
            ytick={-24, -20, ..., -12},
            ylabel={$\log_{10}$ $\dxeirr$ [\si[per-mode=symbol]{\square\metre\per\second}]}
        ]
        \foreach \n in {1, ..., 25}{%
            \addplot[sample, forget plot] table[x index=0, y index=\n] {dat/Turnbull_predictions.dat};
        }
        \addplot[measurement, default error bar] table[x expr=1e4/\thisrowno{0}, y index=1, y error index=2] {dat/Turnbull_measurements.dat};
        \addlegendentry{\emph{Turnbull et al.\ '82, Turnbull et al.\ '89}}
        \nextgroupplot[%
            xmin=4.4, xmax=8.9, ymin=-24, ymax=-12,
            ytick={-24, -20, ..., -12},
            ylabel={$\log_{10}$ $\dxetheq$ [\si[per-mode=symbol]{\square\metre\per\second}]}
        ]
        \foreach \n in {1, ..., 25}{%
            \addplot[sample, forget plot] table[x index=0, y index=\n] {dat/Miekely_Felix_predictions.dat};
        }
        \addplot[measurement, default error bar] table[x expr=1e4/\thisrowno{0}, y index=1, y error index=2] {dat/Miekely_Felix_measurements.dat};
        \addlegendentry{\emph{Miekeley \& Felix '72}}
    \end{groupplot}%
    \begin{groupplot}[%
            group style={%
                group size=2 by 2,
                xticklabels at=edge top,
                xlabels at=edge top,
                vertical sep=\baselineskip,
                horizontal sep=2cm
            },
            default axis,
            axis on top,
            axis y line=none,
            xlabel={temperature $T$ [\si{\kelvin}]},
            xticklabel pos=top,
            axis x line*=top,
            xmin=4.4, xmax=8.9,
            ymin=4.4, ymax=8.9,
            xtick={5, 5.56, 6.25, 7.14, 8.34},
        ]
        \nextgroupplot[%
            xticklabels={2000, 1800, 1600, 1400, 1200},
        ]
        \addplot[opacity=0] coordinates { (8.85,8.9) (8.9,8.85) }; 
        \nextgroupplot[%
            xticklabels={2000, 1800, 1600, 1400, 1200},
        ]
        \addplot[opacity=0] coordinates { (8.85,8.9) (8.9,8.85) }; 
        \nextgroupplot[%
        ]
        \addplot[opacity=0] coordinates { (8.85,8.9) (8.9,8.85) }; 
        \nextgroupplot[%
        ]
        \addplot[opacity=0] coordinates { (8.85,8.9) (8.9,8.85) }; 
    \end{groupplot}
\end{tikzpicture}
	\tikzexternaldisable

    \caption{Measurement values and associated errors for each set of experimental data summaries. Also shown are a set of 25 \Centipede{} predictions of the diffusivity quantity predicted by each experiment at 26 different temperatures, with input parameter values drawn from a uniform distribution between the corresponding lower and upper bounds of each, and under the operating conditions corresponding to each set of experimental data summaries, see \cref{tab:data_sets_overview}.}
    \label{fig:centipede_predictions_diffusivities}
\end{figure*}

The diffusion of xenon in \ce{UO2} nuclear fuel has been studied extensively by both experiments~\cite{matzke1980,turnbull1982,turnbull1989,rest2019} and simulation~\cite{catlow1978,jackson1986,grimes1991,yun2008,govers2010,moore2013,andersson2014,andersson2015,cooper2016,perriot2019}. From these studies, it is understood that fission gas release is a multi-stage process, with the diffusion of individual gas atoms, assisted by the damage produced by fission fragments, as the essential material property that defines the fission gas response of a particular nuclear fuel.

Most existing fission gas release models rely on the analysis for the bulk xenon diffusivity performed in~\cite{turnbull1982}. In this work, the fission gas diffusivity is divided into three different temperature ranges: $D_1$ (temperatures larger than \SI{1600}{\kelvin}), $D_2$ (temperatures between \SI{1600}{\kelvin} and \SI{1200}{\kelvin}) and $D_3$ (temperatures below \SI{1200}{\kelvin}). These three regimes are shown in \cref{fig:diffusivity_regions}.

In the high-temperature $D_1$ or \emph{intrinsic} regime, the diffusivity is dominated by the thermal defect concentrations. It is assumed that, in this regime, defects due to irradiation are quickly annealed and do not impact diffusion. In the intermediate-temperature $D_2$ regime, radiation-induced defect concentrations start to dominate over the intrinsic mechanism. In the low-temperature $D_3$ regime, the xenon diffusivity is driven directly by atomic mixing during radiation damage, exhibiting an athermal behavior. 

Several modelling attempts have been made to explain the behavior of xenon diffusion in \ce{UO2}, see, e.g.,~\cite{cooper2016,perriot2019,matthews2019,matthews2020}. Despite the progress made in these recent works, the precise mechanisms underlying the diffusion process are still being investigated. For example, the cluster dynamics simulations from~\cite{matthews2020} for the prediction of the xenon diffusivity under irradiation in the $D_2$ regime underpin a mechanistic diffusion model that describes the complex interactions between point defects in the \ce{UO2} lattice and individual xenon atoms. This model contains a total of 183 parameters, including reaction energies, binding energies, activation energies and attempt frequencies for all lattice defects. While a first-principles approach was used to develop this model, a significant uncertainty is associated with all of these parameters.

\begin{figure*}[t]
    \centering
    \begin{minipage}[t]{0.38\textwidth}
	\tikzsetnextfilename{centipede_predictions_stoichiometry_1}
	\tikzexternalenable
	\setlength{\figurewidth}{6.5cm}
\setlength{\figureheight}{5cm}
\pgfplotstableread[header=false]{dat/stoichiometry_predictions.dat}{\stoich}
\begin{tikzpicture}[%
	]
	\begin{axis}[%
			default axis,
			axis on top,
			legend style={%
				at={(0.025, 0)},
				anchor=south west,
				/tikz/every odd column/.append style={column sep=1mm}
			},
			clip mode=individual,
			xlabel={$10^4/T$ [\si[per-mode=symbol]{\per\kelvin}]},
			ylabel={$\log_{10}(\pOtwo)$ [\si{\atm}]},
			xtick={4, ..., 11},
			ytick={-25, -20, ..., -5},
			axis x line*=bottom,
			xmin=4, xmax=11,
			ymin=-25, ymax=-5,
			axis on top,
			axis x line*=bottom,
		]
		\addplot[only marks, fill=black, mark options={line width=0.5pt, mark size=1pt}] table[x index=0, y index=1] \stoich;
	\end{axis}
	\begin{axis}[%
			default axis,
			axis on top,
			axis y line=none,
			xlabel={temperature $T$ [\si{\kelvin}]},
			xmin=4, xmax=11,
			ymin=4, ymax=11,
			xticklabel pos=top,
			axis x line*=top,
			xtick={5, 6.25, 7.14, 8.34, 10},
			xticklabels={2000, 1600, 1400, 1200, 1000},
		]
		\addplot[opacity=0] coordinates { (10.95, 11) (11, 10.95) }; 
	\end{axis}
\end{tikzpicture}
	\tikzexternaldisable

    \end{minipage}\hspace{1em}%
    \begin{minipage}[t]{0.55\textwidth}
	\tikzsetnextfilename{centipede_predictions_stoichiometry_2}
	\tikzexternalenable
	\setlength{\figurewidth}{10cm}
\setlength{\figureheight}{5cm}
\begin{tikzpicture}[%
    measurement/.style={black, default marker line, only marks, mark options={line width=0.4pt}},
    sample/.style={black!20, default marker line, only marks, mark options={line width=0.4pt}},
    ]
    \begin{axis}[%
            default axis,
            axis on top,
            xmin=-1, xmax=104,
            legend style={%
                at={(1, 0)},
                anchor=south east,
                /tikz/every odd column/.append style={column sep=1mm},
                font={\normalfont},
            },
            xlabel={index of T/$\log_{10}(\pOtwo{})$ value},
            ylabel={non-stoichiometry $\nonstoich$},
            clip mode=individual,
            ymin=-0.006, ymax=0.006,
            ytick={-0.006, -0.004, ..., 0.006}
        ]
        \addplot[measurement, default error bar, y dir=plus] table[x expr=\coordindex, y index=0, y error index=1] {dat/stoichiometry_measurements.dat};
        \addplot[measurement, forget plot, default error bar, y dir=minus] table[x expr=\coordindex, y index=0, y error index=1] {dat/stoichiometry_measurements.dat};
    \end{axis}
\end{tikzpicture}
	\tikzexternaldisable

    \end{minipage}%
    \vspace{-\baselineskip}
    \caption{Measurement values and associated errors for the stoichiometric data from~\cite{markin1968,wheeler1971,wheeler1972,javed1972,aronson1958,une1983,une1982,hagemark1966}. The figure on the left illustrates how the index of $T$/$\log_{10}(\pOtwo{})$ is mapped to a $(T, \log_{10}(\pOtwo{}))$ pair, see~\cref{tab:stoichiometry_data_set}.}
    \label{fig:centipede_predictions_stoichiometry}
\end{figure*}
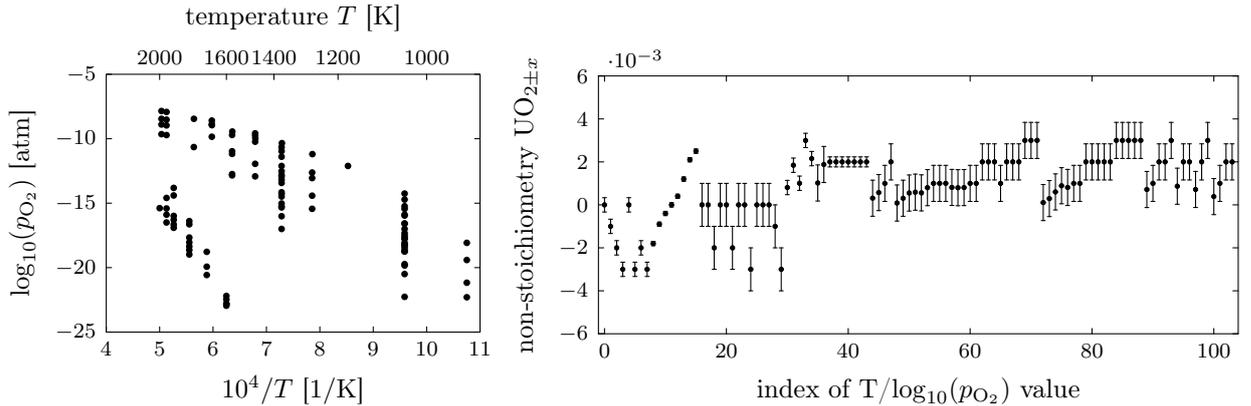

The cluster dynamics simulations in~\cite{matthews2020} were performed using \Centipede{}. \Centipede{} predicts the uranium diffusivity $\dutheq$ and xenon diffusivity $\dxetheq$ under thermal equilibrium conditions, as well as the uranium diffusivity $\duirr$ and xenon diffusivity $\dxeirr$ under irradiation, as a function of temperature, oxygen partial pressure, and fission rate. In order to predict the diffusion coefficients, \Centipede{} also models the fuel non-stoichiometry (i.e., the deviation $x$ of $\nonstoich$ from perfect stoichiometric \ce{UO2}). \Cref{fig:centipede_predictions_diffusivities} contains an illustration of the diffusivity predictions, and \cref{fig:centipede_predictions_stoichiometry} contains an illustration of the non-stoichiometry predictions. Note that the diffusivity predictions are plotted as a function of temperature $T$, while the non-stoichiometry predictions are shown as a function of both $T$ and oxygen partial pressure ($\pOtwo$).

In our calibration setup, the \Centipede{} outputs will be matched with the experimental data summaries taken from previous diffusion, gas release and thermodynamic experiments reported in the literature, in particular~\cite{sabioni1998,davies1963,turnbull1982,turnbull1989,markin1968,wheeler1971,wheeler1972,javed1972,aronson1958,une1983,une1982,hagemark1966}. These data sets are invariably reported as summary statistics, i.e., each data point contains a mean measurement value with an associated error. Again, we refer to \cref{fig:centipede_predictions_diffusivities} for an illustration. An overview of the different data sets is shown in \cref{tab:data_sets_overview}.

A Bayesian calibration methodology for this setting, in which only summary statistics are available, is the \emph{data-free inference} (DFI) approach. DFI has been introduced in~\cite{berry2012}, and was applied in various settings in, amongst others,~\cite{najm2014,chowdhary2016,khalil2017,casey2019}. In a context where data summaries are available, but the original data is not, DFI generates synthetic data on the experimentally observed outputs, that is consistent with the reported summary statistics when used to estimate model parameters using Bayesian inference. The procedure entails a nested inference scheme that evolves in both the data and parameter spaces. The computational complexity of the DFI procedure, and the need to build physically meaningful data-generating models for each experiment, renders its full utilization challenging in the present multi-experiment setting. We introduce a modification of the DFI method discussed in \cite{chowdhary2016}, that simplifies the synthetic data-generation process, and is computationally tractable.

Our approach considers the entire ensemble of data summaries across all experiments to learn the joint distribution of all uncertain parameters in the model. Since there may be a different number of data points in each set of data summaries, the calibration result may be dominated by data sets that contain a large number of measurements, such as the Miekeley \& Felix or stoichiometric data set, see~\cref{tab:data_sets_overview}. In order to avoid this, we propose a weighted approach where the contribution of each data set to the likelihood is normalized by the number of data points it contains.

To further reduce the computational burden, we replace the actual \Centipede{} evaluations by a computationally inexpensive surrogate model. In particular, we choose to fit a polynomial chaos expansion (PCE) surrogate model to the \Centipede{} outputs, see~\cite{ghanem1991,najm2009}.

\begin{table*}
    \centering
    \small
    \begin{tabular}{ccccccc} \toprule
        $d$ & experimental data summaries & reference(s) & measured quantity & $N_d$ & \texttt{Hf\_pO2} [\si{\electronvolt}] & \texttt{T0} [\si{\kelvin}] \\ \midrule
        1 & Sabioni et al. & \cite{sabioni1998} & $\dutheq$ & 10 & 5.10 & 1973 \\
        2 & Davies \& Long & \cite{davies1963} & $\dxetheq$ & 8 & 5.10 & 1973 \\
        3 & Turnbull et al. & \cite{turnbull1982,turnbull1989} & $\dxeirr$ & 16 & 5.10 & 1973 \\
        4 & Miekeley \& Felix & \cite{miekeley1972} & $\dxetheq$ & 32 & 6.11 & 1973 \\
        5 & stoichiometry & \cite{markin1968,wheeler1971,wheeler1972,javed1972,aronson1958,une1983,une1982,hagemark1966} & $\nonstoich$ & 104 & - & - \\ \bottomrule
    \end{tabular}%
    \vspace{-\baselineskip}
    \caption{Overview of the $D = 5$ different data sets in the calibration setup. The operating conditions \texttt{Hf\_pO2} and \texttt{T0} vary between the different experiments, and will be estimated in the present study. The listed values are current best estimates.}
    \label{tab:data_sets_overview}
\end{table*}

The remainder of this paper is organized as follows. First, in \cref{sec:simulation_of_fission_gas_diffusivity_in_UO2}, we provide more details on the cluster dynamics simulation code \Centipede{}. Next, in \cref{sec:bayesian_calibration_for_summary_statistics}, we outline our parameter estimation strategy. More details on the surrogate construction are provided in \cref{sec:polynomial_chaos_surrogate_construction_and_parameter_reduction}. In \cref{sec:results_and_discussion}, we report the results obtained by applying our calibration framework to characterize the uncertainty in the atomistic-scale model parameters of \Centipede{}, and provide an interpretation of these results. Finally, a conclusion and pointers to future work are given in \cref{sec:conclusion_and_further_work}.

\section{Simulation of fission gas diffusivity in \texorpdfstring{\ce{UO2}}{UO2}}\label{sec:simulation_of_fission_gas_diffusivity_in_UO2}

To compute the diffusivities under thermal equilibrium in the $D_1$ regime, an analytical point defect model, based on density functional theory (DFT) calculations for the energies and (semi-)empirical potential (EP) calculations for the entropies, was proposed in~\cite{perriot2019}. It was shown that, in this regime, the active diffusion mechanism is a vacancy mechanism, i.e., a single xenon atom occupying a cluster consisting of two uranium vacancies and one oxygen vacancy. The predicted xenon diffusivities depend on the non-stoichiometry of the fuel ($\nonstoich$), which, in turn, is governed by the prescribed $\pOtwo{}$. The resulting xenon diffusivity agrees reasonably well with the experimental data summaries due to Davies \& Long, see~\cite{davies1963}, which is considered to be the most accurate data for application in fuel performance codes in this high-temperature regime, see~\cite{turnbull1982,turnbull1989}.

The athermal diffusivity in the low-temperature $D_3$ regime has been estimated from molecular dynamics (MD) simulations for the atomic mixing induced by electronic stopping of fission fragments causing thermal spikes in~\cite{cooper2016}.

A model for the evolution of point defects and xenon clusters under irradiation in \ce{UO2} in the $D_2$ regime has been introduced in~\cite{matthews2020}, as the analytical point defect model that is valid under the thermal equilibrium conditions in the $D_1$ regime cannot be used under irradiation. The model is based on the free-energy cluster dynamics framework from~\cite{matthews2019}. To capture the non-equilibrium response due to irradiation, the creation of point defects due to irradiation, as well as the interaction of point defects or clusters of point defects with other defects (including their self-interactions) and the interaction with lattice sinks, must be modelled. \Centipede{}, the cluster dynamics simulator that implements the model from~\cite{matthews2019}, solves a set of coupled ordinary differential equations (ODEs) that determine the atom fraction or concentration $X_\delta$ of a defect $\delta$ at a given temperature $T$. The atom fraction $X_\delta$ of a defect $\delta$ satisfies
\begin{align}
    \frac{\mathrm{d} X_\delta}{\mathrm{d} t} &= \underbrace{\vphantom{\sum_{\gamma}}\;\;\;\;\;\;\dot{B}_\delta\;\;\;\;\;\;}_{\textrm{defect production}} + \underbrace{\sum_{\gamma} \dot{R}_{\delta, \gamma}(X_\delta, X_\gamma, T, G)}_{\textrm{interactions with other defects } \gamma} \label{eq:coupled_odes}\\
    &- \underbrace{\sum_{\sigma} \dot{S}_{\delta, \sigma}(X_\delta, X_\sigma, T, G)}_{\textrm{interactions with sinks } \sigma}, 
\end{align}
where $\dot{R}_{\delta, \gamma}$ is the reaction rate, $\dot{S}_{\delta, \sigma}$ is the sink rate, and $G$ is the free energy in the system, see~\cite[equation (1)]{matthews2020}. The free energy governs the direction of the reaction and provides a natural way to account for non-stoichiometry and other thermodynamic considerations. \Centipede{} solves the coupled set of reaction equations in~\eqref{eq:coupled_odes} for the condition where $\mathrm{d}X_\delta/\mathrm{d}t = 0$ for all $\delta$, i.e., the pseudo-steady state condition, which provides the concentration of all point defects, clusters of point defects, and xenon clusters. Note that each defect species is dependent on all other point defects, resulting in a system of coupled ODEs that rapidly grows as the number of different species used to describe the system is increased. It should also be noted that, without the presence of irradiation, the solution to~\eqref{eq:coupled_odes} reduces to the solution of the analytical model that is used to describe the thermal equilibrium case. The sought-after diffusivities may be calculated by considering the concentration and mobility of each individual cluster, see~\cite[equation (4)]{matthews2020}.

The reaction rates $\dot{R}_{\delta, \gamma}$ depend on the change in the chemical potential (or \emph{driving force}). This driving force can be formulated as a change in the free energy $G$. In order to calculate the necessary reaction rates, an extensive set of atomistic input parameters is required. This includes thermodynamic (binding energies \texttt{DFT} [\si{\electronvolt}] and entropies \texttt{S} [$k_B$]) and kinetic (activation energies \texttt{Q} [\si{\electronvolt}] and attempt frequencies \texttt{w} [\si{\THz}]) properties of xenon-vacancy clusters and interstitial defects. Lower and upper bounds for these parameters were determined based on the results reported in \cite{matthews2020,perriot2019}, and have been collected in \cref{tab:all_parameters_overview}.

Additionally, there are uncertainties in the \texttt{DFT} binding energies originating from corrections applied to the values obtained from DFT calculations of charged supercells, see~\cite{perriot2019}. In particular, two correction terms are added to each binding energy, that scale quadratically with the charge of the defect. These correction terms depend on the parameters \texttt{charge\_correction\_DFT} and \texttt{charge\_sq\_correction\_DFT}, respectively, see \cref{tab:all_parameters_overview}. The correction terms allow us to capture the systematic error in the binding energies.

Note that the formation energies of point defects are dependent on the $\pOtwo$ of the system. In~\cite{matthews2019}, a simple model was proposed for the temperature-dependent $\pOtwo$, that depends on two experimentally defined parameters: the enthalpy for the reaction controlling the oxygen potential (\texttt{Hf\_pO2}) and the temperature at which \ce{UO2} is assumed stoichiometric (\texttt{T0}) for the particular value of \texttt{Hf\_pO2}, see~\cite[equation (34)]{matthews2019}. Both \texttt{Hf\_pO2} and \texttt{T0} change the degree of non-stoichiometry $\nonstoich$, the former through the enthalpy and the latter via the entropy. When combined with the point defect formation energies, \texttt{Hf\_pO2} and \texttt{T0} describe an Arrhenius relation for $\pOtwo$ as function of temperature. This relation is controlled by the active oxygen buffering reaction in each experiment and, by extension, the details of the experimental setup, which are rarely reported. In our numerical results in~\cref{sec:results_and_discussion}, we will incorporate these two data-set-dependent parameters, and they are to be estimated along with the other parameters in the model.

Finally, we mention that \Centipede{} focuses on accurately capturing the intrinsic diffusivity of fission gas in the chemistry and irradiation response of point defects and relatively small defect clusters with up to about 10 uranium and 20 oxygen vacancies interacting with a single xenon atom. Any cluster larger than roughly 10 uranium vacancies, or containing more than one xenon atom, acts as an immobile sink, rather than as a mobile cluster, and consequently does not contribute to diffusion. This distinguishes it from the \Xolotl{} cluster dynamics code from~\cite{xolotl}, which simulates clusters of xenon atoms containing up to millions of atoms, needed to describe the full intra-granular behavior of fission gas. The coupling of \Centipede{} and \Xolotl{} to describe the full xenon-vacancy phase space in a single simulation is currently ongoing.

\section{Bayesian calibration for summary statistics}\label{sec:bayesian_calibration_for_summary_statistics}

In this section, we outline our Bayesian calibration strategy. Our goal is to perform Bayesian inference, that is, given the experimental data summaries $\calD$ and a prior $p(\bsnu)$ on the \Centipede{} model parameters $\bsnu \in \bbR^s$, with $s$ the number of parameters, we want to compute the posterior distribution $p(\bsnu | \calD)$, i.e., the probability density on the parameters $\bsnu$ given that we observe the experimental data summaries $\calD$. According to Bayes' rule, the prior and posterior are related through the likelihood function $\calL_\calD(\bsnu)$ as
\begin{equation}\label{eq:bayes}
    p(\bsnu | \calD) \propto \calL_\calD(\bsnu) p(\bsnu),
\end{equation}
where $\calL(\bsnu) \coloneqq p(\calD | \bsnu)$ expresses the likelihood of observing the data given the parameter values $\bsnu$. In \cref{sec:bayesian_calibration}, we will briefly discuss the different components of \cref{eq:bayes} in more detail, before outlining our calibration strategy in \cref{sec:bayesian_calibration_with_summary_statistics}, and formulating an expression for the likelihood in \cref{sec:full_likelihood_construction}.

\subsection{Bayesian calibration}\label{sec:bayesian_calibration}

We start by formalizing the available information shown in \cref{fig:centipede_predictions_diffusivities,fig:centipede_predictions_stoichiometry}. The data summaries $\calD$ are composed of $D = 5$ sets of experimental data summaries $\calD_d$, $d = 1, 2, \ldots, D$, with $N_d$ measurement stations $\bsx_d^{(n)}$, $n = 1, 2, \ldots, N_d$, in each set. These measurement stations $\bsx_d^{(n)}$ correspond to different temperatures $T$ and/or $\pOtwo{}$ values. An overview of these data summaries is shown in \cref{tab:data_sets_overview}. Each set $\calD_d$ consists of mean values $y_d^{(n)}$ with associated uncertainties $s_d^{(n)}$, defined at the $N_d$ measurement stations $\bsx_d^{(n)}$, $n = 1, 2, \ldots, N_d$, i.e.,
\begin{equation}\label{eq:experimental_data_sets}
    \calD_d = \{ \bsx_d^{(n)}, y_d^{(n)}, s_d^{(n)} \}_{n=1}^{N_d}, \quad d = 1, 2, \ldots, D
\end{equation}
and $\calD = \{ \calD_d \}_{d=1}^D$.

Let $f^{\textrm{true}}_d(\bsx)$ be the true model that provides the exact values of the quantity being measured in $\calD_d$. The true values $f^{\textrm{true}}_d(\bsx_d^{(n)})$ are not directly accessible, as observations are corrupted by noise. For example, in the case of additive Gaussian noise, one may assume that the measured data $q_d^{(n)}$ are available as
\begin{equation}\label{eq:model_output_with_noise}
    q_d^{(n)} \coloneqq f^{\textrm{true}}_d(\bsx_d^{(n)}) + \sigma_d^{(n)} \eta_d^{(n)},
\end{equation}
where $\sigma_d^{(n)}$, $n=1, 2, \ldots, N_d$, is the standard deviation of the noise of the measurement at station $\bsx_d^{(n)}$, and where the $\eta_d^{(n)}$ are drawn from a standard normal distribution, i.e., $\eta_d^{(n)} \sim \calN(0, 1)$. Given the noisy observations $q_d^{(n)}$ of $f^{\textrm{true}}_d(\bsx_d^{(n)})$, we may use this information, along with a prior distribution $p(\bsnu)$, to estimate the parameters $\bsnu$ in an \emph{assumed model} $f_d(\bsx, \bsnu)$. In our setting, this model $f_d(\bsx, \bsnu)$ will be the \Centipede{} predictions of the quantity being measured in the $d$th data set.

In particular, suppose we are given a set of observations $\calQ_d \coloneqq \{ q_d^{(n)} \}_{n = 1}^{N_d}$ at $\bsx_d^{(n)}$, $n=1, 2, \ldots, N_d$. If we assume that the observation errors $\sigma_d^{(n)} \eta_d^{(n)}$ are independent, then the likelihood takes the form
\begin{align}\label{eq:gaussian_likelihood}
    \calL_{\calQ_d}(\bsnu) &\coloneqq p(\calQ_d | \bsnu) \\ &= \prod_{n=1}^{N_d} \frac{1}{\sigma_d^{(n)}\sqrt{2\pi}} \exp\mathopen{}\left(-\frac{\left(q_d^{(n)} - f_d(\bsx_d^{(n)}, \bsnu)\right)^2}{2{\sigma_d^{(n)}}^2}\mathclose{}\right).
\end{align}
Equivalently, the log-likelihood becomes
\begin{align}
    \log \calL_{\calQ_d}(\bsnu) &= \\ &\hspace{-0.5cm}-\frac{1}{2} \sum_{n=1}^{N_d} \left[ \log(2\pi{\sigma_d^{(n)}}^2) + \frac{\left(q_d^{(n)} - f_d(\bsx_d^{(n)}, \bsnu)\right)^2}{{\sigma_d^{(n)}}^2} \right].
\end{align}

A full log-likelihood for the calibration problem that involves observations $\calQ \coloneqq \{\calQ_d\}_{d=1}^D$ from $D$ independent experiments can simply be constructed as
\begin{equation}
    \log \calL_\calQ(\bsnu) = \sum_{d = 1}^D \log \calL_{\calQ_d}(\bsnu),
\end{equation}
and samples of the posterior $p(\bsnu | \calQ)$ can then be obtained by Markov chain Monte Carlo (MCMC), see, e.g.,~\cite{brooks2011} and \cref{sec:markov_chain_monte_carlo}. 

In our current setup, however, the observations $\calQ$ are generally not available. Instead, we are given \emph{summary statistics}, such as a mean or a standard deviation of processed observations, as indicated in \cref{eq:experimental_data_sets}. A DFI framework where data summaries involving measurement values and associated error bars on model outputs are available has been proposed in \cite{chowdhary2016}. In this work, two interpretations of error bars were discussed. In a first interpretation, the error bars are considered as quantifying the degree of error or scatter in the measurements, while, in the second interpretation, they are considered as quantifying the resulting uncertainty in the measured quantity. Here, we interpret the error bars as the latter type, which we believe to be a more natural interpretation of the reported measurements. 

The DFI method provides a joint posterior density on both the data and model parameters by enforcing consistency between the reported summary statistics and the statistics of the data~\cite{jaynes1957, jaynes1957a}. The algorithm entails a nested sampling procedure, with an outer   MCMC chain on the data space, and an inner MCMC chain on the parameter space. At each step of the outer chain, a new data set is proposed, after which the inner chain is executed in full to provide samples of the associated posterior, which are then used to check for consistency of the proposed data set with the reported summary statistics. Each accepted data set provides a consistent posterior on the model parameters. The final \emph{pooled posterior} is ultimately obtained by combining the posteriors for all consistent data sets.

We propose a simplified DFI construction that avoids the nested MCMC chain construction of the original scheme, thus increasing the computational efficiency. We presume a Gaussian distribution for the synthetic data on the quantity of interest at each measurement station. For each data set, these distributions are scaled such that statistics derived from the model predictions approximate the reported error bars. We then define a consistent data set as one for which the statistics computed from the data are, up to a tolerance, equal to the reported summary statistics. We illustrate below that this consistency metric, along with the presumed Gaussian distribution for the synthetic data, replaces the outer inference problem in DFI by an optimization procedure, which is more computationally tractable. The use of this presumed distribution distinguishes our approach from~\cite{chowdhary2016}.

\subsection{Bayesian calibration with summary statistics}\label{sec:bayesian_calibration_with_summary_statistics}

\begin{figure}
    \centering
	\tikzsetnextfilename{synthetic_data_set}
	\tikzexternalenable
	\setlength{\figurewidth}{8cm}
\setlength{\figureheight}{6cm}
\begin{tikzpicture}[%
        measurement/.style={black, default marker line, only marks},
        synthetic/.style={only marks, mark size=1pt},
    ]
    \begin{axis}[%
            default axis,
            axis on top,
            xmin=4.4, xmax=8.9, ymin=-24, ymax=-14,
            xtick={5, ..., 8},
            xlabel={$10^4/T$ [\si[per-mode=symbol]{\per\kelvin}]},
            ylabel={$\log_{10}$ $\dxetheq$ [\si[per-mode=symbol]{\square\metre\per\second}]},
            legend style={%
                at={(0.025, 0)},
                anchor=south west,
                /tikz/every odd column/.append style={column sep=1mm}
            },
            clip mode=individual,
            cycle list name=color list,
            enlargelimits=false,
            axis on top,
            axis x line*=bottom,
        ]
        \addplot[measurement, default error bar] table[x expr=1e4/\thisrowno{0}, y index=1, y error index=2] {dat/Davies_Long_measurements.dat};
        \addlegendentry{\emph{Davies \& Long '63}}
        \foreach \color [count=\n] in {ForestGreen, Red, CornflowerBlue, RoyalPurple, YellowOrange} {
            \edef\temp{%
                \noexpand\addplot[synthetic, color=\color, x filter/.code=\noexpand\pgfmathparse{\noexpand\pgfmathresult+0.1}] table[x index=0, y index=3\n] {dat/synthetic_data_set.dat};
            }
            \temp
        }
    \end{axis}
	\begin{axis}[%
            default axis,
            axis on top,
            axis y line=none,
            xlabel={temperature $T$ [\si{\kelvin}]},
            xmin=4, xmax=11,
            ymin=4, ymax=11,
            xticklabel pos=top,
            axis x line*=top,
            xtick={5, 6.25, 7.14, 8.34, 10},
            xticklabels={2000, 1600, 1400, 1200, 1000},
        ]
        \addplot[opacity=0] coordinates { (10.95, 11) (11, 10.95) }; 
    \end{axis}
\end{tikzpicture}
	\tikzexternaldisable

    \caption{Example of synthetic data sets for the Davies \& Long experimental data summaries using $K_d = 5$ and $\beta_d = 1$. Different colors indicate different values for $k$, $1 \leq k \leq K_d$.}
    \label{fig:synthetic_data_set}
\end{figure}
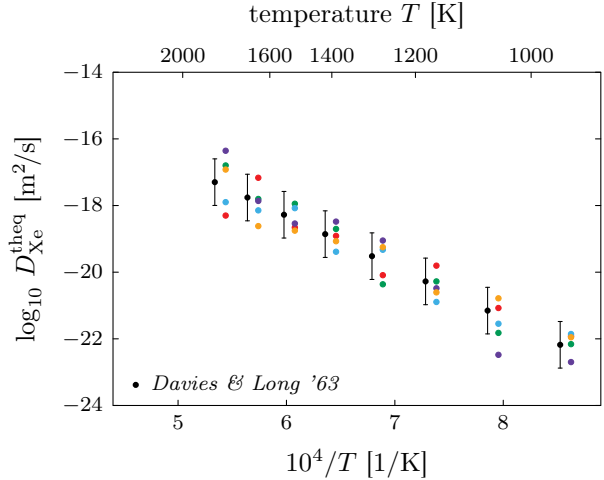

We start by defining a collection of $K_d$ synthetic data sets $\calZ_d^{(k)} \coloneqq \{ z_d^{(n, k)} \}_{n=1}^{N_d}$, $k=1, 2, \ldots, K_d$, where
\begin{equation}\label{eq:synthetic_data_sets}
    z_d^{(n, k)} \sim \calN\left(y_d^{(n)}, \beta_d {s_d^{(n)}}^2 \right),
\end{equation}
with $y_d^{(n)}$ and $s_d^{(n)}$ the measurement value and error respectively, see~\eqref{eq:experimental_data_sets}, and where $\beta_d > 0$ is a scale factor for the variance. Hence, the data set $\calZ_d^{(k)}$ contains synthetic observations that are sampled from a Gaussian distribution centered at $y_d^{(n)}$ for each $n=1, 2, \ldots, N_d$, and with a variance that can be tuned by choosing appropriate values for $\beta_d$. An example of a collection of synthetic data sets $\calZ_d^{(k)}$ for the Davies \& Long experimental data summaries with $K_d = 5$ and $\beta_d = 1$ is shown in \cref{fig:synthetic_data_set}.

Each synthetic data set $\calZ_d^{(k)}$, $k = 1, 2, \ldots, K_d$, represents an opinion about the true posterior through its corresponding log-likelihood
\begin{align}
    \log \calL_{\calZ_d}^{(k)}(\bsnu) &\coloneqq \\ &\hspace{-1cm}-\frac{1}{2} \sum_{n=1}^{N_d} \left[ \log(2\pi\beta_d{s_d^{(n)}}^2) + \frac{\left(z_d^{(n, k)} - f_d(\bsx_d^{(n)}, \bsnu)\right)^2}{\beta_d{s_d^{(n)}}^2} \right].
\end{align}

These $K_d$ different opinions can be combined using \emph{logarithmic pooling}, see~\cite{berry2012}. This can be accomplished by gathering all synthetic data sets $\calZ_d^{(k)}$ into a single data set $\calZ_d \coloneqq \{ \calZ_d^{(k)} \}_{k=1}^{K_d}$, and setting up an inference problem that uses an averaged log-likelihood
\begin{align}
    \log \calL_{\calZ_d}(\bsnu) &\coloneqq \frac{1}{K_d} \sum_{k=1}^{K_d} \log \calL_{\calZ_d}^{(k)}(\bsnu) \\
    &= -\frac{1}{2} \sum_{n=1}^{N_d} \left[ \vphantom{\frac{1}{K_d \beta_d{s_d^{(n)}}^2 } \sum_{k=1}^{K_d}} \log(2\pi\beta_d{s_d^{(n)}}^2) \;+ \right. \\ & \hspace{0.75cm} \left. \frac{1}{K_d \beta_d{s_d^{(n)}}^2 } \sum_{k=1}^{K_d} \left(z_d^{(n, k)} - f_d(\bsx_d^{(n)}, \bsnu)\right)^2 \right].
\end{align}
Logarithmic average pooling has a number of desirable properties over simple linear pooling. The latter uses an arithmetic averaging of the posteriors $p(\bsnu | \calZ_d^{(k)})$. We refer to~\cite{genest1986} for more details.

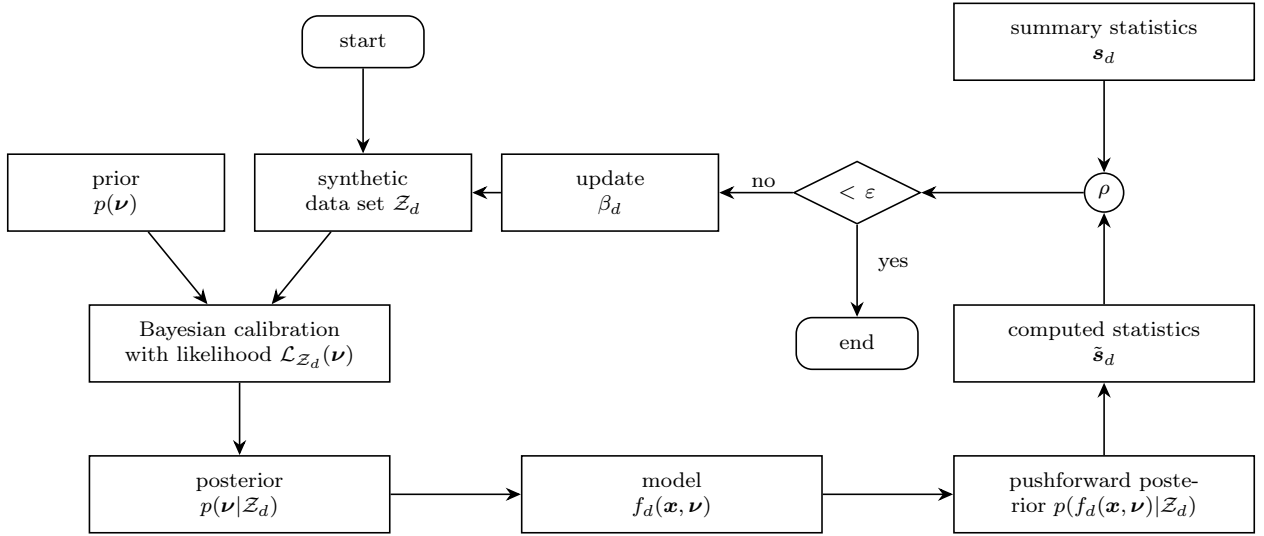
\begin{figure*}
    \centering
    \small
	\tikzsetnextfilename{calibration_framework}
	\tikzexternalenable
	\newlength\nodehdist
\setlength{\nodehdist}{3.25cm}
\newlength\nodevdist
\setlength{\nodevdist}{2cm}
\begin{tikzpicture}[%
        node distance=\nodevdist and \nodehdist,
        on grid,
        every node/.style={%
            text width=3.6cm,
            align=center,
            draw,
            semithick,
            inner sep=5pt,
            execute at end node={\vphantom{\strut}},
            execute at begin node={\vphantom{\strut}},
            font=\scriptsize
        },
        startstop/.style={%
            rounded corners=6pt,
            text width=1.25cm
        },
        narrower/.style={%
            text width=2.5cm
        },
        arrow/.style={%
            ->,
            semithick
        },
        small/.style={%
            text width=1cm,
            draw=none,
            inner sep=-1pt
        },
    ]
    \node[startstop] (start) {start};
    \node[below=of start, narrower] (data) {synthetic data set $\calZ_d$};
    \node[left=of data, narrower] (prior) {prior \\ $p(\bsnu)$};

    \node[right=2\nodehdist of data, diamond, aspect=2, text width=1cm, inner sep=0pt] (consistent) {$< \varepsilon$};
    \node[below=of consistent, startstop] (stop) {end};
    \node[above right=of consistent] (summary) {summary statistics \\ $\bss_d$};
    \node[below right=of consistent] (computed) {computed statistics \\ $\tilde\bss_d$};
    \node[right=of consistent, circle, text width=0.25cm, inner sep=1pt] (-) {\raisebox{1.5pt}{$\rho$}};
    \node[below=of computed] (pushforward) {pushforward posterior $p(f_d(\bsx, \bsnu) | \calZ_d)$};
    \node[right=of data, narrower] (beta) {update \\ $\beta_d$};

    \coordinate (halfway) at ($(prior)!0.5!(data)$);
    \coordinate (bayesian) at (halfway |- computed);
    \node[] (likelihood) at (bayesian) {Bayesian calibration with likelihood $\calL_{\calZ_d}(\bsnu)$};
    \node[below=of likelihood] (posterior) {posterior \\ $p(\bsnu | \calZ_d)$};
    \node[] (centipede) at ($(posterior)!0.5!(pushforward)$) {model \\ $f_d(\bsx, \bsnu)$};

    \draw[arrow] (start) -- (data);
    \draw[arrow] (data) -- (likelihood);
    \draw[arrow] (prior) -- (likelihood);
    \draw[arrow] (likelihood) -- (posterior);
    \draw[arrow] (posterior) -- (centipede);
    \draw[arrow] (centipede) -- (pushforward);
    \draw[arrow] (pushforward) -- (computed);
    \draw[arrow] (computed) -- (-);
    \draw[arrow] (summary) -- (-);
    \draw[arrow] (-) -- (consistent);
    \draw[arrow] (consistent) -- node[anchor=west, small,pos=0.4] {yes} (stop);
    \draw[arrow] (consistent) -- node[anchor=south, small, pos=0.4] {no} (beta);
    \draw[arrow] (beta) -- (data);
\end{tikzpicture}
	\tikzexternaldisable

    \caption{Bayesian calibration framework used to generate a consistent synthetic data set $\calZ_d$ according to \cref{alg:simplified_dfi}.}
    \label{fig:calibration_framework}
\end{figure*}

Once we obtain the posterior density $p(\bsnu | \calZ_d)$ on the model parameters, samples from the posterior can be propagated through the assumed forward model $f_d(\bsx, \bsnu)$, in order to obtain samples from the \emph{pushforward posterior} density $p(f_d(\bsx, \bsnu) | \calZ_d)$. The pushforward posterior is the target of the forward uncertainty quantification process given the parameter posterior $p(\bsnu | \calZ_d)$.

From $p(f_d(\bsx, \bsnu) | \calZ_d)$, we can estimate statistics on model outputs which can be compared to the reported summary statistics $s_d^{(n)}$ to decide whether the proposed data set $\calZ_d$ is consistent. In what follows, let $\bss_d \coloneqq \{s_d^{(n)}\}_{n=1}^{N_d}$ be the reported summary statistics in the $d$th experiment, and $\tilde\bss_d \coloneqq \{\tilde{s}_d^{(n)}\}_{n=1}^{N_d}$ be the corresponding statistics computed from the pushforward posterior density of the $d$th set of experimental data summaries at each measurement location. We may be interested in, for example, the standard deviation of the pushforward posterior, in which case the $\tilde{s}_d^{(n)}$ may correspond to the sample standard deviations at the measurement stations $\bsx_d^{(n)}$. We define a consistent data set $\calZ_d$ to be a data set that satisfies
\begin{equation}\label{eq:dfi_constraint}
    \rho(\bss_d, \tilde\bss_d) \le \varepsilon
\end{equation}
for a given distance metric $\rho$ and given tolerance $\varepsilon > 0$. Consistency may be satisfied by choosing an appropriate value for the scale factor $\beta_d$ in~\eqref{eq:synthetic_data_sets}. Hence, a consistent data set can be found by solving a one-dimensional optimization problem where we look for a value of $\beta_d$ and corresponding data set $\calZ_d$ that generates a pushforward posterior density for which the computed statistics $\tilde\bss_d$ satisfy~\eqref{eq:dfi_constraint}. The complete process for generating a consistent data set $\calZ_d$ is given in \cref{alg:simplified_dfi} and shown schematically in \cref{fig:calibration_framework}.

In each step of the iterative procedure, we generate a proposed synthetic data set $\calZ_d$ according to \cref{eq:synthetic_data_sets} using the current value of $\beta_d$. Next, we set up an inference problem to compute samples from the posterior $p(\bsnu | \calZ_d)$. These samples are propagated through the forward model $f_d(\bsx, \bsnu)$ for each measurement station $\bsx_d^{(n)}$. After that, we compute the desired statistic from the set of pushed forward samples, and compare the computed statistics to the reported summary statistics. This process is repeated until the statistics extracted from the data are consistent with the reported summary statistics in the sense of \cref{eq:dfi_constraint}. A crucial step in the algorithm is the update of the scaling factor $\beta_d$. Since the computed statistics $\tilde\bss_d$ depend on the chosen set of posterior samples, it is a random quantity. Hence, in order for \cref{alg:simplified_dfi} to converge, we propose to use a stochastic optimizer to update the value of $\beta_d$, see, e.g.,~\cite{spall2012}. However, we find numerically that, in our application, the objective function in~\eqref{eq:dfi_constraint} changes only mildly with a change in the choice for the set of posterior samples, provided that $K_d$ and $M$ in \cref{alg:simplified_dfi} are large enough. Therefore, a reasonable approximation for a consistent data set that satisfies \cref{eq:dfi_constraint} can be obtained by evaluating the objective function for a set of appropriately-chosen scaling parameters $\beta_d$, and by selecting the value of $\beta_d$ that resulted in the smallest value of $\rho(\bss_d, \tilde\bss_d)$ across all candidates. This is the strategy we adopt in our numerical results below.

\begin{algorithm*}
    \small
    \begin{algorithmic}[1]
        \Statex \textbf{input:} prior $p(\bsnu)$, number of synthetic data sets $K_d$, starting value for $\beta_d$,
        \Statex \hspace*{\WidthOfInput}MCMC chain starting values $\bsnu_0$, number of MCMC iterations $M$, statistic $w$, 
        \Statex \hspace*{\WidthOfInput}distance metric $\rho$, target tolerance $\varepsilon$, reported summary statistics $\bss_d$
        \Statex \textbf{output:} a set of consistent synthetic data sets $\calZ_d^{(k)}, k = 1, 2, \ldots, K_d$
        \Statex
        \Procedure{\textsf{generate\_consistent\_data}}{$p(\bsnu), K_d, \beta_d, \bsnu_0, M, w, \varepsilon, \bss_d$}
            \Repeat
                \State generate synthetic data $\calZ_d^{(k)}, k = 1, 2, \ldots, K_d$, according to \cref{eq:synthetic_data_sets} 
                \State $\{\bsnu_m\}_{m=1}^M \gets \textsf{MCMC}(p(\bsnu), \calL_{\calZ_d}(\bsnu), \bsnu_0, M)$ \Comment{obtain samples from the posterior $p(\bsnu | \calZ_d)$} 
                \For {$n = 1, 2, \ldots, N_d$} \Comment{loop over all measurement stations}
                    \For {$m = 1, 2, \ldots, M$} \Comment{loop over all posterior samples}
                        \State $\mathsf{f}_m \gets f(\bsx_d^{(n)}, \bsnu_m)$ \Comment{evaluate the forward model at the posterior samples}
                    \EndFor
                    \State $\tilde{s}_d^{(n)} \gets w(\mathsf{f}_1, \mathsf{f}_2, \ldots, \mathsf{f}_M)$ \Comment{compute the desired statistic}
                \EndFor
                \State update the value for $\beta_d$
            \Until $\rho(\bss_d, \tilde\bss_d) \le \varepsilon$ \Comment{evaluate \cref{eq:dfi_constraint}}
        \EndProcedure
    \end{algorithmic}
    \caption{Generating consistent data}
    \label{alg:simplified_dfi}
\end{algorithm*}

\subsection{Full likelihood construction}\label{sec:full_likelihood_construction}

Once we have obtained consistent synthetic data sets $\calZ_d$ for each experiment $d = 1, 2, \ldots, D$, we combine them in a single data set $\calZ = \{ \calZ_d \}_{d=1}^D$ and set up a final inference problem with log-likelihood
\allowdisplaybreaks
\begin{align}
    \log \calL_{\calZ}(\bsnu) &\coloneqq \sum_{d=1}^{D} \log \calL_{\calZ_d}(\bsnu) \label{eq:dfi_likelihood}\\
    &= -\frac{1}{2} \sum_{d=1}^D \sum_{n=1}^{N_d} \left[ \vphantom{\frac{1}{K_d \beta_d{s_d^{(n)}}^2} \sum_{k=1}^{K_d}} \log(2\pi\beta_d{s_d^{(n)}}^2) \;+ \right. \\ & \hspace{0.75cm}\left. \frac{1}{K_d \beta_d{s_d^{(n)}}^2} \sum_{k=1}^{K_d} \left(z_d^{(n, k)} - f_d(\bsx_d^{(n)}, \bsnu)\right)^2 \right].
\end{align}

It is possible to generalize our proposed likelihood in~\eqref{eq:dfi_likelihood} by using different \emph{weights} for each data set. These weights allow us to express various degrees of confidence in the respective experiments. Suppose we have a set of nonzero positive weights $\bsalpha = \{\alpha_d\}_{d=1}^D$ that sum to 1. These weights can be used to update the final log-likelihood as
\begin{align}
    \log \calL_{\calZ, \bsalpha}(\bsnu) &\coloneqq -\frac{D}{2} \sum_{d=1}^D \alpha_d \sum_{n=1}^{N_d} \left[ \vphantom{\frac{1}{K_d \beta_d{s_d^{(n)}}^2} \sum_{k=1}^{K_d}} \log(2\pi\beta_d{s_d^{(n)}}^2) \;+ \right. \\ &\left. \frac{1}{K_d \beta_d{s_d^{(n)}}^2} \sum_{k=1}^{K_d} \left(z_d^{(n, k)} - f_d(\bsx_d^{(n)}, \bsnu)\right)^2 \right].\label{eq:weighted_likelihood}
\end{align}
We may use these weights, for example, to account for the different number of measurement stations $N_d$ in each data set. In that case, the weights could be chosen as
\begin{equation}\label{eq:weights}
    \alpha_d \coloneqq \frac{N_d^{-1}}{\sum_{d=1}^D N_d^{-1}}.
\end{equation}

\section{Polynomial chaos surrogate construction and parameter reduction}\label{sec:polynomial_chaos_surrogate_construction_and_parameter_reduction}

The calibration approach outlined in \cref{sec:bayesian_calibration_for_summary_statistics} relies heavily on the ability to evaluate the likelihood function $\calL(\bsnu)$ in~\eqref{eq:dfi_likelihood} or~\eqref{eq:weighted_likelihood}, and thus also on the ability to evaluate the assumed model $f_d(\bsx, \bsnu)$ which, in our setting, is \Centipede{}. To avoid excessive computational costs, we propose to use a surrogate model that is inexpensive to evaluate and replaces \Centipede{} in the calibration loop. In this work, we focus on \emph{polynomial chaos expansion} (PCE) surrogate models. Given the prohibitively large number of samples required for constructing accurate surrogates in high dimensions, we use a \emph{global sensitivity analysis} (GSA) to identify a reduced set of parameters. This allows us to construct a more accurate surrogate in this lower-dimensional space. The sensitivity analysis will rank the parameters according to their relative effect on the variance of the output, allowing a down-selection based on the fractional contribution of each parameter to the total output variance. 

We will briefly recall the PCE surrogate model construction process in \cref{sec:polynomial_chaos_expansions}, and discuss dimension reduction using GSA in \cref{sec:dimension_reduction_using_global_sensitivity_analysis}.

\subsection{Polynomial chaos expansions}\label{sec:polynomial_chaos_expansions}

A PCE surrogate model $\tilde{f}_d^{(n)}(\bsnu)$ for the \Centipede{} prediction $f_d(\bsx_d^{(n)}, \bsnu)$ at measurement station $\bsx_d^{(n)}$ can be defined as
\begin{equation}\label{eq:pce}
    \tilde{f}_d^{(n)}(\bsnu) \coloneqq \sum_{\bsu \in \calI_d^{(n)}} c_{d, \bsu}^{(n)} \Phi_\bsu(\bsxi),
\end{equation}
where $\bsu = (u_1, u_2, \ldots, u_s) \in \N_0^s$ is a multi-index of length $s$, $\calI_d^{(n)}$ is a set of multi-indices, $\Phi_\bsu$ is a multivariate orthogonal polynomial expressed in terms of the i.i.d.\ random variables $\bsxi = (\xi_1, \xi_2, \ldots, \xi_s)$, and $c_{d, \bsu}^{(n)}$ is a deterministic coefficient that needs to be determined, see, e.g.,~\cite{ghanem1991,ghanem1999,lemaitre2001,reagan2003,najm2009,ernst2012}. The basis functions $\Phi_\bsu$ are defined as
\begin{equation}
    \Phi_\bsu(\bsxi) \coloneqq \prod_{j=1}^s \phi_{u_j}(\xi_j),
\end{equation}
where $\phi_{u_j}$ are one-dimensional polynomials of degree $u_j$, $j=1, 2, \ldots, s$. By convention, the \emph{order} $|u|$ of the multivariate polynomial $\Phi_\bsu$ is given as the sum of all degrees, i.e., $|u| \coloneqq u_1 + u_2 + \ldots + u_s$.

In our numerical experiments in \cref{sec:results_and_discussion}, and for the purpose of surrogate construction, we define the input parameters $\nu_j$ as uniformly distributed on $[a_j,b_j]$. In this case, the polynomials $\phi_{u_j}$ are the normalized Legendre orthogonal polynomials, see~\cite{sargsyan2016}, and the random variables $\xi_j$ correspond to the model parameter values rescaled to $[-1, 1]$, i.e.,
\begin{equation}\label{eq:mapping}
    \xi_j = 2 \frac{\nu_j - a_j}{b_j - a_j} - 1, \quad j = 1, 2, \ldots, s.
\end{equation}

There are various options for finding the coefficients $c_{d, \bsu}^{(n)}$, as well as the index set $\calI_d^{(n)}$, based on a set of input-output evaluations, see, e.g.,~\cite{ghanem1991,najm2009,debusschere2004,sargsyan2016}. We will use the iterative Bayesian compressive sensing approach outlined in~\cite{sargsyan2014}. 

When evaluating the log-likelihood $\log \calL_{\calZ, \bsalpha}(\bsnu)$ from \cref{eq:weighted_likelihood}, we can now query the computationally cheap surrogate model $\tilde{f}_d^{(n)}(\bsnu)$ instead of the actual model $f_d(\bsx_d^{(n)}, \bsnu)$, at specific parameter values $\bsnu$. This avoids the need to run \Centipede{} in the likelihood evaluation. In particular, the log-likelihood $\log \calL_{\calZ, \bsalpha}(\bsnu)$ can be approximated as
\begin{align}
   \log \calL_{\calZ, \bsalpha}^{\textrm{surr}}(\bsnu) &\coloneqq -\frac{D}{2} \sum_{d=1}^D \alpha_d \sum_{n=1}^{N_d} \left[ \vphantom{\frac{1}{K_d \beta_d{s_d^{(n)}}^2} \sum_{k=1}^{K_d}} \log(2\pi\beta_d{s_d^{(n)}}^2) \; + \right. \\ &\hspace{0.5cm} \left. \frac{1}{K_d \beta_d{s_d^{(n)}}^2} \sum_{k=1}^{K_d} \left(z_d^{(n, k)} - \tilde{f}_d^{(n)}(\bsnu)\right)^2 \right]. \label{eq:likelihood_with_surrogate}
\end{align}

\subsection{Dimension reduction using global sensitivity analysis}\label{sec:dimension_reduction_using_global_sensitivity_analysis}

A natural way to order the input parameters according to their relative importance is provided through the computation of the Sobol' sensitivity indices, see, e.g.,~\cite{sobol2001}. The Sobol' indices measure fractional contributions of each parameter to the total output variance. The indices can be obtained from a variance-based sensitivity analysis, using a set of randomly-chosen input-output evaluations of the model, see, e.g.,~\cite{crestaux2009, saltelli2008}. However, when a PCE surrogate model is available, the Sobol' sensitivity indices can be extracted directly from the coefficients of the expansion, exploiting the orthogonality of the basis functions. For example, the \emph{total-effect} Sobol' sensitivity indices are defined as
\begin{align}\label{eq:total_effect_sensitivity_index}
    S^{(n)}_{d, j} &\approx \frac{\sum_{\bsu \in \calJ_{d, j}^{(n)}} {c_{d, \bsu}^{(n)}}^2}{\sum_{\bsu \in \calI_d^{(n)} \setminus \{\bszero\}} {c_{d, \bsu}^{(n)}}^2},
\end{align}
with $\calJ_{d, j}^{(n)} = \{\bsu \in \calI_d^{(n)} : u_j > 0 \}$. The total-effect sensitivity index is a measure of sensitivity describing which share of the total variance of the model output can be attributed to the $j$th parameter, including its interaction with other input variables~\cite{sobol2001}. Parameters with small total-effect indices have an overall small contribution to uncertainty in model outputs, and can thus be treated as deterministic, thereby decreasing the dimensionality of the uncertain input space, and the corresponding dimensionality of the surrogate. Having constructed a PCE surrogate model, one can easily evaluate the sensitivity indices by gathering the (square of the) appropriate coefficients. Note that
\begin{equation}
    \sum_{j=1}^s S^{(n)}_{d, j} \geq 1,
\end{equation}
due to the fact that the interaction effects are counted multiple times in the index set $\calJ_{d, j}^{(n)}$.

\section{Results and discussion}\label{sec:results_and_discussion}

In this section, we present our main results obtained by using the calibration framework outlined in \cref{sec:bayesian_calibration_for_summary_statistics,sec:polynomial_chaos_surrogate_construction_and_parameter_reduction} to estimate the parameters of \Centipede{}. Overall, our calibration strategy consists of three steps:

\begin{enumerate}
    \item First, we create a set of PCE surrogates for \Centipede{} in the full, 183-dimensional parameter space. We use GSA to identify a set of 24 important parameters.
    \item Next, we reconstruct the set of PCE surrogates in the reduced, 24-dimensional parameter space. These surrogates are more accurate, and can be used to replace the actual \Centipede{} predictions in the evaluation of the likelihood.
    \item Finally, we generate a consistent synthetic data set for each experiment, and use these data sets to perform Bayesian calibration using both the unweighted and weighted likelihood formulations.
\end{enumerate}

The remainder of this section is organized as follows. First, in \cref{sec:experimental_setup}, we provide more details on the experimental setup. Afterwards, in \cref{sec:surrogate_construction_and_sensitivity_analysis,sec:dimension_reduction,sec:generating_consistent_synthetic_data_sets}, we discuss the three steps of our calibration strategy in more detail. The result of the calibration effort is reported in \cref{sec:calibration_results,sec:effect_of_weights_in_the_likelihood}, and a discussion of these results is provided in \cref{sec:discussion}.

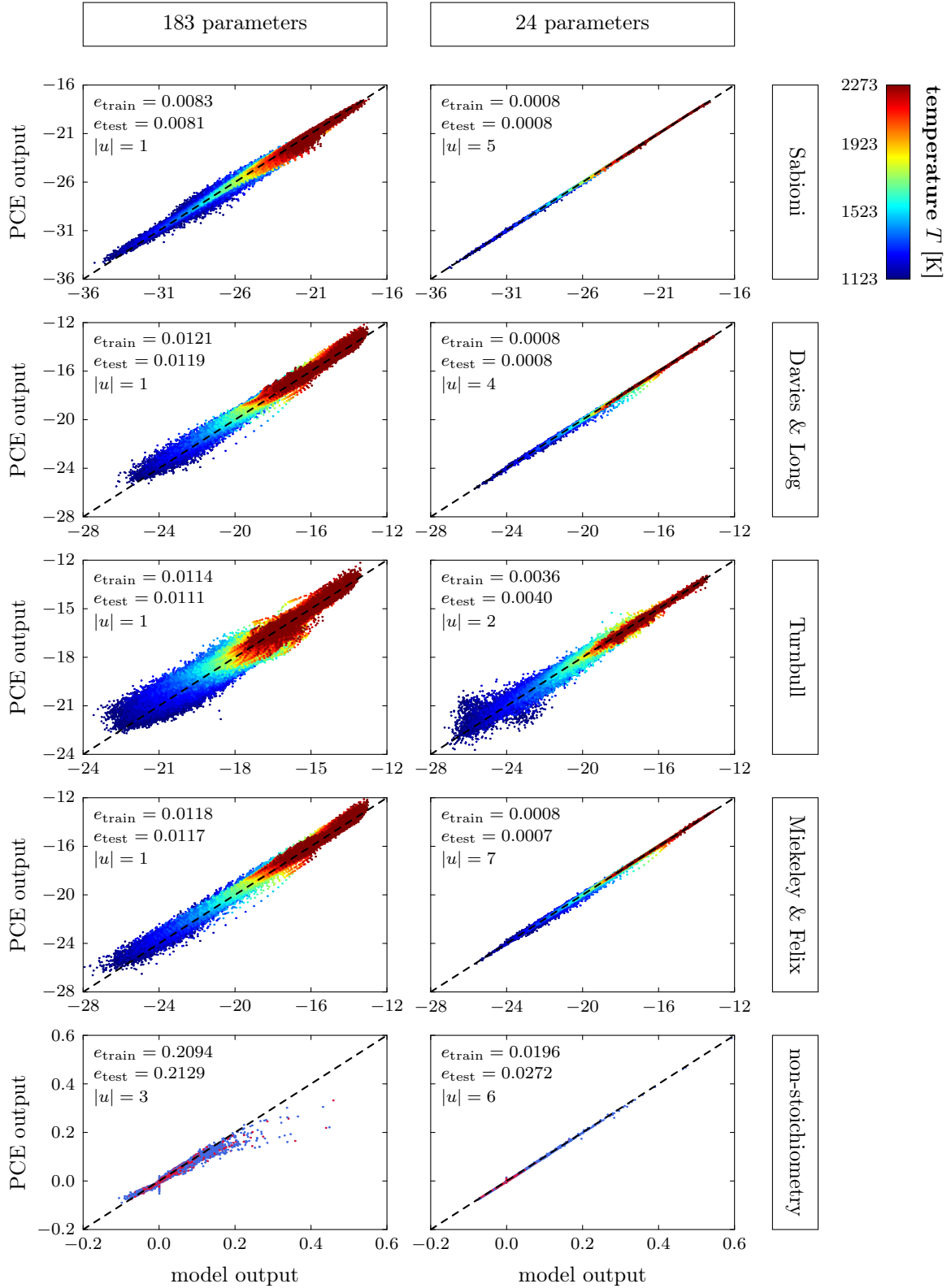
\begin{figure*}
    \centering
	\tikzsetnextfilename{parity_plots}
	\tikzexternalenable
	\setlength{\figurewidth}{6.5cm}
\setlength{\figureheight}{4.75cm}
\begin{tikzpicture}[%
        line/.style={default dashed line},
        label/.style={anchor=north west, font=\scriptsize, xshift=1pt},
    ]
    \begin{groupplot}[%
            group style={%
                group size=2 by 5,
                horizontal sep=1.85em,
                vertical sep=1.85em,
                xlabels at=edge bottom,
                ylabels at=edge left,
                yticklabels at=edge left,
            },
            default axis,
            axis on top,
            legend style={%
                at={(1, 1)},
                anchor=north east,
                /tikz/every odd column/.append style={column sep=1mm}
            },
            clip mode=individual,
            xlabel={model output},
            ylabel={PCE output},
            ylabel style={at={(-0.15,0.5)}}
        ]

        
        \nextgroupplot[%
            xmin=-36, xmax=-16,
            ymin=-36, ymax=-16,
            xtick={-36, -31, -26, -21, -16},
            ytick={-36, -31, -26, -21, -16},
            execute at end axis = {%
                \draw[black] (rel axis cs:0,1.2) rectangle node[pos=0.5] {\small\strut 183 parameters} (rel axis cs:1,1.425);
            }
        ]
        \addplot graphics[xmin=-36, xmax=-16, ymin=-36, ymax=-16] {plt/parity_Sabioni_183_params.png};
        \addplot[line] coordinates {(-36, -36) (-16, -16)};
        \node[label, yshift=-1pt] at (rel axis cs:0, 1) {$e_{\mathrm{train}} = 0.0083$};
        \node[label, yshift=-11pt] at (rel axis cs:0, 1) {$e_{\mathrm{test}} = 0.0081$};
        \node[label, yshift=-21pt] at (rel axis cs:0, 1) {$|u| = 1$};

        \nextgroupplot[%
        xmin=-36, xmax=-16,
        ymin=-36, ymax=-16,
        xtick={-36, -31, -26, -21, -16},
        ytick={-36, -31, -26, -21, -16},
            execute at end axis = {%
                \draw[black] (rel axis cs:0,1.2) rectangle node[pos=0.5] {\small\strut 24 parameters} (rel axis cs:1,1.425);
                \draw[black] (rel axis cs:1.125, 0) rectangle node[pos=0.5, rotate=-90] {\small\strut Sabioni} (rel axis cs:1.275, 1);
            },
            colormap/jet,
            colorbar,
            point meta min=1123, point meta max=2273,
            colorbar style={%
                ylabel={\textbf{temperature} $T$ [\si{\kelvin}]},
                ytick={1123, 1523, 1923, 2273},
                width=1em,
                at={(1.5, 0.5)},
                anchor=west,
                ytick style={draw=none},
                yticklabel pos=lower,
                ylabel style={align=center, at={(1.125, 0.5)}, anchor=south, rotate=180},
            },
        ]
        \addplot graphics[xmin=-36, xmax=-16, ymin=-36, ymax=-16] {plt/parity_Sabioni_24_params.png};
        \addplot[line] coordinates {(-36, -36) (-16, -16)};
        \node[label, yshift=-1pt] at (rel axis cs:0, 1) {$e_{\mathrm{train}} = 0.0008$};
        \node[label, yshift=-11pt] at (rel axis cs:0, 1) {$e_{\mathrm{test}} = 0.0008$};
        \node[label, yshift=-21pt] at (rel axis cs:0, 1) {$|u| = 5$};

        
        \nextgroupplot[%
            xmin=-28, xmax=-12,
            ymin=-28, ymax=-12,
            xtick={-28, -24, -20, -16, -12},
            ytick={-28, -24, -20, -16, -12},
        ]
        \addplot graphics[xmin=-28, xmax=-12, ymin=-28, ymax=-12] {plt/parity_Davies_Long_183_params.png};
        \addplot[line] coordinates {(-28, -28) (-12, -12)};
        \node[label, yshift=-1pt] at (rel axis cs:0, 1) {$e_{\mathrm{train}} = 0.0121$};
        \node[label, yshift=-11pt] at (rel axis cs:0, 1) {$e_{\mathrm{test}} = 0.0119$};
        \node[label, yshift=-21pt] at (rel axis cs:0, 1) {$|u| = 1$};

        \nextgroupplot[%
        xmin=-28, xmax=-12,
        ymin=-28, ymax=-12,
        xtick={-28, -24, -20, -16, -12},
        ytick={-28, -24, -20, -16, -12},
            execute at end axis = {%
                \draw[black] (rel axis cs:1.125, 0) rectangle node[pos=0.5, rotate=-90] {\small\strut Davies \& Long} (rel axis cs:1.275, 1);
            }
        ]
        \addplot graphics[xmin=-28, xmax=-12, ymin=-28, ymax=-12] {plt/parity_Davies_Long_24_params.png};
        \addplot[line] coordinates {(-28, -28) (-12, -12)};
        \node[label, yshift=-1pt] at (rel axis cs:0, 1) {$e_{\mathrm{train}} = 0.0008$};
        \node[label, yshift=-11pt] at (rel axis cs:0, 1) {$e_{\mathrm{test}} = 0.0008$};
        \node[label, yshift=-21pt] at (rel axis cs:0, 1) {$|u| = 4$};

        
        \nextgroupplot[%
            xmin=-24, xmax=-12,
            ymin=-24, ymax=-12,
            xtick={-24, -21, -18, -15, -12},
            ytick={-24, -21, -18, -15, -12},
        ]
        \addplot graphics[xmin=-24, xmax=-12, ymin=-24, ymax=-12] {plt/parity_Turnbull_183_params.png};
        \addplot[line] coordinates {(-24, -24) (-12, -12)};
        \node[label, yshift=-1pt] at (rel axis cs:0, 1) {$e_{\mathrm{train}} = 0.0114$};
        \node[label, yshift=-11pt] at (rel axis cs:0, 1) {$e_{\mathrm{test}} = 0.0111$};
        \node[label, yshift=-21pt] at (rel axis cs:0, 1) {$|u| = 1$};

        \nextgroupplot[%
            xmin=-28, xmax=-12,
            ymin=-28, ymax=-12,
            xtick={-28, -24, -20, -16, -12},
            ytick={-28, -24, -20, -16, -12},
            execute at end axis = {%
                \draw[black] (rel axis cs:1.125, 0) rectangle node[pos=0.5, rotate=-90] {\small\strut Turnbull} (rel axis cs:1.275, 1);
            }
        ]
        \addplot graphics[xmin=-28, xmax=-12, ymin=-28, ymax=-12] {plt/parity_Turnbull_24_params.png};
        \addplot[line] coordinates {(-28, -28) (-12, -12)};
        \node[label, yshift=-1pt] at (rel axis cs:0, 1) {$e_{\mathrm{train}} = 0.0036$};
        \node[label, yshift=-11pt] at (rel axis cs:0, 1) {$e_{\mathrm{test}} = 0.0040$};
        \node[label, yshift=-21pt] at (rel axis cs:0, 1) {$|u| = 2$};

        
        \nextgroupplot[%
            xmin=-28, xmax=-12,
            ymin=-28, ymax=-12,
            xtick={-28, -24, -20, -16, -12},
            ytick={-28, -24, -20, -16, -12},
        ]
        \addplot graphics[xmin=-28, xmax=-12, ymin=-28, ymax=-12] {plt/parity_Miekely_Felix_183_params.png};
        \addplot[line] coordinates {(-28, -28) (-12, -12)};
        \node[label, yshift=-1pt] at (rel axis cs:0, 1) {$e_{\mathrm{train}} = 0.0118$};
        \node[label, yshift=-11pt] at (rel axis cs:0, 1) {$e_{\mathrm{test}} = 0.0117$};
        \node[label, yshift=-21pt] at (rel axis cs:0, 1) {$|u| = 1$};

        \nextgroupplot[%
            xmin=-28, xmax=-12,
            ymin=-28, ymax=-12,
            xtick={-28, -24, -20, -16, -12},
            ytick={-28, -24, -20, -16, -12},
            execute at end axis = {%
                \draw[black] (rel axis cs:1.125, 0) rectangle node[pos=0.5, rotate=-90] {\small\strut Miekeley \& Felix} (rel axis cs:1.275, 1);
            }
        ]
        \addplot graphics[xmin=-28, xmax=-12, ymin=-28, ymax=-12] {plt/parity_Miekely_Felix_24_params.png};
        \addplot[line] coordinates {(-28, -28) (-12, -12)};
        \node[label, yshift=-1pt] at (rel axis cs:0, 1) {$e_{\mathrm{train}} = 0.0008$};
        \node[label, yshift=-11pt] at (rel axis cs:0, 1) {$e_{\mathrm{test}} = 0.0007$};
        \node[label, yshift=-21pt] at (rel axis cs:0, 1) {$|u| = 7$};

        
        \nextgroupplot[%
            xmin=-0.2, xmax=0.6,
            ymin=-0.2, ymax=0.6,
            xtick={-0.2,0,0.2,0.4,0.6},
            ytick={-0.2,0,0.2,0.4,0.6},
            xticklabel style={
                /pgf/number format/.cd,
                    fixed,
                    fixed zerofill,
                    precision=1
            },
            yticklabel style={
                /pgf/number format/.cd,
                    fixed,
                    fixed zerofill,
                    precision=1
            }
        ]
        \addplot graphics[xmin=-0.2, xmax=0.6, ymin=-0.2, ymax=0.6] {plt/parity_stoichiometry_183_params.png};
        \addplot[line] coordinates {(-0.2, -0.2) (0.6, 0.6)};
        \node[label, yshift=-1pt] at (rel axis cs:0, 1) {$e_{\mathrm{train}} = 0.2094$};
        \node[label, yshift=-11pt] at (rel axis cs:0, 1) {$e_{\mathrm{test}} = 0.2129$};
        \node[label, yshift=-21pt] at (rel axis cs:0, 1) {$|u| = 3$};

        \nextgroupplot[%
            xmin=-0.2, xmax=0.6,
            ymin=-0.2, ymax=0.6,
            xtick={-0.2,0,0.2,0.4,0.6},
            ytick={-0.2,0,0.2,0.4,0.6},
            xticklabel style={
                /pgf/number format/.cd,
                    fixed,
                    fixed zerofill,
                    precision=1
            },
            yticklabel style={
                /pgf/number format/.cd,
                    fixed,
                    fixed zerofill,
                    precision=1
            },
            execute at end axis = {%
                \draw[black] (rel axis cs:1.125, 0) rectangle node[pos=0.5, rotate=-90] {\small\strut non-stoichiometry} (rel axis cs:1.275, 1);
            }
        ]
        \addplot graphics[xmin=-0.2, xmax=0.6, ymin=-0.2, ymax=0.6] {plt/parity_stoichiometry_24_params.png};
        \addplot[line] coordinates {(-0.2, -0.2) (0.6, 0.6)};
        \node[label, yshift=-1pt] at (rel axis cs:0, 1) {$e_{\mathrm{train}} = 0.0196$};
        \node[label, yshift=-11pt] at (rel axis cs:0, 1) {$e_{\mathrm{test}} = 0.0272$};
        \node[label, yshift=-21pt] at (rel axis cs:0, 1) {$|u| = 6$};
    \end{groupplot}
\end{tikzpicture}
	\tikzexternaldisable

    \caption{Comparison of the predicted outputs from the PCE surrogate model and the actual model outputs for the quantity predicted by each set of data summaries for the 183-parameter model (\emph{left column}) and the 24-parameter model (\emph{right column}). We indicate the relative training error $e_\mathrm{train}$, the relative test error $e_\mathrm{test}$, and the order of the PCE $|u|$ for the surrogate with the largest relative test error across all (combinations of) temperatures (and $\pOtwo$ values). For the diffusivity predictions (first 4 rows), different colors indicate different temperatures, and for the non-stoichiometry predictions (last row), we distinguish between training (\textcolor{NavyBlue}{\emph{blue}}) and test samples (\textcolor{Red}{\emph{red}}).}
    \label{fig:parity_plots}
\end{figure*}

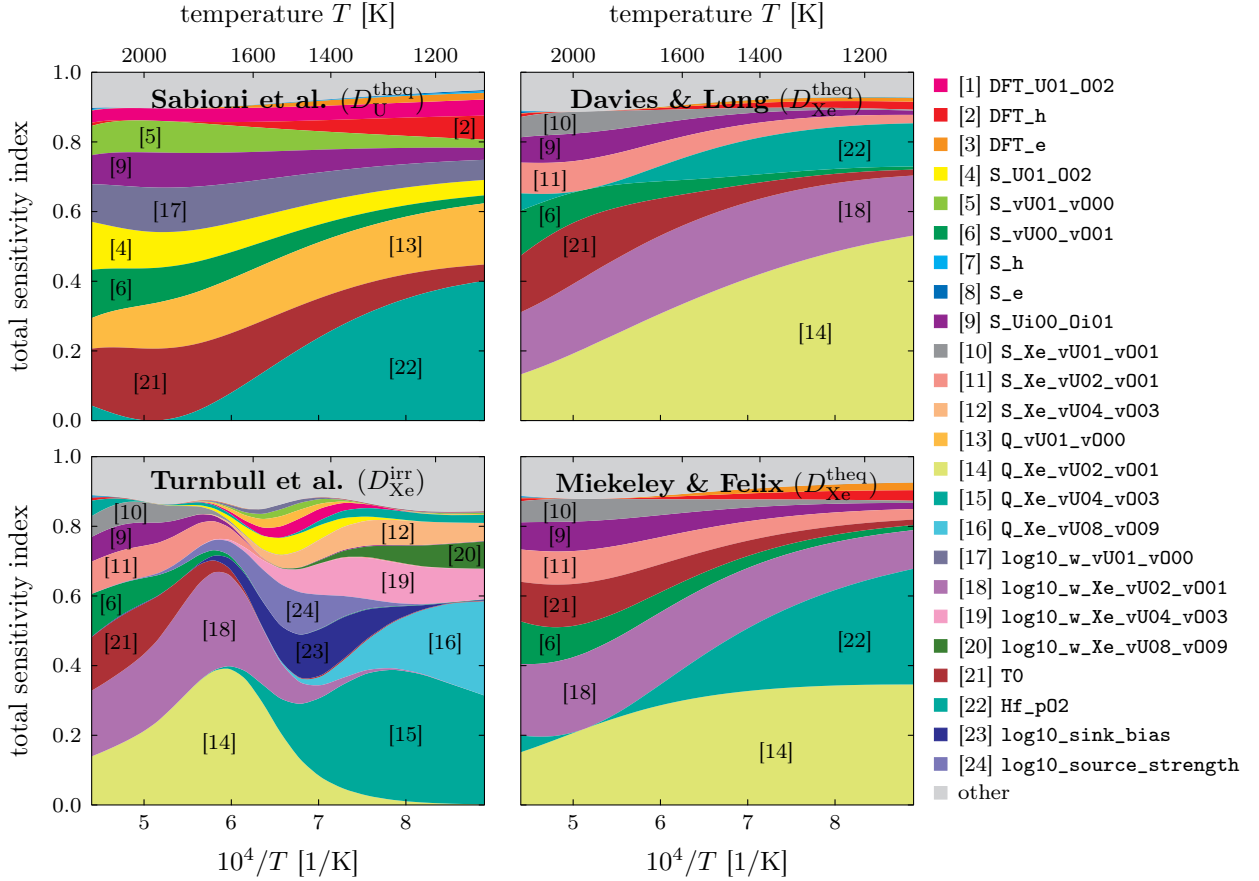
\begin{figure*}[t]
    \centering
	\tikzsetnextfilename{total_sensitivity_diffusivities}
	\tikzexternalenable
	\newcommand{\plotsens}[1]{%
    \pgfplotstableread[]{dat/#1_lower_totsens.dat}{\upp}
    \pgfplotstableread[]{dat/#1_upper_totsens.dat}{\low}
    \pgfplotsinvokeforeach{1, ..., 25}{
        \addplot[name path=bottom, draw=none, smooth, forget plot] table[x index=0, y index=##1] {\low};
        \addplot[name path=top, draw=none, smooth, forget plot] table[x index=0, y index=##1] {\upp};
        \addplot+[%
            draw=none,
        ] fill between[of=bottom and top];
    }
}

\newcommand{\plotsenswithlegend}[1]{%
    \pgfplotstableread[]{dat/#1_lower_totsens.dat}{\upp}
    \pgfplotstableread[]{dat/#1_upper_totsens.dat}{\low}
    \pgfplotsinvokeforeach{1, ..., 24}{
        \addplot[name path=bottom, draw=none, smooth, forget plot] table[x index=0, y index=##1] {\low};
        \addplot[name path=top, draw=none, smooth, forget plot] table[x index=0, y index=##1] {\upp};
        \addplot+[%
            draw=none,
            legend image code/.code={\expandafter\draw [yshift=-0.4ex] (0,0) rectangle (1ex,1ex);},
        ] fill between[of=bottom and top];
        \pgfplotstablegetcolumnnamebyindex{##1}\of\low\to{\name}
        \def\underscore{\pgfplotsutilstrreplace{_}{\_}}%
        \expandafter\underscore\expandafter{\name}%
        \addlegendentryexpanded{[##1] \texttt{\pgfplotsretval}}
    }
    \addplot[name path=bottom, draw=none, smooth, forget plot] table[x index=0, y index=25] {\low};
    \addplot[name path=top, draw=none, smooth, forget plot] table[x index=0, y index=25] {\upp};
    \addplot+[%
        draw=none,
        legend image code/.code={\expandafter\draw [yshift=-0.4ex] (0,0) rectangle (1ex,1ex);},
    ] fill between[of=bottom and top];
    \pgfplotstablegetcolumnnamebyindex{25}\of\low\to{\name}
    \def\underscore{\pgfplotsutilstrreplace{_}{\_}}%
    \expandafter\underscore\expandafter{\name}%
    \addlegendentryexpanded{\pgfplotsretval}
}

\pgfplotscreateplotcyclelist{sens}{
    {RubineRed},
    {Red},
    {BurntOrange},
    {Yellow},
    {LimeGreen},
    {ForestGreen},
    {Cyan},
    {NavyBlue},
    {Plum},
    {Gray},
    {Salmon},
    {Apricot},
    {Dandelion},
    {GreenYellow},
    {JungleGreen},
    {SkyBlue},
    {CadetBlue},
    {Orchid},
    {Lavender},
    {OliveGreen},
    {Maroon},
    {Emerald},
    {Blue},
    {Periwinkle},
    {Gray!40}
}

\setlength{\figurewidth}{6.75cm}
\setlength{\figureheight}{6.2cm}
\begin{tikzpicture}[
	]
	\begin{groupplot}[%
            group style={%
                group size=2 by 2,
                xticklabels at=edge bottom,
                xlabels at=edge bottom,
                yticklabels at=edge left,
                ylabels at=edge left,
                vertical sep=\baselineskip,
                horizontal sep=\baselineskip
            },
            default axis,
            axis on top,
            xmin = 4.4, xmax = 8.9, ymin = 0, ymax=1,
            legend style = {at = {(1.03, 1.02)}, column sep=2pt, row sep=-0.5pt, font=\scriptsize},
            xlabel = {$10^4/T$ [\si[per-mode=symbol]{\per\kelvin}]},
            ylabel = {total sensitivity index},
            ytick = {0, 0.2, ..., 2},
            xtick = {5, 6, 7, 8, 9},
            y tick label style={/pgf/number format/.cd, fixed, zerofill, precision=1},
            cycle list name=sens,
            every axis title/.style={at={(0,1)}, anchor=south west, align=left},
            axis x line*=bottom
        ]
        \nextgroupplot[] 
        \plotsens{Sabioni}
        \node[font=\scriptsize] at (rel axis cs:0.8, 0.15) {[22]};
        \node[font=\scriptsize] at (rel axis cs:0.15, 0.105) {[21]};
        \node[font=\scriptsize] at (rel axis cs:0.8, 0.5) {[13]};
        \node[font=\scriptsize] at (rel axis cs:0.075, 0.375) {[6]};
        \node[font=\scriptsize] at (rel axis cs:0.075, 0.49) {[4]};
        \node[font=\scriptsize] at (rel axis cs:0.2, 0.6) {[17]};
        \node[font=\scriptsize] at (rel axis cs:0.075, 0.725) {[9]};
        \node[font=\scriptsize] at (rel axis cs:0.15, 0.8125) {[5]};
        \node[font=\scriptsize] at (rel axis cs:0.95, 0.8375) {[2]};
        \node[font=\small, anchor=north] at (rel axis cs:0.5, 1) {\strut\textbf{Sabioni et al.} ($\dutheq$)};
        \nextgroupplot[] 
        \plotsenswithlegend{Davies_Long}
        \node[font=\scriptsize] at (rel axis cs:0.75, 0.25) {[14]};
        \node[font=\scriptsize] at (rel axis cs:0.85, 0.6) {[18]};
        \node[font=\scriptsize] at (rel axis cs:0.15, 0.5) {[21]};
        \node[font=\scriptsize] at (rel axis cs:0.075, 0.5865) {[6]};
        \node[font=\scriptsize] at (rel axis cs:0.85, 0.775) {[22]};
        \node[font=\scriptsize] at (rel axis cs:0.075, 0.69) {[11]};
        \node[font=\scriptsize] at (rel axis cs:0.075, 0.78) {[9]};
        \node[font=\scriptsize] at (rel axis cs:0.1, 0.85) {[10]};
        \node[font=\small, anchor=north] at (rel axis cs:0.5, 1) {\strut\textbf{Davies \& Long} ($\dxetheq$)};
        \nextgroupplot[] 
        \plotsens{Turnbull}
        \node[font=\scriptsize] at (rel axis cs:0.325, 0.175) {[14]};
        \node[font=\scriptsize] at (rel axis cs:0.8, 0.2) {[15]};
        \node[font=\scriptsize] at (rel axis cs:0.325, 0.5) {[18]};
        \node[font=\scriptsize] at (rel axis cs:0.9, 0.4625) {[16]};
        \node[font=\scriptsize] at (rel axis cs:0.075, 0.4625) {[21]};
        \node[font=\scriptsize] at (rel axis cs:0.5625, 0.4375) {[23]};
        \node[font=\scriptsize] at (rel axis cs:0.05, 0.575) {[6]};
        \node[font=\scriptsize] at (rel axis cs:0.5375, 0.555) {[24]};
        \node[font=\scriptsize] at (rel axis cs:0.78, 0.6375) {[19]};
        \node[font=\scriptsize] at (rel axis cs:0.075, 0.68) {[11]};
        \node[font=\scriptsize] at (rel axis cs:0.95, 0.715) {[20]};
        \node[font=\scriptsize] at (rel axis cs:0.075, 0.765) {[9]};
        \node[font=\scriptsize] at (rel axis cs:0.78, 0.78) {[12]};
        \node[font=\scriptsize] at (rel axis cs:0.1, 0.835) {[10]};
        \node[font=\small, anchor=north] at (rel axis cs:0.5, 1) {\strut\textbf{Turnbull et al.} ($\dxeirr$)};
        \nextgroupplot[] 
        \plotsens{Miekely_Felix}
        \node[font=\scriptsize] at (rel axis cs:0.65, 0.15) {[14]};
        \node[font=\scriptsize] at (rel axis cs:0.85, 0.47) {[22]};
        \node[font=\scriptsize] at (rel axis cs:0.15, 0.32) {[18]};
        \node[font=\scriptsize] at (rel axis cs:0.075, 0.46) {[6]};
        \node[font=\scriptsize] at (rel axis cs:0.1, 0.57) {[21]};
        \node[font=\scriptsize] at (rel axis cs:0.1, 0.68) {[11]};
        \node[font=\scriptsize] at (rel axis cs:0.1, 0.77) {[9]};
        \node[font=\scriptsize] at (rel axis cs:0.1, 0.84) {[10]};
        \node[font=\small, anchor=north] at (rel axis cs:0.5, 1) {\strut\textbf{Miekeley \& Felix} ($\dxetheq$)};
	\end{groupplot}
    \begin{groupplot}[%
            group style={%
                group size=2 by 2,
                xticklabels at=edge top,
                xlabels at=edge top,
                vertical sep=\baselineskip,
                horizontal sep=\baselineskip
            },
            default axis,
            axis on top,
            axis y line=none,
            xlabel={temperature $T$ [\si{\kelvin}]},
            xticklabel pos=top,
            axis x line*=top,
            xmin=4.4, xmax=8.9,
            ymin=4.4, ymax=8.9,
            xtick={5, 6.25, 7.14, 8.34},
        ]
        \nextgroupplot[%
            xticklabels={2000, 1600, 1400, 1200},
        ]
        \addplot[opacity=0] coordinates { (8.85,8.9) (8.9,8.85) }; 
        \nextgroupplot[%
            xticklabels={2000, 1600, 1400, 1200},
        ]
        \addplot[opacity=0] coordinates { (8.85,8.9) (8.9,8.85) }; 
        \nextgroupplot[%
        ]
        \addplot[opacity=0] coordinates { (8.85,8.9) (8.9,8.85) }; 
        \nextgroupplot[%
        ]
        \addplot[opacity=0] coordinates { (8.85,8.9) (8.9,8.85) }; 
    \end{groupplot}
\end{tikzpicture}
	\tikzexternaldisable

    \caption{Total-effect Sobol' sensitivity indices of the diffusivity predictions in each experiment as a function of temperature. Only the sensitivity indices of the 24 most important parameters are included. Different colors correspond to different parameters.}
    \label{fig:total_sensitivity_diffusivities}
\end{figure*}

\subsection{Experimental setup}\label{sec:experimental_setup}

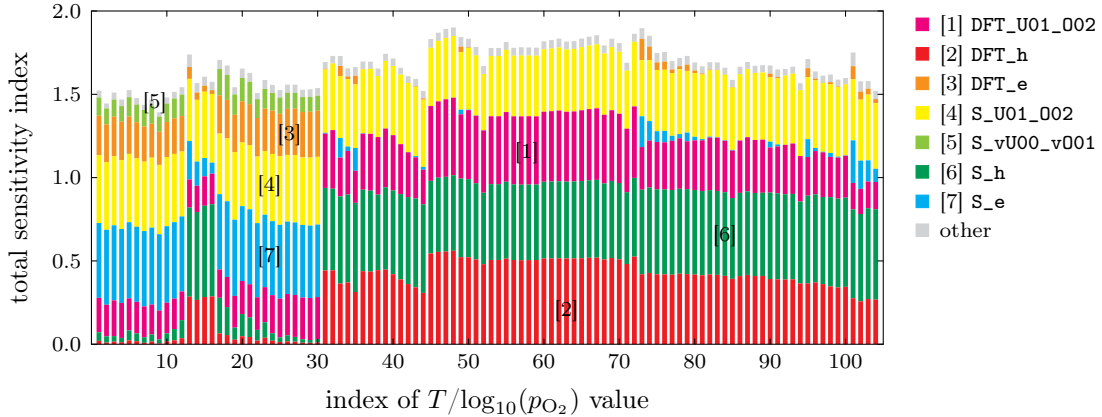
\begin{figure*}[t]
    \centering
	\tikzsetnextfilename{total_sensitivity_stoichiometry}
	\tikzexternalenable
	\pgfplotscreateplotcyclelist{sens}{
    {Red, fill={Red}},
    {ForestGreen, fill={ForestGreen}},
    {RubineRed, fill={RubineRed}},
    {Cyan, fill={Cyan}},
    {Yellow, fill={Yellow}},
    {BurntOrange, fill={BurntOrange}},
    {LimeGreen, fill={LimeGreen}},
    {Gray!40, fill={Gray!40}}
}

\setlength{\figurewidth}{12cm}
\setlength{\figureheight}{6cm}
\begin{tikzpicture}[
	]
	\begin{axis}[%
            default axis,
            axis on top,
            xmin = 0, xmax = 105, ymin = 0, ymax=2,
            legend style = {at = {(1.03, 1.02)}, column sep=2pt, row sep=-0.5pt, font=\scriptsize},
            xlabel={index of $T$/$\log_{10}(\pOtwo)$ value},
            ylabel = {total sensitivity index},
            y tick label style={/pgf/number format/.cd, fixed, zerofill, precision=1},
            every axis title/.style={at={(0,1)}, anchor=south west, font=\scriptsize, align=left},
            ybar stacked,
            cycle list name=sens,
            bar width=1.8pt,
            set layers,
            xtick={10, 20, ..., 100},
        ]
        \addlegendimage{legend image code/.code={\fill [RubineRed, yshift=-0.4ex] (0,0) rectangle (1ex,1ex);}}
        \addlegendentry{[1] \texttt{DFT\_U01\_O02}}
        \addlegendimage{legend image code/.code={\fill [Red, yshift=-0.4ex] (0,0) rectangle (1ex,1ex);}}
        \addlegendentry{[2] \texttt{DFT\_h}}
        \addlegendimage{legend image code/.code={\fill [BurntOrange, yshift=-0.4ex] (0,0) rectangle (1ex,1ex);}}
        \addlegendentry{[3] \texttt{DFT\_e}}
        \addlegendimage{legend image code/.code={\fill [Yellow, yshift=-0.4ex] (0,0) rectangle (1ex,1ex);}}
        \addlegendentry{[4] \texttt{S\_U01\_O02}}
        \addlegendimage{legend image code/.code={\fill [LimeGreen, yshift=-0.4ex] (0,0) rectangle (1ex,1ex);}}
        \addlegendentry{[5] \texttt{S\_vU00\_vO01}}
        \addlegendimage{legend image code/.code={\fill [ForestGreen, yshift=-0.4ex] (0,0) rectangle (1ex,1ex);}}
        \addlegendentry{[6] \texttt{S\_h}}
        \addlegendimage{legend image code/.code={\fill [Cyan, yshift=-0.4ex] (0,0) rectangle (1ex,1ex);}}
        \addlegendentry{[7] \texttt{S\_e}}
        \addlegendimage{legend image code/.code={\fill [Gray!40, yshift=-0.4ex] (0,0) rectangle (1ex,1ex);}}
        \addlegendentry{other}
        \pgfplotsinvokeforeach{1, ..., 8}{
            \addplot+[%
                draw=none,
            ] table[x index=0, y index=#1] {dat/sens_stoich.dat};
        }
	\end{axis}
    \begin{axis}[%
        default axis,
        axis on top,
        xmin = 0, xmax = 105, ymin = 0, ymax=0.7,
        legend style = {at = {(1.03, 1.02)}, column sep=2pt, row sep=-0.5pt, font=\tiny},
        xlabel={index of $T$/$\log_{10}(\pOtwo)$ value},
        ylabel = {total sensitivity index},
        ytick = {0, 0.1, ..., 0.7},
        y tick label style={/pgf/number format/.cd, fixed, zerofill, precision=1},
        every axis title/.style={at={(0,1)}, anchor=south west, font=\scriptsize, align=left},
        ybar stacked,
        cycle list name=sens,
        bar width=1.8pt,
        axis x line=none,
        axis y line=none,
    ]
    \node[font=\scriptsize] at (rel axis cs:0.6, 0.1) {[2]};
    \node[font=\scriptsize] at (rel axis cs:0.8, 0.325) {[6]};
    \node[font=\scriptsize] at (rel axis cs:0.55, 0.575) {[1]};
    \node[font=\scriptsize] at (rel axis cs:0.225, 0.2625) {[7]};
    \node[font=\scriptsize] at (rel axis cs:0.225, 0.475) {[4]};
    \node[font=\scriptsize] at (rel axis cs:0.25, 0.63) {[3]};
    \node[font=\scriptsize] at (rel axis cs:0.08, 0.725) {[5]};
\end{axis}
\end{tikzpicture}
	\tikzexternaldisable

    \caption{Total-effect Sobol' sensitivity indices of the non-stoichiometry predictions $\nonstoich$ as a function of the $T$/$\log_{10}(\pOtwo)$ index. We only included the 12 thermodynamic parameters in the sensitivity analysis, as the kinetic parameters have no effect on the non-stoichiometry predictions. Only the sensitivity indices of the 7 most important parameters are shown. Different colors correspond to different parameters. Indices 1--30 correspond to stoichiometric predictions at high temperatures and low $\pOtwo$, while indices 31--104 correspond to low temperatures and high $\pOtwo$.}
    \label{fig:total_sensitivity_stoichiometry}
\end{figure*}

\Centipede{} implements the Free Energy Cluster Dynamics (FECD) method from~\cite{matthews2019} within the \Moose{} framework~\cite{permann2020}. \Moose{} provides access to the Finite Element code \libMesh{}~\cite{kirk2006}, and the partial differential equation (PDE) solver \PETSc{}~\cite{petsc-web-page}. The latter is a crucial component for the efficient solution of the system of nonlinear ODEs in~\eqref{eq:coupled_odes} that represent the cluster dynamics physics.

The uranium and xenon diffusivities under thermal equilibrium, $\dutheq$ and $\dxetheq$, and xenon diffusivity under irradiation conditions, $\dxeirr$, vary with temperature. The \Centipede{} predictions of these diffusivities are available at a set of 26 temperatures provided in \cref{tab:measurement_locations_for_the_diffusivity_predictions}. These temperatures span the range of experimental conditions where the data summaries are available. Evaluating \Centipede{} at all temperatures where the data summaries are available would lead to excessive computational requirements, as one would have to solve \cref{eq:coupled_odes} at each temperature. Consequentially, the PCE surrogate model predictions for the uranium and xenon diffusivities will be available only at these 26 temperatures. Since the evaluation of the likelihood in \cref{eq:likelihood_with_surrogate} requires access to a set of surrogate models $f_d^{(n)}$, $n = 1, 2, \ldots, N_d$ defined at each temperature $\bsx_d^{(n)}$, we propose to use a suitable interpolation scheme. In what follows, we assume a linear interpolation scheme, and remark that linear interpolation of the PCE outputs can be accomplished by linear interpolation of the PCE coefficients.

The fuel non-stoichiometry $\nonstoich{}$ is a function of both $T$ and $\pOtwo{}$. \Centipede{} predictions for $\nonstoich{}$ are available at the same 104 combinations of $T$ and $\pOtwo$ where the data summaries are available, so no interpolation of the PCE surrogate models is required in this case.

As indicated in \cref{tab:data_sets_overview}, the operating conditions \texttt{Hf\_pO2} and \texttt{T0} are unique for each experiment that predicts the diffusivity quantities. As such, we will allow them to vary independently. This can be achieved by enriching the set of parameters used in the calibration with experiment-specific copies of these operating conditions. This increases the number of parameters to calibrate from 24 to 30.

The lower bounds $a_j$ and upper bounds $b_j$ used to construct the PCE surrogate models are defined in \cref{tab:all_parameters_overview}. The PCE surrogates and subsequent sensitivity analysis are performed using the uncertainty quantification toolkit (\texttt{UQTk}), see~\cite{debusschere2004, debusschere2017}. We also implemented the DFI method outlined in \cref{sec:bayesian_calibration} in \texttt{UQTk}.

We remark that not all parameter combinations yield valid output samples of the stoichiometry and/or diffusivity, due to a lack of convergence of the underlying ODE solver, or because the sampled set of parameters resulted in unphysical responses. In our experiments, the number of code failures is about 3\% for the 183-parameter model, and 2\% for the 24-parameter model. These code failures seem to happen because of convoluted interaction effects between parameters, as we were unable to attribute the failures to certain regions of the parameter space. A careful analysis of these code failures is left for future work.

During calibration, we employ Gaussian priors on these parameters, 
\begin{equation}
    p(\bsnu) = \prod_{j=1}^s p(\nu_j)
\end{equation}
with
\begin{equation}
    p(\nu_j) \sim \calN\left(\frac{a_j + b_j}{2},\frac{ \left(b_j - a_j\right)^2}{36}\right).
\end{equation}
This choice means that $99\%$ of the probability mass of the prior is contained between the given lower and upper bounds.

\subsection{Surrogate construction and sensitivity analysis}\label{sec:surrogate_construction_and_sensitivity_analysis}

\begin{table*}[t]
    \centering
    \colorlet{colorA}{Plum}
    \colorlet{colorB}{Cyan}
    \colorlet{colorC}{ForestGreen}
    \colorlet{colorD}{GreenYellow}
    \colorlet{colorE}{BurntOrange}
    \colorlet{colorF}{RubineRed}
    \tiny
    \begin{tabularx}{\textwidth}{>{\ttfamily}X>{\ttfamily}X>{\ttfamily}X>{\ttfamily}X>{\ttfamily}X} \toprule
        \normalfont\scriptsize Sabioni et al. & \normalfont\scriptsize Davies \& Long & \normalfont\scriptsize Turnbull et al. & \normalfont\scriptsize Miekeley \& Felix & \normalfont\scriptsize non-stoichiometry \\ \midrule
        \textcolor{colorF}{Hf\_pO2} & \textcolor{colorF}{Q\_Xe\_vU02\_vO01} & \textcolor{colorF}{Q\_Xe\_vU02\_vO01} & \textcolor{colorF}{Q\_Xe\_vU02\_vO01} & \textcolor{colorF}{DFT\_h}\\
\textcolor{colorF}{T0} & \textcolor{colorF}{log10\_w\_Xe\_vU02\_vO01} & \textcolor{colorF}{Q\_Xe\_vU04\_vO03} & \textcolor{colorF}{Hf\_pO2} & \textcolor{colorF}{S\_h}\\
\textcolor{colorF}{Q\_vU01\_vO00} & \textcolor{colorF}{T0} & \textcolor{colorF}{log10\_w\_Xe\_vU02\_vO01} & \textcolor{colorF}{log10\_w\_Xe\_vU02\_vO01} & \textcolor{colorF}{DFT\_U01\_O02}\\
\textcolor{colorF}{S\_vU00\_vO01} & \textcolor{colorF}{S\_vU00\_vO01} & \textcolor{colorF}{Q\_Xe\_vU08\_vO09} & \textcolor{colorF}{S\_vU00\_vO01} & \textcolor{colorF}{S\_e}\\
\textcolor{colorF}{S\_U01\_O02} & \textcolor{colorF}{Hf\_pO2} & \textcolor{colorF}{T0} & \textcolor{colorF}{T0} & \textcolor{colorF}{S\_U01\_O02}\\
\textcolor{colorF}{log10\_w\_vU01\_vO00} & \textcolor{colorF}{S\_Xe\_vU02\_vO01} & \textcolor{colorF}{log10\_sink\_bias} & \textcolor{colorF}{S\_Xe\_vU02\_vO01} & \textcolor{colorE}{DFT\_e}\\
\textcolor{colorF}{S\_Ui00\_Oi01} & \textcolor{colorE}{S\_Ui00\_Oi01} & \textcolor{colorF}{S\_vU00\_vO01} & \textcolor{colorE}{S\_Ui00\_Oi01} & \textcolor{colorB}{S\_vU00\_vO01}\\
\textcolor{colorD}{S\_vU01\_vO00} & \textcolor{colorC}{S\_Xe\_vU01\_vO01} & \textcolor{colorF}{log10\_source\_strength} & \textcolor{colorC}{S\_Xe\_vU01\_vO01} & \textcolor{black}{DFT\_vU00\_vO01}\\
\textcolor{colorC}{DFT\_h} & \textcolor{colorB}{DFT\_vU00\_vO01} & \textcolor{colorF}{log10\_w\_Xe\_vU04\_vO03} & \textcolor{colorB}{DFT\_vU00\_vO01} & \textcolor{black}{S\_vU01\_vO00}\\
\textcolor{colorB}{DFT\_U01\_O02} & \textcolor{colorB}{DFT\_Xe\_vU02\_vO01} & \textcolor{colorF}{S\_Xe\_vU02\_vO01} & \textcolor{colorB}{DFT\_Xe\_vU02\_vO01} & \textcolor{black}{DFT\_vU01\_vO00}\\
\textcolor{colorB}{DFT\_vU00\_vO01} & \textcolor{colorA}{DFT\_Ui00\_Oi01} & \textcolor{colorF}{log10\_w\_Xe\_vU08\_vO09} & \textcolor{colorA}{DFT\_Ui00\_Oi01} & \textcolor{black}{S\_Ui00\_Oi01}\\
\textcolor{colorA}{DFT\_vU01\_vO00} & \textcolor{colorA}{DFT\_h} & \textcolor{colorF}{S\_Ui00\_Oi01} & \textcolor{colorA}{DFT\_h} & \textcolor{black}{DFT\_Ui00\_Oi01}\\
\textcolor{black}{DFT\_Ui00\_Oi01} & \textcolor{colorA}{DFT\_Xe\_vU01\_vO01} & \textcolor{colorF}{S\_Xe\_vU04\_vO03} & \textcolor{colorA}{DFT\_Xe\_vU01\_vO01} & \\
\textcolor{black}{DFT\_e} & \textcolor{black}{DFT\_Xe\_vU02\_vO02} & \textcolor{colorE}{S\_Xe\_vU01\_vO01} & \textcolor{black}{DFT\_e} & \\
\textcolor{black}{S\_h} & \textcolor{black}{S\_Xe\_vU02\_vO02} & \textcolor{colorE}{log10\_sink\_strength} & \textcolor{black}{DFT\_Xe\_vU02\_vO02} & \\
\textcolor{black}{S\_e} & \textcolor{black}{DFT\_e} & \textcolor{colorD}{DFT\_Xe\_vU04\_vO03} & \textcolor{black}{S\_Xe\_vU02\_vO02} & \\
\textcolor{black}{Q\_vU01\_vO02} & \textcolor{black}{S\_e} & \textcolor{colorD}{S\_U01\_O02} & \textcolor{black}{S\_e} & \\
\textcolor{black}{S\_vU01\_vO02} & \textcolor{black}{S\_h} & \textcolor{colorC}{Hf\_pO2} & \textcolor{black}{S\_h} & \\
\textcolor{black}{log10\_w\_vU01\_vO02} & \textcolor{black}{S\_Xe\_vU01\_vO02} & \textcolor{colorC}{DFT\_vU00\_vO01} & \textcolor{black}{charge\_correction\_DFT} & \\
\textcolor{black}{charge\_sq\_correction\_DFT} & \textcolor{black}{charge\_correction\_DFT} & \textcolor{colorC}{DFT\_Xe\_vU04\_vO02} & \textcolor{black}{S\_Xe\_vU01\_vO02} & \\
\textcolor{black}{DFT\_vU01\_vO02} & \textcolor{black}{charge\_sq\_correction\_DFT} & \textcolor{colorC}{DFT\_U01\_O02} & \textcolor{black}{charge\_sq\_correction\_DFT} & \\
\textcolor{black}{charge\_correction\_DFT} & \textcolor{black}{DFT\_Xe\_vU01\_vO02} & \textcolor{colorB}{S\_Xe\_vU02\_vO02} & \textcolor{black}{DFT\_Xe\_vU01\_vO02} & \\
\textcolor{black}{Q\_vU02\_vO02} & \textcolor{black}{DFT\_vU01\_vO00} & \textcolor{colorB}{S\_Xe\_vU04\_vO02} & \textcolor{black}{Q\_Xe\_vU02\_vO02} & \\
\textcolor{black}{S\_vU02\_vO02} & \textcolor{black}{Q\_Xe\_vU02\_vO02} & \textcolor{colorB}{DFT\_Xe\_vU02\_vO01} & \textcolor{black}{log10\_w\_Xe\_vU02\_vO02} & \\
\textcolor{black}{log10\_w\_vU02\_vO02} & \textcolor{black}{Q\_Xe\_vU02\_vO00} & \textcolor{colorB}{Q\_vU01\_vO00} & \textcolor{black}{log10\_sink\_bias} & \\
\textcolor{black}{log10\_w\_vU02\_vO01} & \textcolor{black}{log10\_w\_Xe\_vU02\_vO02} & \textcolor{colorA}{DFT\_Ui00\_Oi01} & \textcolor{black}{DFT\_Xe\_vU02\_vO00} & \\
\textcolor{black}{Q\_vU02\_vO00} & \textcolor{black}{Q\_Xe\_vU04\_vO03} & \textcolor{colorA}{S\_vU01\_vO00} & \textcolor{black}{S\_vU01\_vO00} & \\
\textcolor{black}{DFT\_vU02\_vO02} & \textcolor{black}{S\_vU01\_vO00} & \textcolor{colorA}{DFT\_Xe\_vU01\_vO01} & \textcolor{black}{log10\_w\_Xe\_Ui01\_Oi00} & \\
\textcolor{black}{S\_vU02\_vO01} & \textcolor{black}{DFT\_Xe\_vU02\_vO00} & \textcolor{colorA}{DFT\_Xe\_vU02\_vO02} & \textcolor{black}{log10\_sink\_strength} & \\
\textcolor{black}{Q\_vU02\_vO01} & \textcolor{black}{log10\_w\_Xe\_vU02\_vO00} & \textcolor{black}{log10\_w\_vU01\_vO00} & \textcolor{black}{log10\_w\_Xe\_vU02\_vO00} & \\
\multicolumn{1}{c}{$\vdots$\qquad\qquad\qquad} & \multicolumn{1}{c}{$\vdots$\qquad\qquad\qquad} & \multicolumn{1}{c}{$\vdots$\qquad\qquad\qquad} & \multicolumn{1}{c}{$\vdots$\qquad\qquad\qquad} & \\
\bottomrule
    \end{tabularx}%
    \caption{Parameters of \Centipede{} ordered according to their maximum Sobol' total sensitivity index across all temperatures and $\pOtwo{}$ values for each experiment. Different colors indicate which parameters will be retained when we want to capture a certain percentage of the total output variance: 70\% (\captionsquare{color=colorF}, 21 parameters), 75\% (\captionsquare{color=colorF} + \captionsquare{color=colorE}, 24 parameters), 80\% (\captionsquare{color=colorF} + \captionsquare{color=colorE} + \captionsquare{color=colorD}, 26 parameters), 85\% (\captionsquare{color=colorF} + \captionsquare{color=colorE} + \captionsquare{color=colorD} + \captionsquare{color=colorC}, 28 parameters), 90\% (\captionsquare{color=colorF} + \captionsquare{color=colorE} + \captionsquare{color=colorD} + \captionsquare{color=colorC} + \captionsquare{color=colorB}, 31 parameters), 95\% (\captionsquare{color=colorF} + \captionsquare{color=colorE} + \captionsquare{color=colorD} + \captionsquare{color=colorC} + \captionsquare{color=colorB} + \captionsquare{color=colorA}, 35 parameters).}
    \label{tab:truncation_strategy}
\end{table*}

For the uranium and xenon diffusivities, we construct a set of first-order PCE surrogate models in 183 dimensions, one for each of the 26 temperatures. Similarly, for the stoichiometry predictions, we construct a higher-order PCE surrogate model at each combination of $T$ and $\pOtwo$ where the data $\calD_5$ is available. These surrogates are built using 10,000 input-output evaluations of \Centipede{} as training data, and 1,000 evaluations as test data. The input training and test data is sampled from a uniform distribution between the given lower bounds $a_j$ and upper bounds $b_j$ for each parameter $j = 1, 2, \ldots, 183$. During the construction of the higher-order surrogates for the non-stoichiometry ($\nonstoich$), we used the adaptive procedure described in~\cite{sargsyan2014}. We illustrate the accuracy of these PCE surrogate models by comparing the predicted outputs from the surrogate with the actual model outputs in the left column of \cref{fig:parity_plots}. Each point in this figure represents a single sample. The surrogate model outputs are in perfect agreement with the \Centipede{} outputs if all points fall on the diagonal (\emph{dashed line}). In the plots, we indicate the relative $\ell_2$ training error, $e_\textrm{train}$, where
\begin{equation}\label{eq:training_error}
    e_\textrm{train}^2 = \frac{\sum_{l=1}^{L_\textrm{train}} \sum_{n=1}^{N_d} \left(f_d^{(n)}(\bsnu_\textrm{train}^{(l)}) - \tilde{f}_d^{(n)}(\bsnu_\textrm{train}^{(l)})\right)^2}{\sum_{l=1}^{L_\textrm{train}} \sum_{n=1}^{N_d} f_d^{(n)}(\bsnu_\textrm{train}^{(l)})^2}
\end{equation}
and $\bsnu_\textrm{train}^{(l)}$, $l=1, 2, \ldots, {L_\textrm{train}}$ are the training samples. We also indicate the test error $e_\textrm{test}$, defined similar to \cref{eq:training_error}, and the order of the PCE $|u|$ corresponding to the surrogate with largest relative test error across all (combinations of) temperatures (and $\pOtwo$ values). Note that, because training and test errors are relatively close, we conclude that the PCE surrogate models do not suffer from overfitting. Also note that the accuracy of these 183-dimensional surrogates is poor. The accuracy of the surrogate models for the non-stoichiometry ($\nonstoich$) seems particularly low, despite the higher-order construction scheme. This accuracy concern could be addressed by increasing the number of training samples $L_\textrm{train}$, causing a consequential growth in the computational requirements. However, we deem the accuracy of the surrogates to be sufficient for the estimation of sensitivity indices.

With these surrogates available, we are able to evaluate the sensitivity indices using~\cref{eq:total_effect_sensitivity_index}. The result is shown in \cref{fig:total_sensitivity_diffusivities} for the diffusivity predictions, and \cref{fig:total_sensitivity_stoichiometry} for the stoichiometry predictions.

\Cref{fig:total_sensitivity_diffusivities} shows the total Sobol' sensitivity indices for the diffusivity quantities as a function of temperature. Different colors indicate different parameters, and the parameters are ordered according to their maximum sensitivity index across all temperatures, for the 24 most important parameters only. Note that the names of these parameters correspond to the internal labelling used by \Centipede{}, see also \cref{tab:all_parameters_overview}. The light gray area on top of each axis represents the fraction of unexplained output variance.

For $\dutheq$, predicted by Sabioni et al., the three most important parameters are \texttt{Hf\_pO2}, \texttt{T0}, and the activation energy of uranium (\texttt{Q\_vU01\_vO00}), which is consistent with the active uranium vacancy diffusion mechanism identified in \cite{matthews2020}. It is interesting to note that out of the three most impactful parameters, two refer to the operating conditions of the experiment, and only one refers to the specific properties of uranium vacancies. As expected for a thermal equilibrium experiment on uranium diffusion only, none of the sensitive parameters refers to the response due to irradiation or the properties of xenon.

For $\dxetheq$, predicted by Davies \& Long, the three most important parameters are the activation energy of xenon in the \ce{U2O} cluster (\texttt{Q\_Xe\_vU02\_vO01}), the (log of the) attempt frequency of xenon in the \ce{U2O} vacancy cluster (\texttt{log10\_w\_Xe\_vU02\_vO01}), and \texttt{T0}. Again, these parameters are consistent with the active diffusion mechanism previously identified, and also emphasize the impact of the thermodynamic conditions of the experiment, which govern the non-stoichiometry $\nonstoich{}$.

For $\dxeirr$, predicted by Turnbull et al., the three most important parameters are \texttt{Q\_Xe\_vU02\_vO01}, the activation energy of the xenon defect located at a \ce{U4O3} vacancy cluster (\texttt{Q\_Xe\_vU04\_vO03}), and \texttt{log10\_w\_Xe\_vU02\_vO01}. It is worth pointing out that the significance of $\ce{Xe}_{\ce{U4O3}}$, i.e., the xenon defect located at the \ce{U4O3} vacancy cluster, as well as the significance of $\ce{Xe}_{\ce{U8O9}}$, which appears in the list of important parameters through the activation energy of the xenon defect located at \ce{U8O9} (\texttt{Q\_Xe\_vU08\_vO09}), has been speculated in~\cite{perriot2019}, and is confirmed in our sensitivity analysis. In particular, in previous studies, $\ce{Xe}_{\ce{U2O}}$ was identified as being responsible for diffusion at high temperature, and $\ce{Xe}_{\ce{U4O3}}$ was identified as being responsible for diffusion at intermediate temperatures, and, although $\ce{Xe}_{\ce{U8O9}}$ was not predicted to be the main contributor to diffusion in previous studies, it was close to the dominant $\ce{Xe}_{\ce{U4O3}}$ cluster in the intermediate temperature range, see~\cite{matthews2020}. The change in sensitivity as a function of temperature in \cref{fig:total_sensitivity_diffusivities} confirms these observations. It also emphasizes the competition between $\ce{Xe}_{\ce{UO}}$ and $\ce{Xe}_{\ce{U2O}}$ at high temperatures, and the transition from thermodynamic parameters (\texttt{T0}) to irradiation parameters (\texttt{log10\_sink\_bias} and \texttt{log10\_source\_strength}) with decreasing temperature.

For $\dxetheq$, predicted by Miekeley \& Felix, the three most important parameters are \texttt{Q\_Xe\_vU02\_vO01}, \texttt{Hf\_pO2}, and \texttt{log10\_w\_Xe\_vU02\_vO01}. Although the exact ordering of the parameters is slightly different from the results reported for $\dxetheq$ predicted by Davies \& Long, the parameter set is very similar, which is expected based on the fact that both data sets predict xenon diffusion under thermal equilibrium conditions. However, we expect that the operating conditions may differ between the two experiments, due to the difference in experimental setup.

\Cref{fig:total_sensitivity_stoichiometry} shows the total Sobol' sensitivity indices for the non-stoichiometry $\nonstoich{}$ as a function of the $T$/$\log_{10}(\pOtwo{})$ index. Again, different colors indicate different parameters, and the parameters are ordered according to their maximum sensitivity index across all indices. Note that, in our sensitivity analysis, we took into account that the non-stoichiometry depends only on the 12 thermodynamic parameters, and not on the kinetic parameters that influence the diffusivities. Only those thermodynamic parameters that are part of the set of 24 most important parameters overall are included in the plot. The three most important parameters for the stoichiometric predictions at high temperatures and low partial pressure (indices 1 -- 30) are the entropy of the electrons (\texttt{S\_e}), the entropy of \ce{UO2} (\texttt{S\_U01\_O02}), and the electron formation energy (\texttt{DFT\_e}). The three most important parameters for the stoichiometric predictions at low temperatures and high partial pressure (indices 31 -- 104) are the formation energy of the holes (\texttt{DFT\_h}), the entropy of the holes (\texttt{S\_h}) and the formation energy of \ce{UO2} (\texttt{DFT\_U01\_O02}). Also note that the sum of the total sensitivity indices is larger than 1, indicating that there are significant interaction effects between the thermodynamic parameters.

Finally, we note that the parameters in the correction terms for the binding energies (i.e, \texttt{charge\_correction\_DFT} and \texttt{charge\_sq\_correction\_DFT}), that account for systematic errors in the binding energies, are unimportant for all quantities of interest.

\subsection{Dimension reduction}\label{sec:dimension_reduction}

The total sensitivity indices allow us to rank the 183 parameters according to their contribution to the output variance. By assuming a prescribed fraction of the output variance that needs to be explained across all $D = 5$ data sets, we can obtain a sequence of reduced models that contain subsets of all 183 parameters. This truncation strategy is illustrated in \cref{tab:truncation_strategy}. Ensuring that at least 75\% of the output variance is captured for each predicted quantity, we identify a set of 24 important parameters. Note that the actual fraction of the variance explained for each predicted output individually is slightly larger than the threshold of $75\%$. That is, the set of required parameters to reach the prescribed threshold in output variance explained for each experiment is a subset of the 24 chosen parameters. By including additional parameters, the overall fraction of variance explained increases beyond that threshold. In particular, the fraction of output variance explained using our reduced set of 24 parameters is $90\%$ (Sabioni et al.), $87\%$ (Davies \& Long), $84\%$ (Turnbull et al.), and $88\%$ (Miekeley \& Felix).

Next, we reconstruct the PCE surrogates for both the diffusivity and non-stoichiometry predictions a second time, now using only 24 uncertain parameters, keeping the 159 parameters that were excluded by the sensitivity analysis at their nominal value. Using approximately 25,000 and 2,500 input-output evaluations of \Centipede{} as training and test data, respectively, we construct a set of higher-order PCE surrogates using the iterative procedure described in~\cite{sargsyan2014}. As before, the input training and test data is sampled from uniform distributions between the lower and upper bounds given in \cref{tab:all_parameters_overview} for the 24 uncertain parameters. To assess the accuracy of these new PCE surrogates, we compare the predicted outputs from the surrogate with the actual model outputs in the right column of \cref{fig:parity_plots}. We also indicate training and test errors computed using~\eqref{eq:training_error}. Because the training and test errors are similar, we again conclude that the surrogates are not overfitted. These results clearly indicate the improved surrogate accuracy in the 24-dimensional context, as compared to the 183-dimensional surrogate.

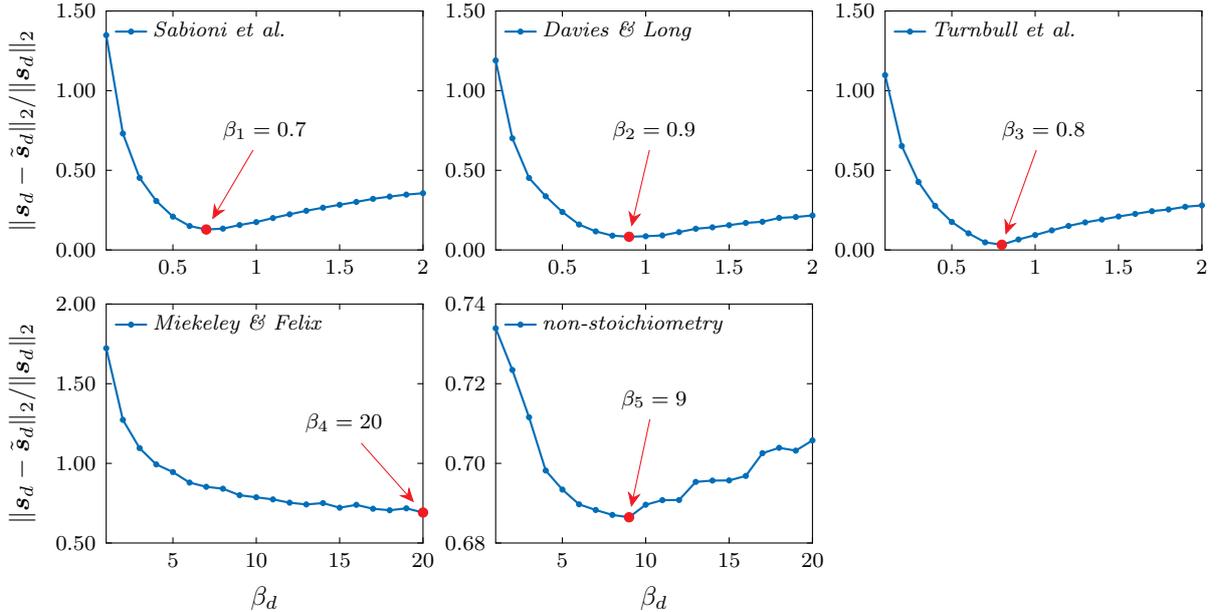
\begin{figure}[t]
    \centering
	\tikzsetnextfilename{consistent_data_sets}
	\tikzexternalenable
	\setlength{\figurewidth}{5.75cm}
\setlength{\figureheight}{4.75cm}
\begin{tikzpicture}[%
        line/.style={default marker line, NavyBlue},
        optimum/.style={only marks, mark=*,mark size=1.75pt, Red},
        arrow/.style={->, Red} 
    ]
    \begin{groupplot}[%
            group style={%
                group size=3 by 2,
                xlabels at=edge bottom,
                ylabels at=edge left,
                vertical sep=1.5\baselineskip,
                horizontal sep=2\baselineskip
            },
            default axis,
            xlabel={$\beta_d$},
            legend style={%
                at={(0, 1)},
                anchor=north west,
            },
            clip mode=individual,
            ylabel={$\|{\bss}_d - {\tilde\bss_d}\|_2/\|{\bss}_d\|_2$},
            y tick label style={
                /pgf/number format/.cd,
                    fixed,
                    fixed zerofill,
                    precision=2,
                /tikz/.cd
            },
            shorter legend
        ]
        \nextgroupplot[%
            xmin=0.1, xmax=2,
            ymin=0, ymax=1.5,
        ]
        \addplot[line] table[] {dat/Sabioni_consistency_check.dat};
        \addlegendentry{\emph{Sabioni et al.}}
        \addplot[optimum] coordinates {(0.7, 0.1284518720219032)} node[] (opt1) {};
        \node[] (beta1) at (rel axis cs:0.5, 0.5) {\scriptsize $\beta_1 = 0.7$};
        \draw[arrow] (beta1) -- (opt1);
        \nextgroupplot[%
            xmin=0.1, xmax=2,
            ymin=0, ymax=1.5,
        ]
        \addplot[line] table[] {dat/Davies_Long_consistency_check.dat};
        \addlegendentry{\emph{Davies \& Long}}
        \addplot[optimum] coordinates {(0.9, 0.08301570214742847)} node[] (opt2) {};
        \node[] (beta2) at (rel axis cs:0.5, 0.5) {\scriptsize $\beta_2 = 0.9$};
        \draw[arrow] (beta2) -- (opt2);
        \nextgroupplot[%
            xmin=0.1, xmax=2,
            ymin=0, ymax=1.5,
        ]
        \addplot[line] table[] {dat/Turnbull_consistency_check.dat};
        \addlegendentry{\emph{Turnbull et al.}}
        \addplot[optimum] coordinates {(0.8, 0.03272060559952702)} node[] (opt3) {};
        \node[] (beta3) at (rel axis cs:0.5, 0.5) {\scriptsize $\beta_3 = 0.8$};
        \draw[arrow] (beta3) -- (opt3);
        \nextgroupplot[%
            xmin=1, xmax=20,
            ymin=0.5, ymax=2,
        ]
        \addplot[line] table[] {dat/Miekely_Felix_consistency_check.dat};
        \addlegendentry{\emph{Miekeley \& Felix}}
        \addplot[optimum] coordinates {(20.0, 0.6913232488149672)} node[] (opt4) {};
        \node[] (beta4) at (rel axis cs:0.75, 0.5) {\scriptsize $\beta_4 = 20$};
        \draw[arrow] (beta4) -- (opt4);
        \nextgroupplot[%
            xmin=1, xmax=20,
            ymin=0.68, ymax=0.74,
        ]
        \addplot[line] table[] {dat/stoichiometry_consistency_check.dat};
        \addlegendentry{\emph{non-stoichiometry}}
        \addplot[optimum] coordinates {(9.0, 0.6864580255827575)} node[] (opt5) {};
        \node[] (beta5) at (rel axis cs:0.5, 0.6) {\scriptsize $\beta_5 = 9$};
        \draw[arrow] (beta5) -- (opt5);
    \end{groupplot}
\end{tikzpicture}
	\tikzexternaldisable

    \caption{Relative $\ell_2$-norm of the difference between the reported summary statistics $\bss_d$ and the computed summary statistics from the pushforward posterior $\tilde\bss_d$ as a function of the synthetic data scaling parameter $\beta_d$ for each set of experimental data summaries. The computed summary statistics $\tilde\bss_d$ correspond to three times the standard deviation ($3\sigma$). Values of the scaling parameter $\beta_d$ used in the final calibration are indicated by \protect\raisebox{-0.5pt}{\protect\tikz{\protect\node[circle, fill=Red, draw=none, inner sep=1.75pt] {};}}.}
    \label{fig:consistent_data_sets}
\end{figure}

\subsection{Generating consistent synthetic data sets}\label{sec:generating_consistent_synthetic_data_sets}

The first step in our calibration framework outlined in \cref{sec:bayesian_calibration_for_summary_statistics} is the generation of consistent synthetic data sets. Having prescribed the synthetic data sets as in~\cref{eq:synthetic_data_sets}, this requires us to find an appropriate value for the scaling factor $\beta_d$ in the variance of the data-generating distribution. These scaling factors can be found by matching the statistics of the pushforward posterior $\tilde\bss_d$, to the given measurement errors $\bss_d$, such that the criterion in~\eqref{eq:dfi_constraint} is satisfied. In our experiments, we will use the $3\sigma$ statistic of the pushforward posterior, and use the relative $\ell_2$ norm as distance metric. We evaluate the statistics $\tilde\bss_d$ for each data set $d=1, 2, \ldots, D$ based on samples from the pushforward posterior, obtained by MCMC, with the log-pooled likelihood defined in~\cref{eq:likelihood_with_surrogate}. We use a total of $M = 10^6$ MCMC steps, $K = 100$ synthetic data sets, a proposal jump size of $0.5$, and evaluate the pushforward posterior with a burn-in of $100\,000$ and subsampling rate of $5$, i.e., we keep 1 out of every 5 samples in order to decorrelate the Markov chain. In order to obtain good starting values for the chain, we performed a few iterations with the deterministic optimization method L-BFGS, see~\cite{liu1989}. We evaluate the metric in~\eqref{eq:dfi_constraint} for a set of 20 judiciously chosen values of the scaling factor $\beta_d$.

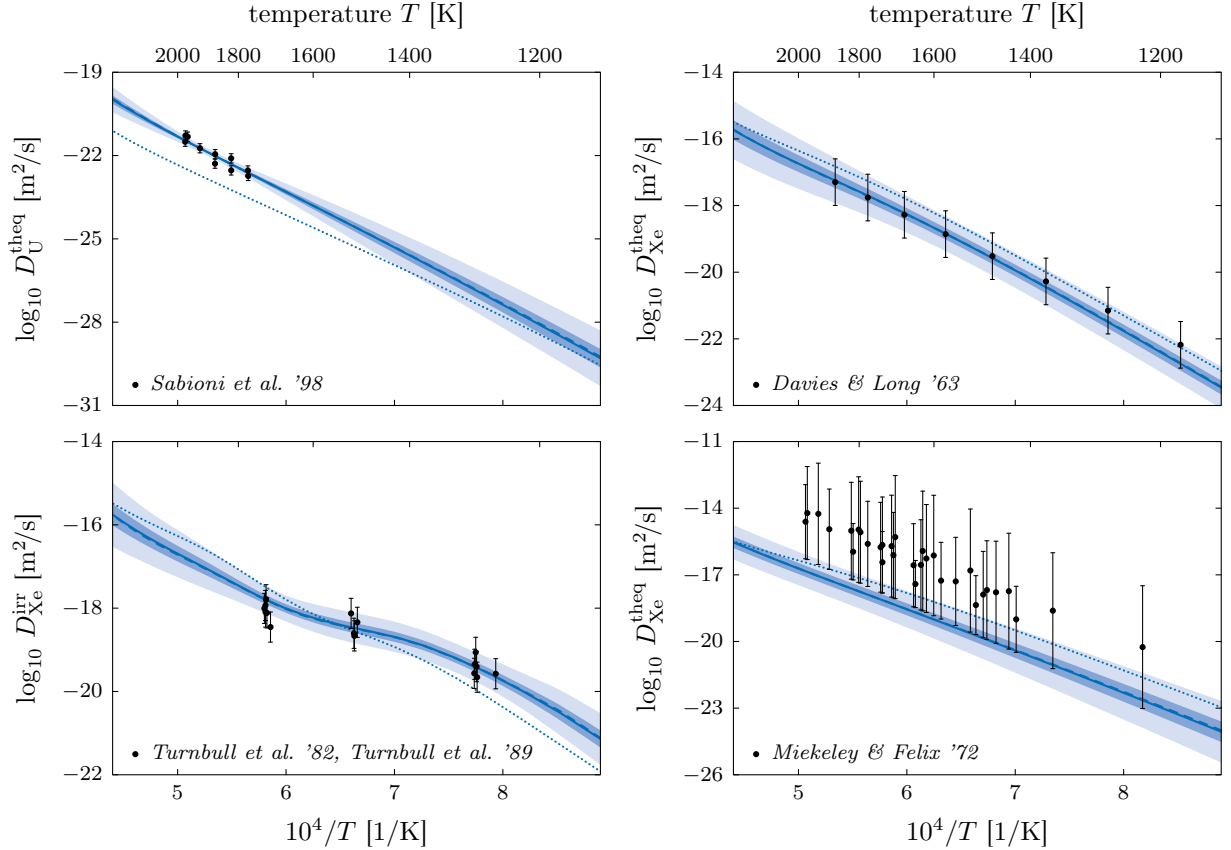
\begin{figure*}[t]
    \centering
	\tikzsetnextfilename{pushforward_posterior_diffusivities}
	\tikzexternalenable
	\newcommand{\plotcalibration}[1]{%
    \pgfplotstableread[header=false]{dat/#1_calibration.dat}{\data}
    \pgfplotstableread[header=false]{dat/#1_measurements.dat}{\meas}
        \addplot[name path=bottom, draw=none, smooth, forget plot] table[x index=0, y expr=\thisrow{1} - 3*\thisrow{2}] \data;
        \addplot[name path=top, draw=none, smooth, forget plot] table[x index=0, y expr=\thisrow{1} + 3*\thisrow{2}] \data;
        \addplot+[%
            draw=none,
            threesigma,
            forget plot
        ] fill between[of=bottom and top];
        \addplot[name path=bottom, draw=none, smooth, forget plot] table[x index=0, y expr=\thisrow{1} - \thisrow{2}] \data;
        \addplot[name path=top, draw=none, smooth, forget plot] table[x index=0, y expr=\thisrow{1} + \thisrow{2}] \data;
        \addplot+[%
            draw=none,
            sigma,
            forget plot
        ] fill between[of=bottom and top];
        \addplot[nominal, forget plot] table[x index=0, y index=4] \data;
        \addplot[mean, forget plot] table[x index=0, y index=1] \data;
        \addplot[map, forget plot] table[x index=0, y index=3] \data;
        \addplot[measurement, default error bar] table[x expr=1e4/\thisrowno{0}, y index=1, y error index=2] \meas;
}

\setlength{\figurewidth}{8cm}
\setlength{\figureheight}{6cm}
\begin{tikzpicture}[%
        sample/.style={default marker line, black!20},
        measurement/.style={black, default marker line, only marks},
        mean/.style={default line, NavyBlue},
        map/.style={default dashed line, NavyBlue},
        threesigma/.style={NavyBlue!15},
        sigma/.style={NavyBlue!50},
        nominal/.style={default dotted line, NavyBlue}
    ]
    \begin{groupplot}[%
            group style={%
                group size=2 by 2,
                xticklabels at=edge bottom,
                xlabels at=edge bottom,
                vertical sep=\baselineskip,
                horizontal sep=1.75cm
            },
            default axis,
            axis on top,
            xtick={5, ..., 8},
            xlabel={$10^4/T$ [\si[per-mode=symbol]{\per\kelvin}]},
            legend style={%
                at={(0.025, 0)},
                anchor=south west,
                /tikz/every odd column/.append style={column sep=1mm}
            },
            clip mode=individual,
            axis x line*=bottom
        ]
        \nextgroupplot[%
            xmin=4.4, xmax=8.9, ymin=-31, ymax=-19,
            ytick={-31, -28, ..., -19},
            ylabel={$\log_{10}$ $\dutheq$ [\si[per-mode=symbol]{\square\metre\per\second}]}
        ]
        \plotcalibration{Sabioni}
        \addlegendentry{\emph{Sabioni et al.\ '98}}
        \nextgroupplot[%
            xmin=4.4, xmax=8.9, ymin=-24, ymax=-14,
            ytick={-24, -22, ..., -14},
            ylabel={$\log_{10}$ $\dxetheq$ [\si[per-mode=symbol]{\square\metre\per\second}]}
        ]
        \plotcalibration{Davies_Long}
        \addlegendentry{\emph{Davies \& Long '63}}
        \nextgroupplot[%
            xmin=4.4, xmax=8.9, ymin=-22, ymax=-14,
            ytick={-14, -16, ..., -22},
            ylabel={$\log_{10}$ $\dxeirr$ [\si[per-mode=symbol]{\square\metre\per\second}]}
        ]
        \plotcalibration{Turnbull}
        \addlegendentry{\emph{Turnbull et al.\ '82, Turnbull et al.\ '89}}
        \nextgroupplot[%
            xmin=4.4, xmax=8.9, ymin=-26, ymax=-11,
            ytick={-26, -23, ..., -11},
            ylabel={$\log_{10}$ $\dxetheq$ [\si[per-mode=symbol]{\square\metre\per\second}]}
        ]
        \plotcalibration{Miekely_Felix}
        \addlegendentry{\emph{Miekeley \& Felix '72}}
    \end{groupplot}
    \begin{groupplot}[%
            group style={%
                group size=2 by 2,
                xticklabels at=edge top,
                xlabels at=edge top,
                vertical sep=\baselineskip,
                horizontal sep=1.75cm
            },
            default axis,
            axis on top,
            axis y line=none,
            xlabel={temperature $T$ [\si{\kelvin}]},
            xticklabel pos=top,
            axis x line*=top,
            xmin=4.4, xmax=8.9,
            ymin=4.4, ymax=8.9,
            xtick={5, 5.56, 6.25, 7.14, 8.34},
        ]
        \nextgroupplot[%
            xticklabels={2000, 1800, 1600, 1400, 1200},
        ]
        \addplot[opacity=0] coordinates { (8.85,8.9) (8.9,8.85) }; 
        \nextgroupplot[%
            xticklabels={2000, 1800, 1600, 1400, 1200},
        ]
        \addplot[opacity=0] coordinates { (8.85,8.9) (8.9,8.85) }; 
        \nextgroupplot[%
        ]
        \addplot[opacity=0] coordinates { (8.85,8.9) (8.9,8.85) }; 
        \nextgroupplot[%
        ]
        \addplot[opacity=0] coordinates { (8.85,8.9) (8.9,8.85) }; 
    \end{groupplot}
\end{tikzpicture}
	\tikzexternaldisable

    \caption{Mean value (\emph{solid line}), mean value $\pm$ standard deviation (\emph{dark shaded area}), mean value $\pm$ three times the standard deviation (\emph{light shaded area}), and MAP (\emph{dashed line}) of the pushforward posterior predictions for $\dutheq$, $\dxetheq$, $\dxeirr$ at the particular operating conditions corresponding to each experiment. We also indicate the diffusivity predictions using the nominal parameter values (\emph{dotted line}) from~\cite{matthews2020}. The data summaries are shown as error bars for comparison.}
    \label{fig:pushforward_posterior_diffusivities}
\end{figure*}

The values of $\|\bss_d - \tilde\bss_d\|_2$ as a function of $\beta_d$ are reported in \cref{fig:consistent_data_sets}. Note that, in order to generate these plots, we fixed the random seeds used to generate the synthetic data set $\calZ_d$ for various choices of $\beta_d$, because we found numerically that this provides slightly more stable results, and $K=100$ synthetic data sets appears to be sufficient to avoid any significant difference in the pushforward posterior predictions due to variations in the number of synthetic data sets $K$. For each experiment, except Miekeley \& Felix, there is a well-defined optimum, where the difference between the reported measurement errors and the $3\sigma$ statistic of the pushforward posterior are in good agreement. For the Miekeley \& Felix experiment, the relative error between the predicted statistics and the reported data summaries is large, and this does not seem to improve by including even larger values of $\beta_d$. This can probably be explained by the relatively large values of the reported uncertainties.

\subsection{Calibration results}\label{sec:calibration_results}

We combine the 5 consistent synthetic data sets $\calZ_d$ obtained in \cref{sec:generating_consistent_synthetic_data_sets} into a single synthetic data set $\calZ = \{ \calZ_d \}_{d=1}^5$. This allows us to construct the full posterior $p(\bsnu | \calZ)$ using MCMC, with a log-likelihood given by \cref{eq:likelihood_with_surrogate}, assuming $\alpha_d=1$, $d=1, 2, \ldots, D$. As before, we use a total of $M = 10^6$ MCMC steps, a proposal jump size of $0.5$, and obtain samples from the posterior assuming a burn-in of $100\,000$ and subsampling rate of $5$. A trace plot of the Markov chain illustrating good mixing is shown in \cref{fig:parameter_chains}.

\begin{figure*}[t]
    \centering
	\tikzsetnextfilename{pushforward_posterior_stoichiometry}
	\tikzexternalenable
	\pgfdeclareplotmark{left}
{%
  \pgfpathmoveto{\pgfqpoint{1.5\pgfplotmarksize}{0pt}}%
  \pgfpathlineto{\pgfqpoint{0\pgfplotmarksize}{0pt}}%
  \pgfpathmoveto{\pgfqpoint{-1.5\pgfplotmarksize}{0pt}}%
  \pgfusepathqstroke
}%

\pgfdeclareplotmark{right}
{%
  \pgfpathmoveto{\pgfqpoint{-1.5\pgfplotmarksize}{0pt}}%
  \pgfpathlineto{\pgfqpoint{0\pgfplotmarksize}{0pt}}%
  \pgfpathmoveto{\pgfqpoint{1.5\pgfplotmarksize}{0pt}}%
  \pgfusepathqstroke
}%

\setlength{\figurewidth}{15cm}
\setlength{\figureheight}{6.5cm}
\begin{tikzpicture}[%
    calibrated/.style={NavyBlue, default marker line, only marks, mark options={fill=white, line width=0.4pt}},
    measurement/.style={default marker line, only marks, mark options={line width=0.4pt}}
    ]
    \begin{axis}[%
            default axis,
            axis on top,
            xmin=0, xmax=105,
            legend style={%
                at={(0.97, 0.03)},
                anchor=south east,
                /tikz/every odd column/.append style={column sep=1mm},
            },
            xlabel={index of T/$\log_{10}(\pOtwo{})$ value},
            ylabel={non-stoichiometry ($\nonstoich$)},
            clip mode=individual,
            ymin=-0.006, ymax=0.006,
            ytick={-0.006, -0.004, ..., 0.006},
            xtick={10, 20, ..., 100},
        ]
        \addplot[measurement, default error bar, error mark={left}, y dir=plus] table[x expr=\coordindex+1, y index=0, y error index=1] {dat/stoichiometry_measurements.dat};
        \addplot[measurement, forget plot, default error bar, error mark={right}, y dir=minus] table[x expr=\coordindex+1, y index=0, y error index=1] {dat/stoichiometry_measurements.dat};
        \addlegendentry{stoichiometric data summaries}
        \addplot[calibrated, default error bar, error mark={right}, y dir=plus] table[x expr=\coordindex+1, y index=1, y error index=2] {dat/stoichiometry_calibration.dat};
        \addplot[calibrated, forget plot, default error bar, error mark={left}, y dir=minus] table[x expr=\coordindex+1, y index=1, y error index=2] {dat/stoichiometry_calibration.dat};
        \addlegendentry{estimated non-stoichiometry}
    \end{axis}
\end{tikzpicture}
	\tikzexternaldisable

    \caption{Mean and standard deviation of the pushforward posterior predictions for $x$ in $\nonstoich$ shown as error bars (\emph{hollow markers}). The data summaries are shown as error bars (\emph{full markers}) for comparison.}
    \label{fig:pushforward_posterior_stoichiometry}
\end{figure*}
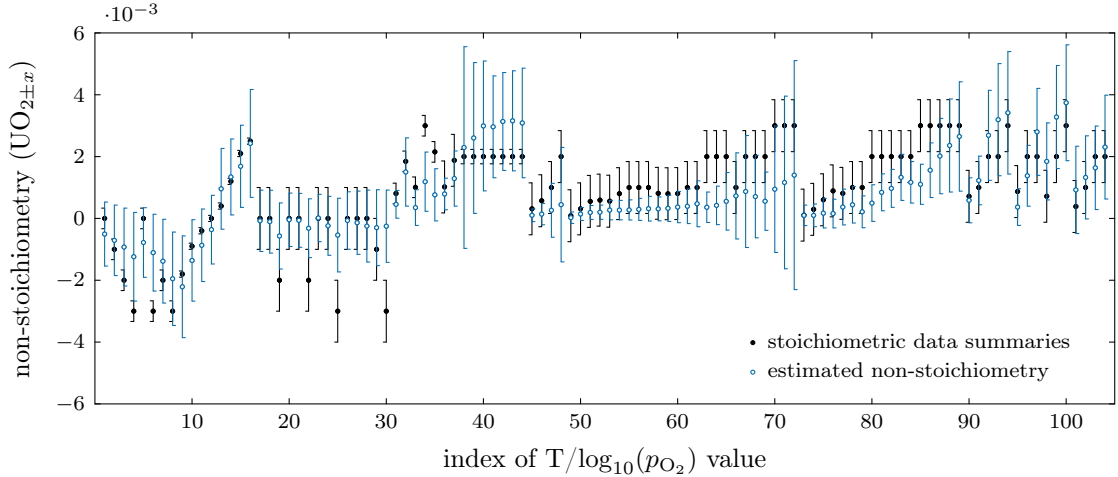

In \cref{fig:pushforward_posterior_diffusivities}, we plot the pushforward posterior of the diffusivity quantities, obtained by evaluating the PCE surrogates for the diffusivities in the posterior samples. We indicate the mean (\emph{full line}), standard deviation (\emph{dark shaded area}), and three times the standard deviation (\emph{light shaded area}) of the pushforward posterior, as well as the pushforward \emph{maximum a posteriori} (MAP) prediction (\emph{dashed line}). The MAP values represent the most probable parameter values after the parameter estimation procedure. Diffusivity predictions using the nominal parameter values from~\cite{matthews2020} are indicated by the \emph{dotted line}. Note that the mean of the pushforward posterior as well as the MAP prediction are in better agreement with the majority of the experimental data summaries than predictions with the original nominal parameters, confirming the efficacy of our calibration framework. Furthermore, the $\pm 3\sigma$ statistic of the pushforward posterior is in good agreement with the reported uncertainties in the majority of the data summaries, as desired. Notably, for the Miekeley \& Felix data set (\emph{bottom right}), there is a discrepancy between the pushforward posterior prediction and the data summaries. This is in agreement with our expectations, as we deem the Miekeley \& Felix experiment to be slightly less reliable. Alternatively, it is possible that the conditions of this experiment are sufficiently different from the assumed approximately stoichiometric \ce{UO2} conditions in \Centipede{}, such that new diffusion mechanisms involving, for example, different cluster charge states could be activated. \Centipede{} would not be able to capture such behavior. The present study helps to identify this issue, and points to potential directions for model improvement. 

Similarly, \cref{fig:pushforward_posterior_stoichiometry} shows a comparison between the pushforward posterior of the non-stoichiometry ($\nonstoich$) and the reported summary statistics. We note that there is a reasonable agreement between both. Our results indicate that the $3\sigma$ statistic of the non-stoichiometry is strongly dependent on the value of $T$ and $\pOtwo$. However, the reported data summaries appear to have mostly constant uncertainties.

\subsection{Effect of weights in the likelihood}\label{sec:effect_of_weights_in_the_likelihood}

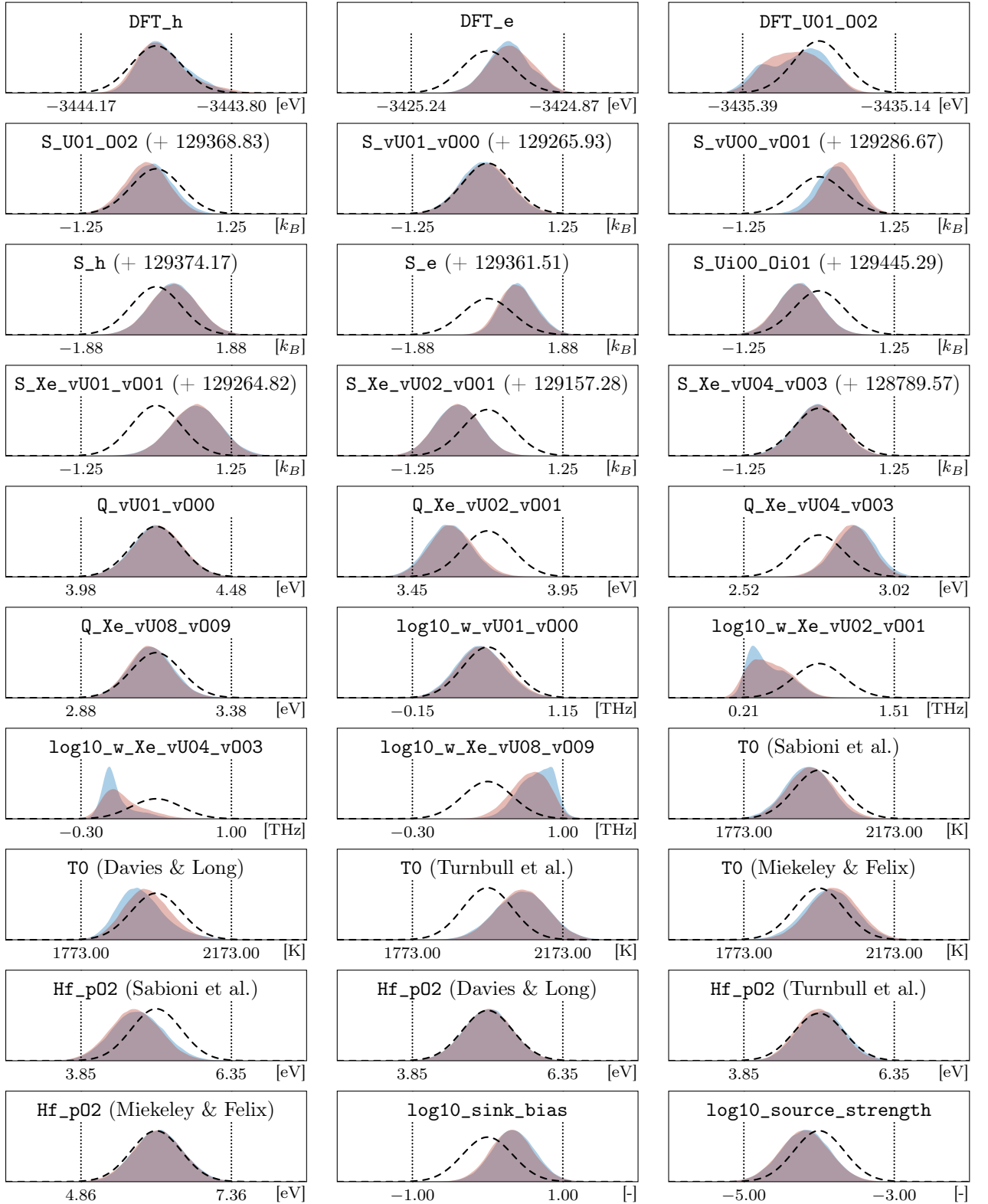
\begin{figure*}
    \centering
	\tikzsetnextfilename{parameter_posteriors}
	\tikzexternalenable
	\newcommand{\plotposterior}[5]{%
    \nextgroupplot[%
        title={\strut #2},
        xtick={#3, #4},
        xlabel={#5}
    ]
    \pgfmathtruncatemacro{\xvals}{#1}
    \pgfmathtruncatemacro{\prior}{#1 + 1}
    \pgfmathtruncatemacro{\unweighted}{#1 + 2}
    \pgfmathtruncatemacro{\weighted}{#1 + 3}
    \addplot[name path=bottom, draw=none, forget plot] table[x index=\xvals, y expr=0] \posteriors;
    \addplot[prior] table[x index=\xvals, y index=\prior] \posteriors;
    \addplot[unweighted, smooth, name path=unweighted] table[x index=\xvals, y index=\unweighted] \posteriors;
    \addplot[weighted, smooth, name path=weighted] table[x index=\xvals, y index=\weighted] \posteriors;
    \addplot+[%
        draw=none,
        unweighted,
        forget plot
    ] fill between[of=bottom and unweighted];
    \addplot+[%
        draw=none,
        weighted,
        forget plot
    ] fill between[of=bottom and weighted];
    \pgfplotsextra{
        \pgfkeysgetvalue{/pgfplots/ymin}{\ymin}
        \pgfkeysgetvalue{/pgfplots/ymax}{\ymax}
        \draw[bounds] (#3, \ymin) -- (#3, 0.65*\ymax);
        \draw[bounds] (#4, \ymin) -- (#4, 0.65*\ymax);
    }
}

\setlength{\figurewidth}{6.75cm}
\setlength{\figureheight}{3.15cm}
\pgfplotstableread[header=false]{dat/parameter_posteriors.dat}{\posteriors}
\begin{tikzpicture}[
        prior/.style={default dashed line, black, smooth},
        filled area/.style={draw=none, opacity=0.33, smooth},
        unweighted/.style={filled area, NavyBlue},
        weighted/.style={filled area, BrickRed},
        bounds/.style={default dotted line, black}
	]
	\begin{groupplot}[%
            group style={%
                group size=3 by 10,
                vertical sep=1.1\baselineskip,
                horizontal sep=1.1\baselineskip,
            },
            default axis,
            axis on top,
            ymajorticks=false,
            enlarge x limits=0,
            enlarge y limits={abs=1.75em, upper},
            major tick length={0pt},
            every axis title/.style={at={(0.5,1)}, anchor=north, font=\small},
            x tick label style={
                /pgf/number format/.cd,
                    fixed,
                    fixed zerofill,
                    precision=2,
                /tikz/.cd
            },
            every axis x label/.style={
                at={(rel axis cs:1,-0.04)},
                anchor=north east,
                font=\scriptsize,
                inner sep=0pt
            }
        ]
        \plotposterior{0}{\small\texttt{DFT\_h}}{-3444.1726670000003}{-3443.7976670000003}{[\si{\eV}]}
        \plotposterior{4}{\small\texttt{DFT\_e}}{-3425.243167}{-3424.868167}{[\si{\eV}]}
        \plotposterior{8}{\small\texttt{DFT\_U01\_O02}}{-3435.3895}{-3435.1395}{[\si{\eV}]}
        \plotposterior{12}{\small\texttt{S\_U01\_O02} (+ 129368.83)}{-1.25}{1.25}{[$k_B$]}
        \plotposterior{16}{\small\texttt{S\_vU01\_vO00} (+ 129265.93)}{-1.25}{1.25}{[$k_B$]}
        \plotposterior{20}{\small\texttt{S\_vU00\_vO01} (+ 129286.67)}{-1.25}{1.25}{[$k_B$]}
        \plotposterior{24}{\small\texttt{S\_h} (+ 129374.17)}{-1.875}{1.875}{[$k_B$]}
        \plotposterior{28}{\small\texttt{S\_e} (+ 129361.51)}{-1.875}{1.875}{[$k_B$]}
        \plotposterior{32}{\small\texttt{S\_Ui00\_Oi01} (+ 129445.29)}{-1.25}{1.25}{[$k_B$]}
        \plotposterior{36}{\small\texttt{S\_Xe\_vU01\_vO01} (+ 129264.82)}{-1.25}{1.25}{[$k_B$]}
        \plotposterior{40}{\small\texttt{S\_Xe\_vU02\_vO01} (+ 129157.28)}{-1.25}{1.25}{[$k_B$]}
        \plotposterior{44}{\small\texttt{S\_Xe\_vU04\_vO03} (+ 128789.57)}{-1.25}{1.25}{[$k_B$]}
        \plotposterior{48}{\small\texttt{Q\_vU01\_vO00}}{3.9831699999999994}{4.483169999999999}{[\si{\eV}]}
        \plotposterior{52}{\small\texttt{Q\_Xe\_vU02\_vO01}}{3.44567683}{3.94567683}{[\si{\eV}]}
        \plotposterior{56}{\small\texttt{Q\_Xe\_vU04\_vO03}}{2.52072157}{3.02072157}{[\si{\eV}]}
        \plotposterior{60}{\small\texttt{Q\_Xe\_vU08\_vO09}}{2.87745711}{3.37745711}{[\si{\eV}]}
        \plotposterior{64}{\small\texttt{log10\_w\_vU01\_vO00}}{-0.14874165128092473}{1.1522883443830565}{[\si{\THz}]}
        \plotposterior{68}{\small\texttt{log10\_w\_Xe\_vU02\_vO01}}{0.21085336531489318}{1.5118833609788744}{[\si{\THz}]}
        \plotposterior{72}{\small\texttt{log10\_w\_Xe\_vU04\_vO03}}{-0.3010299956639812}{1.0}{[\si{\THz}]}
        \plotposterior{76}{\small\texttt{log10\_w\_Xe\_vU08\_vO09}}{-0.3010299956639812}{1.0}{[\si{\THz}]}
        \plotposterior{80}{\small\texttt{T0} (Sabioni et al.)}{1773.0}{2173.0}{[\si{\K}]}
        \plotposterior{84}{\small\texttt{T0} (Davies \& Long)}{1773.0}{2173.0}{[\si{\K}]}
        \plotposterior{88}{\small\texttt{T0} (Turnbull et al.)}{1773.0}{2173.0}{[\si{\K}]}
        \plotposterior{92}{\small\texttt{T0} (Miekeley \& Felix)}{1773.0}{2173.0}{[\si{\K}]}
        \plotposterior{96}{\small\texttt{Hf\_pO2} (Sabioni et al.)}{3.8499999999999996}{6.35}{[\si{\eV}]}
        \plotposterior{100}{\small\texttt{Hf\_pO2} (Davies \& Long)}{3.8499999999999996}{6.35}{[\si{\eV}]}
        \plotposterior{104}{\small\texttt{Hf\_pO2} (Turnbull et al.)}{3.8499999999999996}{6.35}{[\si{\eV}]}
        \plotposterior{108}{\small\texttt{Hf\_pO2} (Miekeley \& Felix)}{4.86}{7.36}{[\si{\eV}]}
        \plotposterior{112}{\small\texttt{log10\_sink\_bias}}{-1.0}{1.0}{[-]}
        \plotposterior{116}{\small\texttt{log10\_source\_strength}}{-5.0}{-3.0}{[-]}
	\end{groupplot}
\end{tikzpicture}
	\tikzexternaldisable

    \caption{Probability density function of the prior (\captionlegend{default dashed line, black}), marginal posterior (\captionsquare{NavyBlue, opacity=0.33}) and marginal posterior with weighted likelihood (\captionsquare{BrickRed, opacity=0.33}) for all parameters. The lower and upper bounds used to construct the PCE surrogate models are indicated by the vertical lines (\captionlegend{default dotted line, black}). A number next to a parameter name indicates that the corresponding value needs to be added to the shown parameter bounds. Note that the \texttt{DFT} parameters represent total energies from DFT calculations as used by \Centipede{}, and that entropies are presented in a reduced form suitable for \Centipede{} and should not be interpreted as an actual entropy, see~\cite{davies1963,cooper2016,liu2022}.}
    \label{fig:parameter_posteriors}
\end{figure*}

Next, we investigate the effect of weights in the likelihood to compensate for the number of measurements in each reported set of data summaries. To this end, we repeat the calibration with a weighted likelihood, where weights are chosen according to \cref{eq:weights}. We compare the obtained marginal posterior densities on the parameters for both the unweighted and weighted likelihoods in \cref{fig:parameter_posteriors}. Overall, the marginal posteriors are in good agreement, although there are some quantitative differences for the formation energy of \ce{UO2} and the attempt frequencies of $\ce{Xe}_{\ce{U2O}}$, $\ce{Xe}_{\ce{U4O3}}$ and $\ce{Xe}_{\ce{U8O9}}$. The numerical values of the MAP predictions of each parameter in the unweighted and weighted case have been included in \cref{tab:uncertain_parameters}.

\subsection{Discussion}\label{sec:discussion}

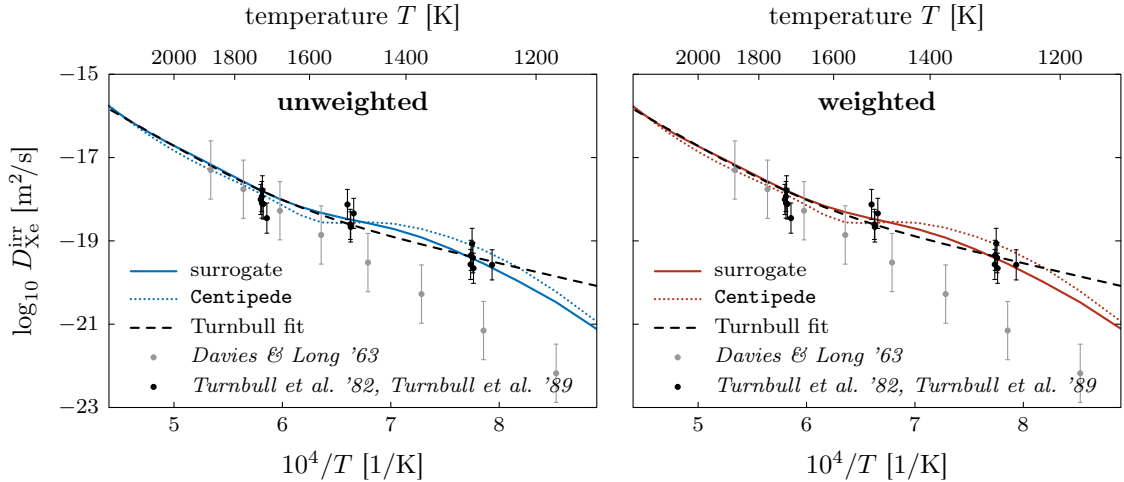
\begin{figure*}[t]
    \centering
	\tikzsetnextfilename{turnbull_diffusivities}
	\tikzexternalenable
	\setlength{\figurewidth}{8cm}
\setlength{\figureheight}{6cm}
\begin{tikzpicture}[%
        sample/.style={default marker line, black!20},
        measurement/.style={black, default marker line, only marks}
    ]
    \begin{groupplot}[%
            group style={%
                group size=2 by 1,
                xticklabels at=edge bottom,
                xlabels at=edge bottom,
                yticklabels at=edge left,
                ylabels at=edge left,
                vertical sep=\baselineskip,
                horizontal sep=\baselineskip,
            },
            default axis,
            axis on top,
            xtick={5, ..., 8},
            legend style={%
                font={\scriptsize},
                at={(0.025, 0)},
                anchor=south west,
                /tikz/every odd column/.append style={column sep=1mm}
            },
            clip mode=individual,
            xlabel={$10^4/T$ [\si[per-mode=symbol]{\per\kelvin}]},
            axis x line*=bottom,
            every axis title/.style={at={(0.5,0.99)}, anchor=north, font=\small},
            xmin=4.4, xmax=8.9, ymin=-23, ymax=-15,
        ]
        \nextgroupplot[%
            ytick={-15, -17, ..., -23},
            ylabel={$\log_{10}$ $\dxeirr$ [\si[per-mode=symbol]{\square\metre\per\second}]},
            title={\strut\textbf{unweighted}}
	    ]
        \addplot[default line, NavyBlue] table[x index=0, y index=1] {dat/unweighted_turnbull_calibration.dat};
        \addlegendentry{surrogate}
        \addplot[default dotted line, NavyBlue] table[x index=0, y index=2] {dat/unweighted_turnbull_calibration.dat};
        \addlegendentry{\Centipede{}}
        \addplot[default dashed line, black] table[x index=0, y index=1] {dat/turnbull.dat};
        \addlegendentry{Turnbull fit}
        \addplot[measurement, black!40, default error bar] table[x expr=1e4/\thisrowno{0}, y index=1, y error index=2] {dat/Davies_Long_measurements.dat};
        \addlegendentry{\emph{Davies \& Long '63}}
        \addplot[measurement, default error bar] table[x expr=1e4/\thisrowno{0}, y index=1, y error index=2] {dat/Turnbull_measurements.dat};
        \addlegendentry{\emph{Turnbull et al.\ '82, Turnbull et al.\ '89}}
        \nextgroupplot[%
            xmin=4.4, xmax=8.9, ymin=-23, ymax=-15,
            ytick={-15, -17, ..., -23},
            title={\strut\textbf{weighted}},
        ]
        \addplot[default line, BrickRed] table[x index=0, y index=1] {dat/weighted_turnbull_calibration.dat};
        \addlegendentry{surrogate}
        \addplot[default dotted line, BrickRed] table[x index=0, y index=2] {dat/weighted_turnbull_calibration.dat};
        \addlegendentry{\Centipede{}}
        \addplot[default dashed line, black] table[x index=0, y index=1] {dat/turnbull.dat};
        \addlegendentry{Turnbull fit}
        \addplot[measurement, black!40, default error bar] table[x expr=1e4/\thisrowno{0}, y index=1, y error index=2] {dat/Davies_Long_measurements.dat};
        \addplot[measurement, default error bar] table[x expr=1e4/\thisrowno{0}, y index=1, y error index=2] {dat/Turnbull_measurements.dat};
        \addlegendentry{\emph{Davies \& Long '63}}
        \addlegendentry{\emph{Turnbull et al.\ '82, Turnbull et al.\ '89}}
    \end{groupplot}%
    \begin{groupplot}[%
            group style={%
                group size=2 by 1,
                xticklabels at=edge top,
                xlabels at=edge top,
                yticklabels at=edge left,
                ylabels at=edge left,
                vertical sep=\baselineskip,
                horizontal sep=\baselineskip
            },
            default axis,
            legend style={%
                font={\scriptsize},
                at={(1, 0.5)},
                anchor=east,
                /tikz/every odd column/.append style={column sep=1mm}
            },
            axis on top,
            axis y line=none,
            xlabel={temperature $T$ [\si{\kelvin}]},
            xticklabel pos=top,
            axis x line*=top,
            xtick={5, 5.56, 6.25, 7.14, 8.34},
            xmin=4.4, xmax=8.9, ymin=-23, ymax=-15,
        ]
        \nextgroupplot[%
            xticklabels={2000, 1800, 1600, 1400, 1200},
        ]
        \addplot[opacity=0] coordinates { (8.85,8.9) (8.9,8.85) }; 
        \nextgroupplot[%
            xticklabels={2000, 1800, 1600, 1400, 1200},
        ]
        \addplot[opacity=0] coordinates { (8.85,8.9) (8.9,8.85) }; 
    \end{groupplot}
\end{tikzpicture}
	\tikzexternaldisable

    \caption{Pushforward MAP prediction for the xenon diffusivity under irradiation with the surrogate (\emph{solid line}) and pushforward MAP prediction of \Centipede{} (\emph{dotted line}) for both the unweighted likelihood (\emph{left}) and the weighted likelihood from \cref{eq:weighted_likelihood}. Also shown are the Turnbull et al.\  measurements for the diffusivity under irradiation from~\cite{turnbull1982,turnbull1989} and the Davies \& Long measurements for the intrinsic diffusivity from~\cite{davies1963}, together with the fit by Turnbull et al.\ (\emph{dashed line}). Note the excellent agreement between the predicted xenon diffusivity and the Turnbull et al.\ fit from the data for high temperatures.}
    \label{fig:turnbull_diffusivities}
\end{figure*}

In \cref{fig:turnbull_diffusivities}, we compare the pushforward MAP prediction of the xenon diffusivity under irradiation conditions computed using the surrogate (\emph{full line}) with the pushforward MAP prediction computed using \Centipede{} (\emph{dotted line}), for the unweighted (\emph{left}) and weighted (\emph{right}) posterior, and using the particular values for the operating conditions estimated for the Turnbull et al.\ experiments. We also show the Turnbull et al.\ experimental data summaries for the xenon diffusivity under irradiation conditions, and the Davies \& Long experimental data summaries for the intrinsic xenon diffusivity, together with the Turnbull et al.\ fit from~\cite{turnbull1982} (\emph{dashed line}). The Turnbull et al.\ fit for xenon diffusion consists of two terms, a $D_1$ contribution from Davies \& Long and a $D_2$ contribution from Turnbull et al., see~\cref{fig:diffusivity_regions}, and is commonly used in the literature, because it gives good results for integral fuel rod tests (not shown here), and because it captures the Davies \& Long and Turnbull et al.\ data that was used to rationlize its form.

The surrogate prediction with the MAP parameters, as well as the corresponding \Centipede{} prediction, are in excellent agreement with the Turnbull fit for temperatures above \SI{1600}{\kelvin}. The fit is slightly worse for low temperatures because the athermal $D_3$ term contributing to diffusion below about \SI{1200}{\kelvin} is not predicted by \Centipede{} (note that we preprocessed the Turnbull et al.\ data, as well as the fit from the data, to remove the athermal $D_3$ term). The excellent agreement of the calibrated diffusivity with the (extrapolation of the) Davies \& Long data is a consequence of the estimated operating conditions for the Davies \& Long data ($\texttt{T0} \approx \SI{2070.1338}{\K}$ and $\texttt{Hf\_pO2} \approx \SI{5.0941}{\electronvolt}$ in the weighted setting) being very similar to those estimated for the Turnbull et al.\ data ($\texttt{T0} \approx \SI{1942.0619}{\K}$ and $\texttt{Hf\_pO2} \approx \SI{5.1278}{\electronvolt}$ in the weighted setting).

\begin{figure*}[t]
    \centering
	\tikzsetnextfilename{diffusion_mechanism}
	\tikzexternalenable
	\setlength{\figurewidth}{7.75cm}
\setlength{\figureheight}{6cm}
\pgfplotstableread[header=false]{dat/diffusion_mechanism_nominal.dat}{\nominal}
\pgfplotstableread[header=false]{dat/diffusion_mechanism_weighted.dat}{\weighted}
\begin{tikzpicture}[%
        defect/.style={default line, smooth},
        key/.style={circle, fill=white, inner sep=1pt, draw},
    ]
    \begin{groupplot}[%
            group style={%
                group size=2 by 1,
                xticklabels at=edge bottom,
                xlabels at=edge bottom,
                yticklabels at=edge left,
                ylabels at=edge left,
                vertical sep=\baselineskip,
                horizontal sep=\baselineskip,
            },
            default axis,
            axis on top,
            xtick={5, ..., 8},
            legend style={%
                font={\scriptsize},
                at={(1, 1)},
                anchor=north west,
                /tikz/every odd column/.append style={column sep=1mm}
            },
            clip mode=individual,
            xlabel={$10^4/T$ [\si[per-mode=symbol]{\per\kelvin}]},
            axis x line*=bottom,
            every axis title/.style={at={(0.5,0.99)}, anchor=north, font=\small},
            xmin=4.4, xmax=8.9, ymin=-30, ymax=-15,
            ytick={-15, -20, ..., -30},
            ylabel={$\log_{10}$ xenon diffusivity [\si[per-mode=symbol]{\square\metre\per\second}]},
            every axis title/.style={at={(0.5,0.975)}, anchor=north, align=center},
        ]
        \nextgroupplot[%
        title={\small\textbf{nominal}}
	    ]
        \addplot[defect, NavyBlue] table[x index=0, y index=1] \nominal node[pos=0.9, key] {\tiny $i$};
        \addplot[defect, RubineRed] table[x index=0, y index=2] \nominal node[pos=0.9, key] {\tiny 1};
        \addplot[defect, ForestGreen] table[x index=0, y index=3] \nominal node[pos=0.2, key] {\tiny 2};
        \addplot[defect, Maroon] table[x index=0, y index=4] \nominal node[pos=0.2, key] {\tiny 3};
        \addplot[defect, Goldenrod] table[x index=0, y index=5] \nominal node[pos=0.85, key] {\tiny 4};
        \addplot[defect, Plum] table[x index=0, y index=6] \nominal node[pos=0.55, key] {\tiny 5};
        \addplot[defect, Salmon] table[x index=0, y index=7] \nominal node[pos=0.8, key] {\tiny 6};
        \addplot[defect, SkyBlue] table[x index=0, y index=8] \nominal node[pos=0.9, key] {\tiny 7};
        \addplot[defect, Orchid] table[x index=0, y index=9] \nominal node[pos=0.9, key] {\tiny 8};
        \addplot[defect, Periwinkle] table[x index=0, y index=10] \nominal node[pos=0.7, key] {\tiny 9};
        \nextgroupplot[%
        title={\small\textbf{weighted}}
        ]
        \addplot[defect, forget plot, NavyBlue] table[x index=0, y index=1] \weighted node[pos=0.9, key] {\tiny $i$};
        \addlegendimage{numbered legend={key, inner sep=1.025pt}{\raisebox{-2pt}{$i$}}, defect, NavyBlue}\addlegendentry{$\ce{Xe}_{\ce{U}_i}$}
        \addplot[defect, forget plot, RubineRed] table[x index=0, y index=2] \weighted node[pos=0.95, key] {\tiny 1};
        \addlegendimage{numbered legend={key, inner sep=1.025pt}{\raisebox{-2pt}{1}}, defect, RubineRed}\addlegendentry{$\ce{Xe}_{\ce{U2O1}}$}
        \addplot[defect, forget plot, ForestGreen] table[x index=0, y index=3] \weighted node[pos=0.3, key] {\tiny 2};
        \addlegendimage{numbered legend={key, inner sep=1.025pt}{\raisebox{-2pt}{2}}, defect, ForestGreen}\addlegendentry{$\ce{Xe}_{\ce{U4O3}}$}
        \addplot[defect, forget plot, Maroon] table[x index=0, y index=4] \weighted node[pos=0.1, key] {\tiny 3};
        \addlegendimage{numbered legend={key, inner sep=1.025pt}{\raisebox{-2pt}{3}}, defect, Maroon}\addlegendentry{$\ce{Xe}_{\ce{U2O2}}$}
        \addplot[defect, forget plot, Goldenrod] table[x index=0, y index=5] \weighted node[pos=0.85, key] {\tiny 4};
        \addlegendimage{numbered legend={key, inner sep=1.025pt}{\raisebox{-2pt}{4}}, defect, Yellow}\addlegendentry{$\ce{Xe}_{\ce{U4O4}}$}
        \addplot[defect, forget plot, Plum] table[x index=0, y index=6] \weighted node[pos=0.04, key] {\tiny 5};
        \addlegendimage{numbered legend={key, inner sep=1.025pt}{\raisebox{-2pt}{5}}, defect, Plum}\addlegendentry{$\ce{Xe}_{\ce{U8O9}}$}
        \addplot[defect, forget plot, Salmon] table[x index=0, y index=7] \weighted node[pos=0.6, key] {\tiny 6};
        \addlegendimage{numbered legend={key, inner sep=1.025pt}{\raisebox{-2pt}{6}}, defect, Salmon}\addlegendentry{$\ce{Xe}_{\ce{U8O10}}$}
        \addplot[defect, forget plot, SkyBlue] table[x index=0, y index=8] \weighted node[pos=0.8, key] {\tiny 7};
        \addlegendimage{numbered legend={key, inner sep=1.025pt}{\raisebox{-2pt}{7}}, defect, SkyBlue}\addlegendentry{$\ce{Xe}_{\ce{U4O2}}$}
        \addplot[defect, forget plot, Orchid] table[x index=0, y index=9] \weighted node[pos=0.35, key] {\tiny 8};
        \addlegendimage{numbered legend={key, inner sep=1.025pt}{\raisebox{-2pt}{8}}, defect, Orchid}\addlegendentry{$\ce{Xe}_{\ce{U4O5}}$}
        \addplot[defect, forget plot, Periwinkle] table[x index=0, y index=10] \weighted node[pos=0.185, key] {\tiny 9};
        \addlegendimage{numbered legend={key, inner sep=1.025pt}{\raisebox{-2pt}{9}}, defect, Periwinkle}\addlegendentry{$\ce{Xe}_{\ce{U8O11}}$}
    \end{groupplot}%
    \begin{groupplot}[%
            group style={%
                group size=2 by 1,
                xticklabels at=edge top,
                xlabels at=edge top,
                yticklabels at=edge left,
                ylabels at=edge left,
                vertical sep=\baselineskip,
                horizontal sep=\baselineskip
            },
            default axis,
            legend style={%
                font={\scriptsize},
                at={(1, 0.5)},
                anchor=east,
                /tikz/every odd column/.append style={column sep=1mm}
            },
            axis on top,
            axis y line=none,
            xlabel={temperature $T$ [\si{\kelvin}]},
            xticklabel pos=top,
            axis x line*=top,
            xtick={5, 5.56, 6.25, 7.14, 8.34},
            xmin=4.4, xmax=8.9, ymin=-23, ymax=-15,
        ]
        \nextgroupplot[%
            xticklabels={2000, 1800, 1600, 1400, 1200},
        ]
        \addplot[opacity=0] coordinates { (8.85,8.9) (8.9,8.85) }; 
        \nextgroupplot[%
            xticklabels={2000, 1800, 1600, 1400, 1200},
        ]
        \addplot[opacity=0] coordinates { (8.85,8.9) (8.9,8.85) }; 
    \end{groupplot}
\end{tikzpicture}
	\tikzexternaldisable

    \vspace{-\baselineskip}
    \caption{Contributions of the individual defects to the xenon diffusivity under irradiation, for the 10 most important defects, for both the nominal parameter values (\emph{left}) and the estimated parameters using the weighted likelihood formulation (\emph{right}).}
    \label{fig:diffusion_mechanism}
\end{figure*}
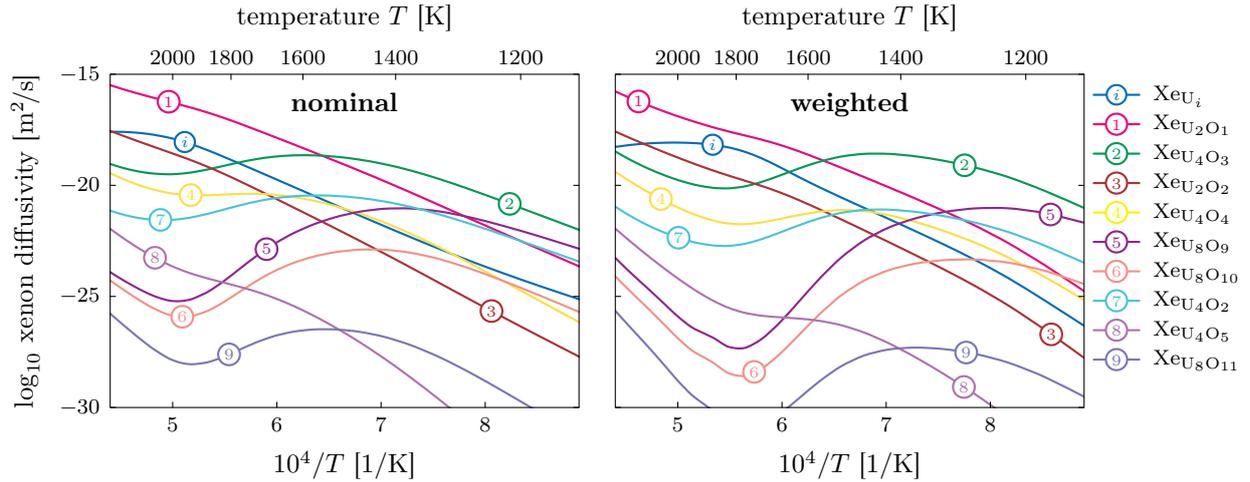

\Cref{fig:diffusion_mechanism} shows the contributions of the individual defects to the xenon diffusivity under irradiation, for the 10 most important defects, for both the nominal parameter values (\emph{left}) and the estimated parameters using the weighted likelihood formulation (\emph{right}). These plots confirm the vacancy-based mechanism for xenon diffusion that was previously identified in~\cite{matthews2020}.

\section{Conclusion and further work}\label{sec:conclusion_and_further_work}

The fission gas diffusivity coefficient is one of the crucial parameters used in many nuclear fuel performance models. Reconciling existing models, such as the cluster dynamics simulation software \Centipede{}, with historical gas release and thermodynamic data is crucial to predict the behavior of fission gas for both current and novel nuclear fuel types. In this work, we have developed a Bayesian calibration framework for \Centipede{}. Our approach provides a posterior distribution on the atomistic parameters in the model underlying the cluster dynamics simulator. A particular challenge in our setup is that only summary statistics (i.e., a measurement value with an associated uncertainty) of the experiments have been reported in the literature. We have employed a DFI approach that uses synthetic data sets in order to match the pushforward predictions of the uranium and xenon diffusivity, as well as the non-stoichiometry of $\nonstoich$, with the available summary statistics. In a departure from the original DFI framework, our construction uses a stochastic optimization argument to enforce consistency of the statistics computed from the pushforward posterior and the reported summary statistics. Further, we reduce computational costs significantly by replacing the expensive evaluations of \Centipede{} with evaluations from a reduced PCE surrogate model. This reduced model was identified using a sensitivity analysis. We also constructed a weighted variant of our method, where we compensate for the number of measurements in each set of experimental data summaries. We found that, in both the unweighted and the weighted case, there is a good agreement between the calibrated atomistic model and the reported summary statistics. Furthermore, our results are in good agreement with the Turnbull et al.\ fit for the xenon diffusivity reported in~\cite{turnbull1982}. The latter analysis has been used as a starting point for many higher-level fission gas release models, as it sets the timescale for the subsequent percolation of gas bubbles at grain boundaries. The excellent agreement between the estimated xenon diffusivity from calibration and the empirical fits based on experimental data confirms that a uranium vacancy-based mechanism can describe the xenon diffusion under irradiation conditions predicted from the experimental data.

\Centipede{} is part of a larger, hierarchical multiscale modelling and simulation framework used to simulate the complete fission gas release process, see, e.g.~\cite{andersson2014}. Uncertainty quantification (UQ) has a critical role to play in connecting these different models at different length and time scales. While our current work provides a demonstration of the application of UQ by calibrating atomistic-scale parameters in one of the lower-level mechanistic models for fission gas release, the same calibration strategy can be used for higher-level codes in the framework, such as \Xolotl{}-\Marmot{}, see~\cite{kim2022,xolotl}, \Bison{}~\cite{williamson2012,williamson2021} and \Nyx{}~\cite{pizzocri2018,pastore2022}.

In future work, we will use our calibration strategy to predict the fission gas evolution at the fuel pin scale, simulated by the fuel performance code \Nyx, see~\cite{kim2022}, and using experimental data from, amongst others,~\cite{baker1977}. The fuel performance code will also use the calibrated xenon diffusivity as input. The simulations performed by \Nyx{} may easily be transferred to a full-fledged fuel performance code such as \Bison{} afterwards. The xenon diffusivity can also be used to determine the impact of fission gas evolution at the grain scale using coupled \Xolotl{}-\Marmot{}. 

\section*{Acknowledgements}

This work was supported by the U.S.\,Department of Energy, Office of Nuclear Energy and Office of Science, Office of Advanced Scientific Computing Research through the Scientific Discovery through Advanced Computing project on Simulation of Fission Gas.

This research made use of the resources of the High Performance Computing Center at Idaho National Laboratory, which is supported by the Office of Nuclear Energy of the U.S. Department of Energy and the Nuclear Science User Facilities under Contract No. DE-AC07-05ID14517.

This article has been co-authored by employees of National Technology \& Engineering Solutions of Sandia, LLC under Contract No. DE-NA0003525 with the U.S. Department of Energy (DOE). The employees co-own right, title and interest in and to the article and are responsible for its contents. The United States Government retains and the publisher, by accepting the article for publication, acknowledges that the United States Government retains a non-exclusive, paid-up, irrevocable, world-wide license to publish or reproduce the published form of this article or allow others to do so, for United States Government purposes. The DOE will provide public access to these results of federally sponsored research in accordance with the DOE Public Access Plan https://www.energy.gov/downloads/doe-public-access-plan. 

Los Alamos National Laboratory, an affirmative action/equal opportunity employer, is operated by Triad National Security, LLC, for the National Nuclear Security Administration of the U.S.\,Department of Energy under Contract No.\,89233218CNA000001.

\bibliographystyle{unsrt}
\bibliography{ref/references}

\begin{appendices}
    \clearpage 

\section{Supplementary data}\label{sec:measurement_locations}

The temperatures where \Centipede{} predicts the diffusivities $\dutheq$, $\dxetheq$ and $\dxeirr$ are included in \cref{tab:measurement_locations_for_the_diffusivity_predictions}. The available experimental data summaries are shown in \cref{tab:sabioni_data_set,tab:davies_long_data_set,tab:turnbull_data_set,tab:miekely_felix_data_set,tab:stoichiometry_data_set}, where $\bsx_d^{(n)}$, $y_d^{(n)}$ and $s_d^{(n)}$ are as defined in \cref{eq:experimental_data_sets}. The nominal values, lower and upper bounds for each parameter are shown in~\cref{tab:all_parameters_overview}. Estimated parameter values obtained after calibration are compiled in \cref{tab:uncertain_parameters}.

\begin{table}[ht]
    \centering
    \begin{minipage}[t]{0.175\textwidth}
        \small
        \begin{tabular}{%
                l@{}
                S[table-format=2]%
                S[table-format=4]%
            }\toprule
            & {\#} & {$T$ [\si{\kelvin}]} \\ \midrule
            & 1 & 1123 \\
& 2 & 1173 \\
& 3 & 1223 \\
& 4 & 1273 \\
& 5 & 1323 \\
& 6 & 1373 \\
& 7 & 1423 \\ \bottomrule
        \end{tabular}
    \end{minipage}%
    \begin{minipage}[t]{0.175\textwidth}
        \small
        \begin{tabular}{%
                l@{}
                S[table-format=2]%
                S[table-format=4]%
            }\toprule
            & {\#} & {$T$ [\si{\kelvin}]} \\ \midrule
            & 8 & 1473 \\
& 9 & 1523 \\
& 10 & 1573 \\
& 11 & 1623 \\
& 12 & 1673 \\
& 13 & 1723 \\
& 14 & 1773 \\ \bottomrule

        \end{tabular}
    \end{minipage}%
    \begin{minipage}[t]{0.175\textwidth}
        \small
        \begin{tabular}{%
                l@{}
                S[table-format=2]%
                S[table-format=4]%
            }\toprule
            & {\#} & {$T$ [\si{\kelvin}]} \\ \midrule
            & 15 & 1823 \\
& 16 & 1873 \\
& 17 & 1923 \\
& 18 & 1948 \\
& 19 & 1973 \\
& 20 & 1998 \\
& 21 & 2023  \\ \bottomrule
        \end{tabular}
    \end{minipage}%
    \begin{minipage}[t]{0.175\textwidth}
        \small
        \begin{tabular}{%
                l@{}
                S[table-format=2]%
                S[table-format=4]%
            }\toprule
            & {\#} & {$T$ [\si{\kelvin}]} \\ \midrule
            & 22 & 2073 \\
& 23 & 2123 \\
& 24 & 2173 \\
& 25 & 2223 \\
& 26 & 2273 \\
& & \\
& &  \\ \bottomrule
        \end{tabular}
    \end{minipage}%
    \caption{Temperatures $T$ [\si{\kelvin}] where the \Centipede{} predictions of $\dutheq$, $\dxetheq$ and $\dxeirr$ are available.}
    \label{tab:measurement_locations_for_the_diffusivity_predictions}
\end{table}

\begin{table}[ht]
    \centering
    \small
    \begin{tabular}{lccc}\toprule
        & $\bsx_1^{(n)}$ & $y_1^{(n)}$ & $s_1^{(n)}$\\
        $n$ & $T$ [\si{\kelvin}] & $\dutheq$ [\si[per-mode=symbol]{\square\metre\per\second}] & $\pm$ [\si[per-mode=symbol]{\square\metre\per\second}] \\ \midrule
        1 & 1971.434 & -21.277 & 0.166 \\
2 & 1963.456 & -21.320 & 0.166 \\
3 & 1972.562 & -21.503 & 0.166 \\
4 & 1920.864 & -21.738 & 0.166 \\
5 & 1870.564 & -21.951 & 0.166 \\
6 & 1871.075 & -22.292 & 0.166 \\
7 & 1820.237 & -22.100 & 0.166 \\
8 & 1820.237 & -22.535 & 0.166 \\
9 & 1770.560 & -22.538 & 0.166 \\
10 & 1769.811 & -22.735 & 0.166 \\ \bottomrule 
    \end{tabular}
    \caption{Sabioni et al.\ data set ($d = 1$) from~\cite{sabioni1998}.}
    \label{tab:sabioni_data_set}
\end{table}

\begin{table}[ht]
    \centering
    \small
    \begin{tabular}{lccc}\toprule
        & $\bsx_2^{(n)}$ & $y_2^{(n)}$ & $s_2^{(n)}$\\
        $n$ & $T$ [\si{\kelvin}] & $\dxetheq$ [\si[per-mode=symbol]{\square\metre\per\second}] & $\pm$ [\si[per-mode=symbol]{\square\metre\per\second}] \\ \midrule
        1 & 1173.000 & -22.179 & 0.699 \\
2 & 1273.000 & -21.153 & 0.699 \\
3 & 1373.000 & -20.277 & 0.699 \\
4 & 1473.000 & -19.519 & 0.699 \\
5 & 1573.000 & -18.858 & 0.699 \\
6 & 1673.000 & -18.276 & 0.699 \\
7 & 1773.000 & -17.760 & 0.699 \\
8 & 1873.000 & -17.298 & 0.699 \\ \bottomrule 
    \end{tabular}
    \caption{Davies \& Long data set ($d = 2$) from~\cite{davies1963}.}
    \label{tab:davies_long_data_set}
\end{table}

\begin{table}[ht]
    \centering
    \small
    \begin{tabular}{lccc}\toprule
        & $\bsx_3^{(n)}$ & $y_3^{(n)}$ & $s_3^{(n)}$\\
        $n$ & $T$ [\si{\kelvin}] & $\dxeirr$ [\si[per-mode=symbol]{\square\metre\per\second}] & $\pm$ [\si[per-mode=symbol]{\square\metre\per\second}] \\ \midrule
        1 & 1724.138 & -18.006 & 0.362 \\
2 & 1721.170 & -17.935 & 0.362 \\
3 & 1720.994 & -18.093 & 0.362 \\
4 & 1719.484 & -17.798 & 0.362 \\
5 & 1717.193 & -18.114 & 0.362 \\
6 & 1707.520 & -18.453 & 0.362 \\
7 & 1515.152 & -18.126 & 0.362 \\
8 & 1509.013 & -18.604 & 0.362 \\
9 & 1508.296 & -18.666 & 0.362 \\
10 & 1502.045 & -18.340 & 0.362 \\
11 & 1292.608 & -19.565 & 0.362 \\
12 & 1291.990 & -19.348 & 0.362 \\
13 & 1290.323 & -19.060 & 0.362 \\
14 & 1289.010 & -19.405 & 0.362 \\
15 & 1288.215 & -19.658 & 0.362 \\
16 & 1260.345 & -19.575 & 0.362 \\ \bottomrule

    \end{tabular}
    \caption{Turnbull et al.\ data set ($d = 3$) from~\cite{turnbull1982,turnbull1989}.}
    \label{tab:turnbull_data_set}
\end{table}

\begin{table}[ht]
    \centering
    \small
    \begin{tabular}{lccc}\toprule
        & $\bsx_4^{(n)}$ & $y_4^{(n)}$ & $s_4^{(n)}$\\
        $n$ & $T$ [\si{\kelvin}] & $\dxetheq$ [\si[per-mode=symbol]{\square\metre\per\second}] & $\pm$ [\si[per-mode=symbol]{\square\metre\per\second}] \\ \midrule
        1 & 1967.849 & -14.220 & 2.094 \\
2 & 1929.348 & -14.254 & 2.280 \\
3 & 1974.416 & -14.608 & 1.669 \\
4 & 1892.324 & -14.946 & 1.807 \\
5 & 1800.203 & -14.963 & 2.376 \\
6 & 1822.382 & -15.014 & 2.179 \\
7 & 1794.742 & -15.082 & 2.294 \\
8 & 1773.227 & -15.606 & 1.917 \\
9 & 1731.707 & -15.656 & 2.159 \\
10 & 1706.731 & -15.707 & 2.291 \\
11 & 1736.791 & -15.758 & 2.021 \\
12 & 1816.786 & -15.961 & 1.269 \\
13 & 1696.941 & -15.301 & 2.770 \\
14 & 1626.948 & -15.927 & 2.694 \\
15 & 1701.822 & -16.113 & 1.922 \\
16 & 1600.541 & -16.130 & 2.711 \\
17 & 1618.049 & -16.265 & 2.430 \\
18 & 1631.434 & -16.552 & 2.032 \\
19 & 1649.628 & -16.569 & 1.869 \\
20 & 1731.707 & -16.434 & 1.381 \\
21 & 1518.392 & -16.806 & 2.768 \\
22 & 1550.218 & -17.296 & 1.985 \\
23 & 1583.408 & -17.262 & 1.725 \\
24 & 1645.042 & -17.414 & 1.061 \\
25 & 1484.114 & -17.685 & 2.219 \\
26 & 1465.731 & -17.786 & 2.300 \\
27 & 1440.747 & -17.735 & 2.607 \\
28 & 1491.597 & -17.887 & 1.942 \\
29 & 1506.791 & -18.361 & 1.323 \\
30 & 1361.196 & -18.614 & 2.608 \\
31 & 1426.849 & -19.003 & 1.486 \\
32 & 1223.294 & -20.254 & 2.763 \\ \bottomrule

    \end{tabular}
    \caption{Miekeley \& Felix data set ($d = 4$) from~\cite{miekeley1972}.}
    \label{tab:miekely_felix_data_set}
\end{table}

\clearpage

\begin{table*}[ht]
    \centering
    \begin{minipage}[t]{0.5\textwidth}
        \footnotesize
        \begin{tabular}{%
                l@{}
                S[table-format=3]%
                S[table-format=4.3]%
                S[table-format=2.3]%
                S[table-format=2.3]%
                S[table-format=2.3]%
            }\toprule
            & & \multicolumn{2}{c}{$\bsx_5^{(n)}\;\;\;\;$} & {$y_5^{(n)}$} & {$s_5^{(n)}$} \\
            & & {$T$} & {$\log_{10}(\pOtwo{})$} & {$\nonstoich{}$} & {$\pm$} \\ 
            & {$n$} & {[\si{\kelvin}]} & {[\si{\atm}]} & {$(\times \; 1,000)$} & {$(\times \; 1,000)$} \\ \midrule
            & 1 & 1800 & -18.043 & 0.000 & 1.000 \\
& 2 & 1800 & -18.359 & -1.000 & 1.000 \\
& 3 & 1800 & -18.638 & -2.000 & 1.000 \\
& 4 & 1800 & -18.979 & -3.000 & 1.000 \\
& 5 & 1900 & -16.288 & 0.000 & 1.000 \\
& 6 & 1900 & -16.656 & -3.000 & 1.000 \\
& 7 & 1900 & -16.910 & -2.000 & 1.000 \\
& 8 & 2000 & -15.386 & -3.000 & 1.000 \\
& 9 & 1950 & -16.498 & -1.800 & 0.300 \\
& 10 & 1950 & -15.893 & -0.900 & 0.300 \\
& 11 & 1950 & -15.411 & -0.400 & 0.300 \\
& 12 & 1950 & -14.593 & 0.000 & 0.300 \\
& 13 & 1950 & -9.717 & 0.400 & 0.300 \\
& 14 & 1950 & -8.967 & 1.200 & 0.300 \\
& 15 & 1950 & -8.519 & 2.100 & 0.300 \\
& 16 & 1950 & -7.924 & 2.500 & 0.300 \\
& 17 & 1900 & -13.815 & 0.000 & 3.000 \\
& 18 & 1900 & -14.401 & 0.000 & 3.000 \\
& 19 & 1900 & -15.987 & -2.000 & 3.000 \\
& 20 & 1800 & -16.391 & 0.000 & 3.000 \\
& 21 & 1800 & -16.638 & 0.000 & 3.000 \\
& 22 & 1800 & -17.666 & -2.000 & 3.000 \\
& 23 & 1700 & -18.775 & 0.000 & 3.000 \\
& 24 & 1700 & -19.928 & 0.000 & 3.000 \\
& 25 & 1700 & -20.572 & -3.000 & 3.000 \\
& 26 & 1600 & -22.197 & 0.000 & 3.000 \\
& 27 & 1600 & -22.470 & 0.000 & 3.000 \\
& 28 & 1600 & -22.812 & 0.000 & 3.000 \\
& 29 & 1600 & -22.955 & -1.000 & 3.000 \\
& 30 & 1600 & -22.824 & -3.000 & 3.000 \\
& 31 & 1273 & -15.438 & 0.807 & 1.000 \\
& 32 & 1273 & -12.623 & 1.845 & 1.000 \\
& 33 & 1573 & -12.845 & 1.007 & 1.000 \\
& 34 & 1573 & -11.179 & 2.999 & 1.000 \\
& 35 & 1773 & -10.648 & 2.153 & 1.000 \\
& 36 & 1273 & -14.419 & 1.019 & 2.520 \\
& 37 & 1273 & -13.054 & 1.879 & 2.520 \\
& 38 & 1172.791 & -12.112 & 2.000 & 0.700 \\
& 39 & 1272.316 & -11.195 & 2.000 & 0.700 \\
& 40 & 1371.829 & -10.343 & 2.000 & 0.700 \\
& 41 & 1472.321 & -9.971 & 2.000 & 0.700 \\
& 42 & 1572.005 & -9.442 & 2.000 & 0.700 \\
& 43 & 1672.690 & -8.951 & 2.000 & 0.700 \\
& 44 & 1772.402 & -8.456 & 2.000 & 0.700 \\
& 45 & 930 & -22.298 & 0.310 & 2.520 \\
& 46 & 930 & -21.171 & 0.572 & 2.520 \\
& 47 & 930 & -19.409 & 1.000 & 2.520 \\
& 48 & 930 & -18.077 & 2.000 & 2.520 \\
& 49 & 1043 & -22.266 & 0.087 & 2.520 \\
& 50 & 1043 & -20.498 & 0.310 & 2.520 \\
& 51 & 1043 & -19.836 & 0.551 & 2.520 \\
& 52 & 1043 & -19.728 & 0.592 & 2.520 \\ \bottomrule
        \end{tabular}
    \end{minipage}%
    \begin{minipage}[t]{0.5\textwidth}
        \footnotesize
        \begin{tabular}{%
                l@{}
                S[table-format=3]%
                S[table-format=4.3]%
                S[table-format=2.3]%
                S[table-format=2.3]%
                S[table-format=2.3]%
            }\toprule
            & & \multicolumn{2}{c}{$\bsx_5^{(n)}\;\;\;\;$} & {$y_5^{(n)}$} & {$s_5^{(n)}$} \\
            & & {$T$} & {$\log_{10}(\pOtwo{})$} & {$\nonstoich{}$} & {$\pm$} \\ 
            & {$n$} & {[\si{\kelvin}]} & {[\si{\atm}]} & {$(\times \; 1,000)$} & {$(\times \; 1,000)$} \\ \midrule
            & 53 & 1043 & -18.740 & 0.558 & 0.252 \\
& 54 & 1043 & -18.745 & 0.801 & 0.252 \\
& 55 & 1043 & -18.660 & 1.000 & 0.252 \\
& 56 & 1043 & -18.444 & 1.000 & 0.252 \\
& 57 & 1043 & -18.200 & 1.000 & 0.252 \\
& 58 & 1043 & -18.270 & 0.812 & 0.252 \\
& 59 & 1043 & -18.080 & 0.799 & 0.252 \\
& 60 & 1043 & -17.719 & 0.805 & 0.252 \\
& 61 & 1043 & -17.526 & 1.000 & 0.252 \\
& 62 & 1043 & -17.000 & 1.000 & 0.252 \\
& 63 & 1043 & -17.772 & 2.000 & 0.252 \\
& 64 & 1043 & -17.323 & 2.000 & 0.252 \\
& 65 & 1043 & -16.593 & 2.000 & 0.252 \\
& 66 & 1043 & -15.889 & 1.000 & 0.252 \\
& 67 & 1043 & -15.441 & 2.000 & 0.252 \\
& 68 & 1043 & -15.963 & 2.000 & 0.252 \\
& 69 & 1043 & -16.572 & 2.000 & 0.252 \\
& 70 & 1043 & -15.235 & 3.000 & 0.252 \\
& 71 & 1043 & -14.719 & 3.000 & 0.252 \\
& 72 & 1043 & -14.255 & 3.000 & 0.252 \\
& 73 & 1373 & -17.000 & 0.107 & 0.252 \\
& 74 & 1373 & -16.013 & 0.294 & 0.252 \\
& 75 & 1373 & -15.249 & 0.603 & 0.252 \\
& 76 & 1373 & -15.330 & 0.891 & 0.252 \\
& 77 & 1373 & -14.484 & 0.811 & 0.252 \\
& 78 & 1373 & -14.289 & 1.000 & 0.252 \\
& 79 & 1373 & -15.025 & 1.000 & 0.252 \\
& 80 & 1373 & -14.155 & 2.000 & 0.252 \\
& 81 & 1373 & -13.429 & 2.000 & 0.252 \\
& 82 & 1373 & -13.186 & 2.000 & 0.252 \\
& 83 & 1373 & -12.526 & 2.000 & 0.252 \\
& 84 & 1373 & -12.821 & 2.000 & 0.252 \\
& 85 & 1373 & -12.939 & 3.000 & 0.252 \\
& 86 & 1373 & -12.113 & 3.000 & 0.252 \\
& 87 & 1373 & -11.398 & 3.000 & 0.252 \\
& 88 & 1373 & -10.967 & 3.000 & 0.252 \\
& 89 & 1373 & -10.654 & 3.000 & 0.252 \\
& 90 & 1473 & -12.921 & 0.715 & 0.252 \\
& 91 & 1473 & -11.947 & 1.000 & 0.252 \\
& 92 & 1473 & -10.232 & 2.000 & 0.252 \\
& 93 & 1473 & -9.764 & 2.000 & 0.252 \\
& 94 & 1473 & -9.583 & 3.000 & 0.252 \\
& 95 & 1573 & -12.745 & 0.869 & 0.252 \\
& 96 & 1573 & -10.975 & 2.000 & 0.252 \\
& 97 & 1573 & -9.708 & 2.000 & 0.252 \\
& 98 & 1673 & -9.848 & 0.717 & 0.252 \\
& 99 & 1673 & -8.873 & 2.000 & 0.252 \\
& 100 & 1673 & -8.592 & 3.000 & 0.252 \\
& 101 & 1986.5 & -9.646 & 0.386 & 0.252 \\
& 102 & 1986.5 & -8.896 & 1.000 & 0.252 \\
& 103 & 1986.5 & -8.456 & 2.000 & 0.252 \\
& 104 & 1986.5 & -7.851 & 2.000 & 0.252 \\ \bottomrule
        \end{tabular}
    \end{minipage}\hfill%
    \caption{Thermodynamic data set ($d = 5$) for the non-stoichiometry from~\cite{markin1968,wheeler1971,wheeler1972,javed1972,aronson1958,une1983,une1982,hagemark1966}.}
    \label{tab:stoichiometry_data_set}
\end{table*}

\begin{table*}[ht]
    \centering
    \footnotesize
    \begin{tabular}{%
            l@{}%
            S[table-format=6.4]%
            S[table-format=6.4]%
            S[table-format=6.4]%
            S[table-format=6.4]%
            S[table-format=6.4]%
            S[table-format=6.4]%
        }\toprule
        {} & \multicolumn{3}{c}{\texttt{DFT} [\si{\eV}]} & \multicolumn{3}{c}{\texttt{S} [\si[per-mode=symbol]{\eV\per\kelvin}]} \\
        {parameter name} & {nominal value} & {lower bound} & {upper bound} & {nominal value} & {lower bound} & {upper bound} \\ \midrule
        \ldots\texttt{\_Xe\_Ui01\_Oi00} & -3425.8334 & -3425.9584 & -3425.7084 & -129446.1222 & -129447.3722 & -129444.8722 \\
\ldots\texttt{\_U01\_O02} & -3435.2645 & -3435.3895 & -3435.1395 & -129368.8308 & -129370.0808 & -129367.5808 \\
\ldots\texttt{\_vU01\_vO00} & -3381.0370 & -3381.1620 & -3380.9120 & -129265.9303 & -129267.1803 & -129264.6803 \\
\ldots\texttt{\_vU00\_vO01} & -3443.2509 & -3443.3759 & -3443.1259 & -129286.6685 & -129287.9185 & -129285.4185 \\
\ldots\texttt{\_h} & -3444.0477 & -3444.1727 & -3443.7977 & -129374.7936 & -129376.0436 & -129372.2936 \\
\ldots\texttt{\_e} & -3425.1182 & -3425.2432 & -3424.8682 & -129362.1380 & -129363.3880 & -129359.6380 \\
\ldots\texttt{\_vU01\_vO01} & -3389.8559 & -3389.9809 & -3389.7309 & -129188.3858 & -129189.6358 & -129187.1358 \\
\ldots\texttt{\_vU01\_vO02} & -3399.7345 & -3399.8595 & -3399.6095 & -129107.4343 & -129108.6843 & -129106.1843 \\
\ldots\texttt{\_vU02\_vO00} & -3324.9130 & -3325.0380 & -3324.7880 & -129160.7825 & -129162.0325 & -129159.5325 \\
\ldots\texttt{\_vU02\_vO01} & -3335.6929 & -3335.8179 & -3335.5679 & -129080.7605 & -129082.0105 & -129079.5105 \\
\ldots\texttt{\_vU02\_vO02} & -3346.1976 & -3346.3226 & -3346.0726 & -129004.2167 & -129005.4667 & -129002.9667 \\
\ldots\texttt{\_Ui01\_Oi00} & -3478.4152 & -3478.5402 & -3478.2902 & -129462.0699 & -129463.3199 & -129460.8199 \\
\ldots\texttt{\_Ui00\_Oi01} & -3423.4492 & -3423.5742 & -3423.3242 & -129445.2888 & -129446.5388 & -129444.0388 \\
\ldots\texttt{\_Ui01\_Oi02} & -3456.3899 & -3456.5149 & -3456.2649 & -129615.3049 & -129616.5549 & -129614.0549 \\
\ldots\texttt{\_Xe\_vU01\_vO00} & -3378.0494 & -3378.1744 & -3377.9244 & -129345.2924 & -129346.5424 & -129344.0424 \\
\ldots\texttt{\_Xe\_vU01\_vO01} & -3388.7748 & -3388.8998 & -3388.6498 & -129264.8194 & -129266.0694 & -129263.5694 \\
\ldots\texttt{\_Xe\_vU01\_vO02} & -3398.1568 & -3398.2818 & -3398.0318 & -129186.5971 & -129187.8471 & -129185.3471 \\
\ldots\texttt{\_Xe\_vU02\_vO00} & -3322.8631 & -3322.9881 & -3322.7381 & -129237.4063 & -129238.6563 & -129236.1563 \\
\ldots\texttt{\_Xe\_vU02\_vO01} & -3335.2667 & -3335.3917 & -3335.1417 & -129157.2840 & -129158.5340 & -129156.0340 \\
\ldots\texttt{\_Xe\_vU02\_vO02} & -3346.5440 & -3346.6690 & -3346.4190 & -129086.9852 & -129088.2352 & -129085.7352 \\
\ldots\texttt{\_Xe\_vU03\_vO01} & -3280.7401 & -3280.8651 & -3280.6151 & -129052.1593 & -129053.4093 & -129050.9093 \\
\ldots\texttt{\_Xe\_vU03\_vO02} & -3292.0453 & -3292.1703 & -3291.9203 & -128979.4136 & -128980.6636 & -128978.1636 \\
\ldots\texttt{\_Xe\_vU03\_vO03} & -3302.3153 & -3302.4403 & -3302.1903 & -128901.4108 & -128902.6608 & -128900.1608 \\
\ldots\texttt{\_Xe\_vU03\_vO04} & -3312.8355 & -3312.9605 & -3312.7105 & -128824.5474 & -128825.7974 & -128823.2974 \\
\ldots\texttt{\_Xe\_vU04\_vO01} & -3225.8450 & -3225.9700 & -3225.7200 & -128933.9056 & -128935.1556 & -128932.6556 \\
\ldots\texttt{\_Xe\_vU04\_vO02} & -3237.7138 & -3237.8388 & -3237.5888 & -128865.7380 & -128866.9880 & -128864.4880 \\
\ldots\texttt{\_Xe\_vU04\_vO03} & -3248.1971 & -3248.3221 & -3248.0721 & -128789.5736 & -128790.8236 & -128788.3236 \\
\ldots\texttt{\_Xe\_vU04\_vO04} & -3258.3206 & -3258.4456 & -3258.1956 & -128712.2956 & -128713.5456 & -128711.0456 \\
\ldots\texttt{\_Xe\_vU04\_vO05} & -3268.5677 & -3268.6927 & -3268.4427 & -128634.4395 & -128635.6895 & -128633.1895 \\
\ldots\texttt{\_Xe\_vU04\_vO06} & -3278.4643 & -3278.5893 & -3278.3393 & -128559.0793 & -128560.3293 & -128557.8293 \\
\ldots\texttt{\_Xe\_vU04\_vO07} & -3288.4526 & -3288.5776 & -3288.3276 & -128483.7539 & -128485.0039 & -128482.5039 \\
\ldots\texttt{\_Xe\_vU05\_vO04} & -3201.5320 & -3201.6570 & -3201.4070 & -128597.6268 & -128598.8768 & -128596.3768 \\
\ldots\texttt{\_Xe\_vU05\_vO05} & -3212.0795 & -3212.2045 & -3211.9545 & -128525.6150 & -128526.8650 & -128524.3650 \\
\ldots\texttt{\_Xe\_vU05\_vO06} & -3224.8106 & -3224.9356 & -3224.6856 & -128454.0513 & -128455.3013 & -128452.8013 \\
\ldots\texttt{\_Xe\_vU05\_vO07} & -3235.4855 & -3235.6105 & -3235.3605 & -128380.5357 & -128381.7857 & -128379.2857 \\
\ldots\texttt{\_Xe\_vU05\_vO08} & -3244.5872 & -3244.7122 & -3244.4622 & -128296.7549 & -128298.0049 & -128295.5049 \\
\ldots\texttt{\_Xe\_vU06\_vO07} & -3182.5388 & -3182.6638 & -3182.4138 & -128272.4893 & -128273.7393 & -128271.2393 \\
\ldots\texttt{\_Xe\_vU06\_vO08} & -3194.7185 & -3194.8435 & -3194.5935 & -128197.0652 & -128198.3152 & -128195.8152 \\
\ldots\texttt{\_Xe\_vU06\_vO09} & -3203.3341 & -3203.4591 & -3203.2091 & -128118.3264 & -128119.5764 & -128117.0764 \\
\ldots\texttt{\_Xe\_vU07\_vO08} & -3140.1520 & -3140.2770 & -3140.0270 & -128087.2454 & -128088.4954 & -128085.9954 \\
\ldots\texttt{\_Xe\_vU07\_vO09} & -3150.4497 & -3150.5747 & -3150.3247 & -128012.6158 & -128013.8658 & -128011.3658 \\
\ldots\texttt{\_Xe\_vU07\_vO10} & -3160.7004 & -3160.8254 & -3160.5754 & -127934.7572 & -127936.0072 & -127933.5072 \\
\ldots\texttt{\_Xe\_vU07\_vO11} & -3170.6988 & -3170.8238 & -3170.5738 & -127856.5141 & -127857.7641 & -127855.2641 \\
\ldots\texttt{\_Xe\_vU08\_vO08} & -3085.2697 & -3085.3947 & -3085.1447 & -127977.1638 & -127978.4138 & -127975.9138 \\
\ldots\texttt{\_Xe\_vU08\_vO09} & -3096.5023 & -3096.6273 & -3096.3773 & -127901.4227 & -127902.6727 & -127900.1727 \\
\ldots\texttt{\_Xe\_vU08\_vO10} & -3107.1922 & -3107.3172 & -3107.0672 & -127826.9061 & -127828.1561 & -127825.6561 \\
\ldots\texttt{\_Xe\_vU08\_vO11} & -3117.1919 & -3117.3169 & -3117.0669 & -127751.2887 & -127752.5387 & -127750.0387 \\
\ldots\texttt{\_Xe\_vU08\_vO12} & -3127.3391 & -3127.4641 & -3127.2141 & -127673.5524 & -127674.8024 & -127672.3024 \\ \bottomrule

    \end{tabular}
    \caption{Nominal values, lower and upper bounds for all parameters. Note that the \texttt{DFT} parameters represent total energies from DFT calculations as used by \Centipede{}, and that entropies are presented in a reduced form suitable for \Centipede{} and should not be interpreted as an actual entropy, see~\cite{davies1963,cooper2016,liu2022}.}
    \label{tab:all_parameters_overview}
\end{table*}

\begin{table*}[ht]
    \ContinuedFloat
    \centering
    \footnotesize
    \begin{tabular}{%
            l@{}%
            S[table-format=6.4]%
            S[table-format=6.4]%
            S[table-format=6.4]%
            S[table-format=6.4]%
            S[table-format=6.4]%
            S[table-format=6.4]%
        }\toprule
        {} & \multicolumn{3}{c}{\texttt{Q} [\si{\eV}]} & \multicolumn{3}{c}{\texttt{log10\_w} [\si{\Hz}]} \\
        {parameter name} & {nominal value} & {lower bound} & {upper bound} & {nominal value} & {lower bound} & {upper bound} \\ \midrule
        \ldots\texttt{\_Xe\_Ui01\_Oi00} & 1.7000 & 1.4500 & 1.9500 & 0 & -0.3010 & 1 \\
\ldots\texttt{\_vU01\_vO00} & 4.2332 & 3.9832 & 4.4832 & 0.1523 & -0.1487 & 1.1523 \\
\ldots\texttt{\_vU01\_vO01} & 4.1171 & 3.8671 & 4.3671 & 0.1523 & -0.1487 & 1.1523 \\
\ldots\texttt{\_vU01\_vO02} & 4.3471 & 4.0971 & 4.5971 & 0.1523 & -0.1487 & 1.1523 \\
\ldots\texttt{\_vU02\_vO00} & 2.4874 & 2.2374 & 2.7374 & 0.6972 & 0.3962 & 1.6972 \\
\ldots\texttt{\_vU02\_vO01} & 3.2656 & 3.0156 & 3.5156 & 0.6972 & 0.3962 & 1.6972 \\
\ldots\texttt{\_vU02\_vO02} & 3.4383 & 3.1883 & 3.6883 & 0.6972 & 0.3962 & 1.6972 \\
\ldots\texttt{\_Ui01\_Oi00} & 4.0845 & 3.8345 & 4.3345 & 0.7364 & 0.4354 & 1.7364 \\
\ldots\texttt{\_Ui01\_Oi02} & 1.9748 & 1.7248 & 2.2248 & 0.7364 & 0.4354 & 1.7364 \\
\ldots\texttt{\_Xe\_vU02\_vO00} & 2.0470 & 1.7970 & 2.2970 & -0.6021 & -0.9031 & 0.3979 \\
\ldots\texttt{\_Xe\_vU02\_vO01} & 3.6957 & 3.4457 & 3.9457 & 0.5119 & 0.2109 & 1.5119 \\
\ldots\texttt{\_Xe\_vU02\_vO02} & 4.3865 & 4.1365 & 4.6365 & 0.2068 & -0.0942 & 1.2068 \\
\ldots\texttt{\_Xe\_vU03\_vO01} & 4.6853 & 4.4353 & 4.9353 & 0 & -0.3010 & 1 \\
\ldots\texttt{\_Xe\_vU03\_vO02} & 4.9853 & 4.7353 & 5.2353 & 0 & -0.3010 & 1 \\
\ldots\texttt{\_Xe\_vU03\_vO03} & 5.2853 & 5.0353 & 5.5353 & 0 & -0.3010 & 1 \\
\ldots\texttt{\_Xe\_vU03\_vO04} & 5.3457 & 5.0957 & 5.5957 & 0 & -0.3010 & 1 \\
\ldots\texttt{\_Xe\_vU04\_vO01} & 3.9635 & 3.7135 & 4.2135 & 0 & -0.3010 & 1 \\
\ldots\texttt{\_Xe\_vU04\_vO02} & 3.4635 & 3.2135 & 3.7135 & 0 & -0.3010 & 1 \\
\ldots\texttt{\_Xe\_vU04\_vO03} & 2.7707 & 2.5207 & 3.0207 & 0 & -0.3010 & 1 \\
\ldots\texttt{\_Xe\_vU04\_vO04} & 3.0707 & 2.8207 & 3.3207 & 0 & -0.3010 & 1 \\
\ldots\texttt{\_Xe\_vU04\_vO05} & 4.5513 & 4.3013 & 4.8013 & 0 & -0.3010 & 1 \\
\ldots\texttt{\_Xe\_vU04\_vO06} & 4.8513 & 4.6013 & 5.1013 & 0 & -0.3010 & 1 \\
\ldots\texttt{\_Xe\_vU04\_vO07} & 5.1513 & 4.9013 & 5.4013 & 0 & -0.3010 & 1 \\
\ldots\texttt{\_Xe\_vU05\_vO04} & 6.7700 & 6.2700 & 7.2700 & 0 & -0.3010 & 1 \\
\ldots\texttt{\_Xe\_vU05\_vO05} & 7.0700 & 6.5700 & 7.5700 & 0 & -0.3010 & 1 \\
\ldots\texttt{\_Xe\_vU05\_vO06} & 7.3700 & 6.8700 & 7.8700 & 0 & -0.3010 & 1 \\
\ldots\texttt{\_Xe\_vU05\_vO07} & 7.6700 & 7.1700 & 8.1700 & 0 & -0.3010 & 1 \\
\ldots\texttt{\_Xe\_vU05\_vO08} & 7.9700 & 7.4700 & 8.4700 & 0 & -0.3010 & 1 \\
\ldots\texttt{\_Xe\_vU06\_vO07} & 9.8300 & 9.3300 & 10.3300 & 0 & -0.3010 & 1 \\
\ldots\texttt{\_Xe\_vU06\_vO08} & 10.1300 & 9.6300 & 10.6300 & 0 & -0.3010 & 1 \\
\ldots\texttt{\_Xe\_vU06\_vO09} & 10.4300 & 9.9300 & 10.9300 & 0 & -0.3010 & 1 \\
\ldots\texttt{\_Xe\_vU07\_vO08} & 5.8600 & 5.3600 & 6.3600 & 0 & -0.3010 & 1 \\
\ldots\texttt{\_Xe\_vU07\_vO09} & 6.1600 & 5.6600 & 6.6600 & 0 & -0.3010 & 1 \\
\ldots\texttt{\_Xe\_vU07\_vO10} & 7.3000 & 6.8000 & 7.8000 & 0 & -0.3010 & 1 \\
\ldots\texttt{\_Xe\_vU07\_vO11} & 7.6000 & 7.1000 & 8.1000 & 0 & -0.3010 & 1 \\
\ldots\texttt{\_Xe\_vU08\_vO08} & 3.1290 & 2.8790 & 3.3790 & 0 & -0.3010 & 1 \\
\ldots\texttt{\_Xe\_vU08\_vO09} & 3.1275 & 2.8775 & 3.3775 & 0 & -0.3010 & 1 \\
\ldots\texttt{\_Xe\_vU08\_vO10} & 3.4347 & 3.1847 & 3.6847 & 0 & -0.3010 & 1 \\
\ldots\texttt{\_Xe\_vU08\_vO11} & 3.7722 & 3.5222 & 4.0222 & 0 & -0.3010 & 1 \\
\ldots\texttt{\_Xe\_vU08\_vO12} & 4.1797 & 3.9297 & 4.4297 & 0 & -0.3010 & 1 \\ \bottomrule
    \end{tabular}
    \vspace{0.5\baselineskip}
    \begin{tabular}{%
            l@{}%
            S[table-format=6.4]%
            c%
            S[table-format=6.4]%
            S[table-format=6.4]%
        }\toprule
        {parameter name} & {nominal value} & {unit} & {lower bound} & {upper bound} \\ \midrule
        \texttt{T0} (Sabioni et al.) & 1973 & [\si{\kelvin}] & 1773 & 2173 \\
\texttt{T0} (Davies \& Long) & 1973 & [\si{\kelvin}] & 1773 & 2173 \\
\texttt{T0} (Turnbull et al.) & 1973 & [\si{\kelvin}] & 1773 & 2173 \\
\texttt{T0} (Miekeley \& Felix) & 1973 & [\si{\kelvin}] & 1773 & 2173 \\
\texttt{Hf\_pO2} (Sabioni et al.) & 5.1 & [\si{\eV}] & 3.85 & 6.35 \\
\texttt{Hf\_pO2} (Davies \& Long) & 5.1 & [\si{\eV}] & 3.85 & 6.35 \\
\texttt{Hf\_pO2} (Turnbull et al.) & 5.1 & [\si{\eV}] & 3.85 & 6.35 \\
\texttt{Hf\_pO2} (Miekeley \& Felix) & 5.1 & [\si{\eV}] & 3.85 & 6.35 \\
\texttt{charge\_correction\_DFT} & 0 & [\si[per-mode=symbol]{\eV\per\square\elementarycharge}] & 0 & 0.0213 \\
\texttt{charge\_sq\_correction\_DFT} & 0 & [\si[per-mode=symbol]{\eV\per\square\elementarycharge}] & 0 & 0.0569 \\
\texttt{log10\_sink\_strength} & -2.2321 & [\si{\per\square\metre}] & -3.2321 & -1.2321 \\
\texttt{log10\_sink\_bias} & 0 & [-] & -1 & 1 \\
\texttt{log10\_source\_strength} & -4 & [\si{\per\second}] & -5 & -3 \\ \bottomrule
    \end{tabular}
    \caption{Nominal values, lower and upper bounds for all parameters in the 183-dimensional model. (\emph{continued})}
\end{table*}

\begin{table*}[p]
    \centering
    \footnotesize
    \begin{tabular}{%
            l@{}%
            S[table-format=6.5]%
            c%
            S[table-format=6.4]%
            S[table-format=6.4]%
            S[table-format=6.4]%
            S[table-format=6.4]%
        }\toprule
        {} & {} & {} & {} & {} & \multicolumn{2}{c}{MAP values} \\
        {parameter name} & {nominal value} & {unit} & {lower bound} & {upper bound} & {unweighted} & {weighted} \\ \midrule
        \texttt{DFT\_h} & -3444.047667 & [\si{\eV}] & -3444.1727 & -3443.7977 & -3443.9877 & -3444.0090 \\
\texttt{DFT\_e} & -3425.118167 & [\si{\eV}] & -3425.2432 & -3424.8682 & -3425.0138 & -3425.0105 \\
\texttt{DFT\_U01\_O02} & -3435.2645 & [\si{\eV}] & -3435.3895 & -3435.1395 & -3435.2646 & -3435.2815 \\
\texttt{S\_U01\_O02} & -129368.8308 & [\si[per-mode=symbol]{\eV\per\kelvin}] & -129370.0808 & -129367.5808 & -129368.9301 & -129368.9920 \\
\texttt{S\_vU01\_vO00} & -129265.9303 & [\si[per-mode=symbol]{\eV\per\kelvin}] & -129267.1803 & -129264.6803 & -129265.9812 & -129265.9563 \\
\texttt{S\_vU00\_vO01} & -129286.6685 & [\si[per-mode=symbol]{\eV\per\kelvin}] & -129287.9185 & -129285.4185 & -129286.3811 & -129286.3336 \\
\texttt{S\_h} & -129374.7936 & [\si[per-mode=symbol]{\eV\per\kelvin}] & -129376.0436 & -129372.2936 & -129373.7106 & -129373.7283 \\
\texttt{S\_e} & -129362.1380 & [\si[per-mode=symbol]{\eV\per\kelvin}] & -129363.3880 & -129359.6380 & -129360.7811 & -129360.7944 \\
\texttt{S\_Ui00\_Oi01} & -129445.2888 & [\si[per-mode=symbol]{\eV\per\kelvin}] & -129446.5388 & -129444.0388 & -129445.5613 & -129445.5858 \\
\texttt{S\_Xe\_vU01\_vO01} & -129264.8194 & [\si[per-mode=symbol]{\eV\per\kelvin}] & -129266.0694 & -129263.5694 & -129264.2927 & -129264.2221 \\
\texttt{S\_Xe\_vU02\_vO01} & -129157.2840 & [\si[per-mode=symbol]{\eV\per\kelvin}] & -129158.5340 & -129156.0340 & -129157.7609 & -129157.7738 \\
\texttt{S\_Xe\_vU04\_vO03} & -128789.5736 & [\si[per-mode=symbol]{\eV\per\kelvin}] & -128790.8236 & -128788.3236 & -128789.5624 & -128789.5619 \\
\texttt{Q\_vU01\_vO00} & 4.233170 & [\si{\eV}] & 3.9832 & 4.4832 & 4.2385 & 4.2393 \\
\texttt{Q\_Xe\_vU02\_vO01} & 3.695676 & [\si{\eV}] & 3.4457 & 3.9457 & 3.5774 & 3.5792 \\
\texttt{Q\_Xe\_vU04\_vO03} & 2.770721 & [\si{\eV}] & 2.5207 & 3.0207 & 2.8886 & 2.8732 \\
\texttt{Q\_Xe\_vU08\_vO09} & 3.127457 & [\si{\eV}] & 2.8775 & 3.3775 & 3.0944 & 3.0949 \\
\texttt{log10\_w\_vU01\_vO00} & 0.152288 & [\si{\Hz}] & -0.1487 & 1.1523 & 0.4609 & 0.4594 \\
\texttt{log10\_w\_Xe\_vU02\_vO01} & 0.511883 & [\si{\Hz}] & 0.2109 & 1.5119 & 0.3125 & 0.2628 \\
\texttt{log10\_w\_Xe\_vU04\_vO03} & 0 & [\si{THz}] & -0.301 & 1 & -0.0867 & -0.0585 \\
\texttt{log10\_w\_Xe\_vU08\_vO09} & 0 & [\si{THz}] & -0.301 & 1 & 0.7639 & 0.6272 \\
\texttt{T0} (Sabioni et al.) & 1973 & [\si{\K}] & 1773 & 2173 & 1946.2082 & 1963.7078 \\
\texttt{T0} (Davies \& Long) & 1973 & [\si{\K}] & 1773 & 2173 & 1930.2581 & 1942.0619 \\
\texttt{T0} (Turnbull et al.) & 1973 & [\si{\K}] & 1773 & 2173 & 2095.7072 & 2070.1338 \\
\texttt{T0} (Miekeley \& Felix) & 1973 & [\si{\K}] & 1773 & 2173 & 2007.1208 & 2014.2658 \\
\texttt{Hf\_pO2} (Sabioni et al.) & 5.1 & [\si{\eV}] & 3.85 & 6.35 & 4.7613 & 4.6997 \\
\texttt{Hf\_pO2} (Davies \& Long) & 5.1 & [\si{\eV}] & 3.85 & 6.35 & 5.1277 & 5.1278 \\
\texttt{Hf\_pO2} (Turnbull et al.) & 5.1 & [\si{\eV}] & 3.85 & 6.35 & 5.1028 & 5.0941 \\
\texttt{Hf\_pO2} (Miekeley \& Felix) & 5.1 & [\si{\eV}] & 4.86 & 7.36 & 6.1647 & 6.1371 \\
\texttt{log10\_sink\_bias} & 0 & [-] & -1 & 1 & 0.2423 & 0.2469 \\
\texttt{log10\_source\_strength} & -4 & [\si{\per\second}] & -5 & -3 & -4.1927 & -4.2061 \\ \bottomrule
    \end{tabular}
    \caption{Nominal values, lower and upper bounds, and MAP values with unweighted and weighted likelihoods for all estimated parameters.}
    \label{tab:uncertain_parameters}
\end{table*}

\clearpage

\begin{algorithm}[t]
    \begin{minipage}{\textwidth}
        \begin{algorithmic}[1]
            \Statex \textbf{input:} prior $p(\bsnu)$, likelihood $\calL(\bsnu)$, starting value $\bsnu_0$,
            \Statex \hspace*{\WidthOfInput} number of samples $M$
            \Statex \textbf{output:} a set of samples $\{\bsnu_m\}_{m=1}^M$ from the posterior $p(\bsnu | \calZ_d)$
            \Statex
            \Procedure{\textsf{MCMC}}{$p(\bsnu), \calL(\bsnu), \bsnu_0, M$}
                \State $\mathsf{p}_0 \gets \calL(\bsnu_0) p(\bsnu_0)$ \Comment{evaluate the posterior at $\bsnu_0$}
                \For {$m$ \textbf{from} 1 \textbf{to} $M$}
                    \State $\bsnu_{\textrm{prop}} \gets \bsnu_{m-1} + \Delta \bsnu$ \Comment{generate a new proposal}
                    \State $\mathsf{p}_{\textrm{prop}} \gets \calL(\bsnu_{\textrm{prop}}) p(\bsnu_{\textrm{prop}})$ \Comment{evaluate the posterior at $\bsnu_{\textrm{prop}}$}
                    \State $\alpha \gets \min(1, \mathsf{p}_{\textrm{prop}} / \mathsf{p}_{m-1})$ \Comment{compute the acceptance ratio}
                    \State $u \gets \calU(0, 1)$ \Comment{sample a uniformly distributed random number}
                    \If {$u < \alpha$} \label{line:metropolis-hastings}
                        \State $\bsnu_m \gets \bsnu_{\textrm{prop}}$ and $\mathsf{p}_m \gets \mathsf{p}_{\textrm{prop}}$ \Comment{accept the new proposal}
                    \Else
                        \State $\bsnu_m \gets \bsnu_{m-1}$ and $\mathsf{p}_m \gets \mathsf{p}_{m-1}$ \Comment{reject the new proposal}
                    \EndIf
                \EndFor
            \EndProcedure
        \end{algorithmic}
        \caption{MCMC procedure}
        \label{alg:mcmc}
    \end{minipage}
\end{algorithm}

\section{Markov chain Monte Carlo}\label{sec:markov_chain_monte_carlo}

A procedure for MCMC sampling is shown in~\cref{alg:mcmc}. The algorithm performs a total of $M$ MCMC iterations. In each step, we propose a jump $\Delta \bsnu$ to explore the $s$-dimensional parameter space. The proposal is accepted when it passes the conventional Metropolis--Hastings acceptance test in step~\ref{line:metropolis-hastings}, here illustrated for the special case of a symmetric proposal distribution, see~\cite{metropolis1953}. The MCMC algorithm works by generating a sequence of candidates, in such a way that the distribution of these candidates approximates the desired distribution for $M \rightarrow \infty$. We used the adaptive MCMC (AMCMC) method proposed in \cite{haario2001,atchade2005}. In AMCMC, the proposed parameter vector $\bsnu_{\textrm{prop}}$ is sampled from a multivariate Gaussian centered at the current state $\bsnu_{m-1}$ with covariance $\cov(\bsnu_0, \bsnu_1, \ldots, \bsnu_{m-1})$. Crucially, the covariance of the proposal distribution depends on the history of the chain. Adaptive formulae that update the estimate for the covariance matrix as more and more proposals are evaluated have been provided in \cite{haario2001}.

\section{Trace plots}\label{sec:trace_plots}

\begin{figure*}[t]
    \centering
	\tikzsetnextfilename{parameter_chains}
	\tikzexternalenable
	\newcommand{\plotchain}[2]{%
    \nextgroupplot[%
        title = {#2}
    ]
    \addplot graphics[xmin=0, xmax=1, ymin=-1.5, ymax=1.5] {plt/chain_#1.png};
    \pgfplotstablegetelem{#1}{0}\of{\map}\edef\mapval{\pgfplotsretval}
    \addplot[map] coordinates {(0, \mapval) (1, \mapval)};
}

\setlength{\figurewidth}{6.5cm}
\setlength{\figureheight}{3cm}
\pgfplotstableread[header=false]{dat/map_params.dat}{\map}
\begin{tikzpicture}[
        map/.style={default dashed line, White}
	]
	\begin{groupplot}[%
            group style={%
                group size=3 by 10,
                xticklabels at=edge bottom,
                xlabels at=edge bottom,
                yticklabels at=edge left,
                ylabels at=edge left,
                vertical sep=1.5\baselineskip,
                horizontal sep=.5\baselineskip
            },
            default axis,
            axis on top,
            xmin = 0, xmax = 1, ymin = -1.5, ymax=1.5,
            xlabel = {\# MCMC iterations $M$ ($\times 10^6$)},
            ytick = {-1, 0, 1},
            every axis title/.style={at={(0.5,0.95)}, anchor=south, font=\small, align=center}
        ]
        \plotchain{0}{\texttt{DFT\_h}}
        \plotchain{1}{\texttt{DFT\_e}}
        \plotchain{2}{\texttt{DFT\_U01\_O02}}
        \plotchain{3}{\texttt{S\_U01\_O02}}
        \plotchain{4}{\texttt{S\_vU01\_vO00}}
        \plotchain{5}{\texttt{S\_vU00\_vO01}}
        \plotchain{6}{\texttt{S\_h}}
        \plotchain{7}{\texttt{S\_e}}
        \plotchain{8}{\texttt{S\_Ui00\_Oi01}}
        \plotchain{9}{\texttt{S\_Xe\_vU01\_vO01}}
        \plotchain{10}{\texttt{S\_Xe\_vU02\_vO01}}
        \plotchain{11}{\texttt{S\_Xe\_vU04\_vO03}}
        \plotchain{12}{\texttt{Q\_vU01\_vO00}}
        \plotchain{13}{\texttt{Q\_Xe\_vU02\_vO01}}
        \plotchain{14}{\texttt{Q\_Xe\_vU04\_vO03}}
        \plotchain{15}{\texttt{Q\_Xe\_vU08\_vO09}}
        \plotchain{16}{\texttt{log10\_w\_vU01\_vO00}}
        \plotchain{17}{\texttt{log10\_w\_Xe\_vU02\_vO01}}
        \plotchain{18}{\texttt{log10\_w\_Xe\_vU04\_vO03}}
        \plotchain{19}{\texttt{log10\_w\_Xe\_vU08\_vO09}}
        \plotchain{20}{\texttt{T0} (Sabioni et al.)}
        \plotchain{21}{\texttt{T0} (Davies \& Long)}
        \plotchain{22}{\texttt{T0} (Turnbull et al.)}
        \plotchain{23}{\texttt{T0} (Miekeley \& Felix)}
        \plotchain{24}{\texttt{Hf\_pO2} (Sabioni et al.)}
        \plotchain{25}{\texttt{Hf\_pO2} (Davies \& Long)}
        \plotchain{26}{\texttt{Hf\_pO2} (Turnbull et al.)}
        \plotchain{27}{\texttt{Hf\_pO2} (Miekeley \& Felix)}
        \plotchain{28}{\texttt{log10\_sink\_bias}}
        \plotchain{29}{\texttt{log10\_source\_strength}}
	\end{groupplot}
\end{tikzpicture}
	\tikzexternaldisable

    \caption{Trace plot of the Markov chain showing the state of each parameter as a function of the iteration number. The MAP value is indicated by the dashed line. Note that the vertical axes represent the parameter value scaled to $[-1, 1]$. To obtain the original parameter values, these must be rescaled to the lower and upper bounds provided in \cref{tab:all_parameters_overview}.}
    \label{fig:parameter_chains}
\end{figure*}
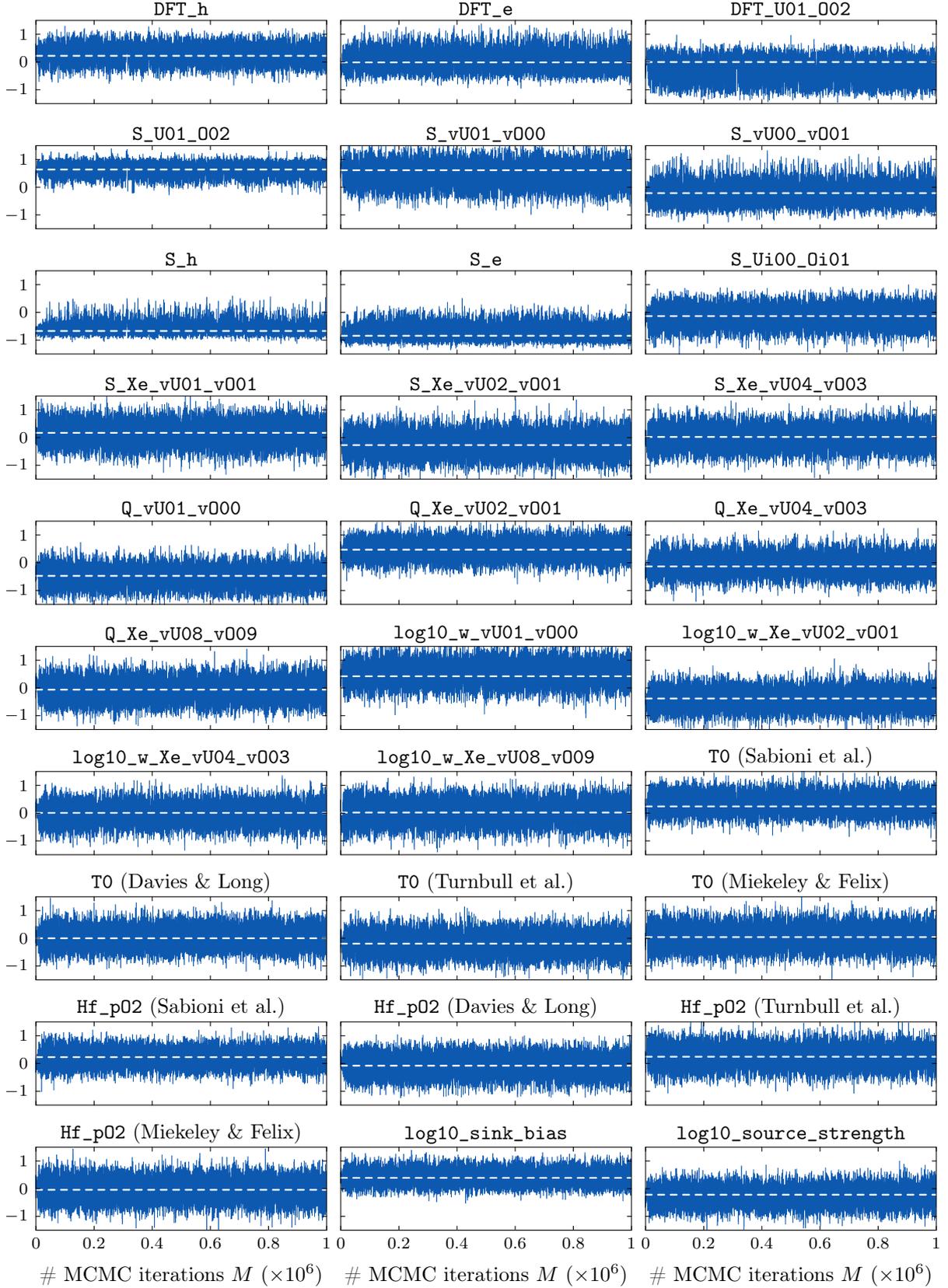

A trace plot of the Markov chain for each of the 30 parameters using the unweighted likelihood is shown in \cref{fig:parameter_chains}. The chain appears to be well-mixed.

\end{appendices}

\end{document}